\documentclass[12pt]{article}

\ifx\pdfoutput\undefined
\usepackage[dvips,bookmarks]{hyperref}
\else
\usepackage{hyperref}
\fi
\hypersetup{colorlinks=false,bookmarksopen,bookmarksnumbered,citecolor=blue,
   pdfstartview=FitH}

\usepackage{latexsym}

\usepackage{amssymb,amsfonts,amsmath}
\usepackage{graphicx} 
\usepackage{indentfirst}

 \usepackage{bbm}

\parskip=\medskipamount

\newcommand {\cC}{{\cal C}}
\newcommand {\cD}{{\cal D}}
\newcommand {\cE}{{\cal E}}
\newcommand {\cF}{{\cal F}}

\newcommand {\cH}{{\cal H}}

\newcommand {\cJ}{{\cal J}}
\newcommand {\cK}{{\cal K}}
\newcommand {\cL}{{\cal L}}
\newcommand {\cM}{{\cal M}}

\newcommand {\cN}{{\cal N}}
\newcommand {\cO}{{\cal O}}

\newcommand {\cS}{{\cal S}}
\newcommand {\cT}{{\cal T}}
\newcommand {\cU}{{\cal U}}

\newcommand{\bL}{{\bf L}}

\newcommand{\bR}{{\bf R}}

\def\a{\alpha}

\def\b{\beta}
\def\c{\chi}
\def\d{\delta}
\def\e{\epsilon}

\def\g{\gamma}
\def\G{\Gamma}

\def\j{\psi}
\def\k{\kappa}
\def\l{\lambda}
\def\m{\mu}
\def\n{\nu}

\def\p{\pi}
\def\q{\theta}
\def\r{\rho}
\def\s{\sigma}
\def\t{\tau}

\def\x{\xi}
\def\z{\zeta}
\def\D{\Delta}
\def\F{\Phi}
\def\J{\Psi}

\def\O{\Omega}

\def\S{\Sigma}
\def\U{\Upsilon}
\def\X{\Xi}
\def\tr{{\rm tr}}
\def\rd{{\rm d}}
\def\ri{{\rm i}}

\newcommand{\ad}{{\dot{\alpha}}}                           
\newcommand{\bd}{{\dot{\beta}}}                            
\newcommand{\ve}{\varepsilon}                            
\newcommand{\cDB}{{\bar\cD}}                            

\newcommand{\pa}{\partial}                           
\newcommand{\hf}{\frac12}

\newcommand{\vf}{\varphi}

\newcommand{\be}{\begin{equation}}
\newcommand{\ee}{\end{equation}}
\newcommand{\bea}{\begin{eqnarray}}
\newcommand{\eea}{\end{eqnarray}}
\newcommand{\non}{\nonumber}
\newcommand{\ba}{\begin{array}}
\newcommand{\ea}{\end{array}}

\newcommand{\1}{\underline{1}}

\newcommand{\mun}{\underline{m}}

\def\dt#1{{\buildrel {\hbox{\LARGE .}} \over {#1}}}    

\newcommand{\bm}[1]{\mbox{\boldmath$#1$}}

\def\double #1{#1{\hbox{\kern-2pt $#1$}}}

\newcommand{\bsubeq}{\begin{subequations}}
\newcommand{\esubeq}{\end{subequations}}

\newcommand{\qb}{{\bar{\theta}}}

\newcommand{\bai}{{\bar i}}
\newcommand{\baj}{{\bar j}}
\newcommand{\bak}{{\bar k}}
\newcommand{\bal}{{\bar l}}

\newcommand{\bau}{{\bar 1}}
\newcommand{\bad}{{\bar 2}}

\newcommand{\rL}{{\rm L}}
\newcommand{\rR}{{\rm R}}

\newcommand{\de}{{\nabla}}
\newcommand{\deb}{{\bar{\nabla}}}

\begin{document}
\begin{titlepage}
\begin{flushright}
UUITP-39/10\\
January, 2011\\
\end{flushright}

\begin{center}
{\Large \bf 
Off-shell supergravity-matter couplings\\
 in three dimensions}\\ 
\end{center}

\begin{center}

{\bf
Sergei M. Kuzenko\footnote{kuzenko@cyllene.uwa.edu.au}${}^{a}$,
Ulf Lindstr\"om\footnote{ulf.lindstrom@physics.uu.se}${}^{b}$ and
Gabriele Tartaglino-Mazzucchelli\footnote{gabriele.tartaglino-mazzucchelli@physics.uu.se}${}^{b}$
} \\
\vspace{5mm}

\footnotesize{
${}^{a}${\it School of Physics M013, The University of Western Australia\\
35 Stirling Highway, Crawley W.A. 6009, Australia}}  
~\\
\vspace{2mm}

\footnotesize{
${}^{b}${\it Theoretical Physics, Department of Physics and Astronomy,
Uppsala University \\ 
Box 516, SE-751 20 Uppsala, Sweden}
}
\vspace{2mm}

\end{center}

\begin{abstract}
\baselineskip=14pt
We develop the superspace geometry of ${\cal N}$-extended conformal supergravity
in three space-time dimensions. General off-shell supergravity-matter couplings
are constructed in the cases ${\cal N} \leq 4$.
\end{abstract}

\vfill
\end{titlepage}

\newpage
\renewcommand{\thefootnote}{\arabic{footnote}}
\setcounter{footnote}{0}

\tableofcontents{}
\vspace{1cm}
\bigskip\hrule


\section{Introduction}
\setcounter{equation}{0}

Supergravity in three space-time dimensions was introduced as early as  1977 \cite{HT,BG}.
Its simplest version with two supercharges  ($\cN=1$) and the corresponding matter couplings
became a textbook subject by 1983 \cite{GGRS}. In the mid-1980s, topologically massive 
$\cN$-extended
supergravity theories were constructed \cite{DK,Deser,vN,RvN,Lindstrom:1989eg} 
in which  supersymmetric
Lorentz Chern-Simons terms were interpreted as extended conformal supergravity.
More recently, (gauged) nonlinear sigma-models with $\cN$ local supersymmetries 
were constructed in the on-shell component  approach \cite{dWNT,dWHS}.
The on-shell approach was also used in \cite{Chu:2009gi,Gran:2008qx}
to construct $\cN=6$ and $\cN=8$ conformal supergravities and their couplings to 
ABJM type theories and BLG M2-branes respectively.
Surprisingly, to the best of our knowledge, no results have appeared on general 
off-shell supergravity-matter 
couplings in the interesting  cases $\cN=3$ and $\cN=4$. It is clear that such results 
should be based on appropriate superspace techniques,  and the latter have not yet been 
developed. 
The present paper is aimed at filling the existing gap.

Recently, there have appeared exciting results on massive 3D supergravity 
\cite{Andringa:2009yc,Bergshoeff:2010mf,Bergshoeff:2010ui} 
which is a supersymmetric extension of the so-called new massive 3D gravity 
\cite{Bergshoeff:2009hq}. A unique feature of this approach is that the (super)gravity action 
is a parity-preserving higher-derivative variant of 3D (super)gravity which respects unitarity. 
So far only the $\cN=1$ massive supergravity version has been fully elaborated 
\cite{Bergshoeff:2010mf}, and linearized results are available, e.g.,  in the 
$\cN=4$ case \cite{Bergshoeff:2010ui}. To go beyond the linearized approximation, one option 
is to develop, as mentioned in \cite{Bergshoeff:2010ui},  an $\cN$-extended superconformal 
tensor calculus.   We believe, however, that developing superspace techniques may lead to a 
more adequate setting, at least in the cases $\cN\leq 4$.

As regards the cases  $\cN=3$ and $\cN=4$, our approach is a natural generalization 
of the off-shell 
formulations for general supergravity-matter theories with eight supercharges 
in five \cite{KT-M5D-1,KT-M5D-2} and four  \cite{KLRT-M1,K-08,KLRT-M2} 
dimensions.\footnote{Similar
ideas have been developed in two dimensions \cite{TartaglinoMazzucchelli:2009ip}.}
These formulations 
build in part on the 
techniques from projective superspace which were originally developed 
for extended Poincar\'e supersymmetry in \cite{KLR,LR-projective,LR-projective2} 
(see also \cite{LR10} for a review).\footnote{The term ``projective superspace'' was coined in 1990
\cite{LR-projective2}.
The modern projective-superspace terminology was introduced in 1998 \cite{G-RRWLvU}.}
The matter couplings in  \cite{KT-M5D-1,KT-M5D-2,KLRT-M1,K-08,KLRT-M2} are described 
in terms of the so-called covariant projective multiplets which are curved-superspace extensions
of the superconformal projective multiplets introduced for the first time 
in  \cite{K-compactified,K-hyper}.\footnote{General superconformal couplings of 
projective multiplets 
has also been given in  \cite{K-compactified,K-hyper}.}
This is in accord with the general principle that matter couplings in Poincar\'e supergravity
can equivalently be described as 
conformal supergravity coupled to superconformal matter \cite{KT,deWvHV,deWPV}.
In three dimensions, therefore, a first step toward developing superspace settings for $\cN=3,~4$
supergravity theories should consist in  a construction of superconformal projective multiplets 
and their self-couplings.
This has recently been achieved as part  of
more general results on off-shell 3D
$\cN \leq 4$ rigid superconformal  sigma-models  \cite{KPT-MvU}.

Projective superspace \cite{KLR,LR-projective,LR-projective2}
is less well  known than  harmonic superspace \cite{GIKOS,GIOS}.
For any number of space-time dimensions in which they exist, ${\rm D}\leq 6$, 
the projective and the harmonic superspace approaches use
the same supermanifold. For instance, in the case of 4D $\cN=2$ supersymmetry, 
they make use of the isotwistor superspace ${\mathbb R}^{4|8} \times {\mathbb C}P^1$ 
introduced originally by Rosly \cite{Rosly}. 
The relationship between the {\it rigid} harmonic and projective superspace formulations
is spelled out in \cite{K-double} (see also \cite{Kuzenko:2010bd} for a recent review).
Essentially, they differ in  (i) the structure of  off-shell   supermultiplets used; 
and (ii) the  supersymmetric action principle chosen. This makes the two approaches 
rather complementary. As emphasized in \cite{KT-M5D-1,KT-M5D-2,KLRT-M1,KLRT-M2},
the difference deepens in the context of supergravity.
Projective superspace is suitable  for developing covariant geometric formulations 
for supergravity-matter systems  \cite{KT-M5D-1,KT-M5D-2,KLRT-M1,KLRT-M2},  
similar to the famous Wess-Zumino approach for 
 4D $\cN=1$ supergravity \cite{WZ,Zumino78}. 
Harmonic superspace offers prepotential formulations \cite{SUGRA-har,GIOS},
similar to the Ogievetsky-Sokatchev approach to 4D $\cN=1$ supergravity 
\cite{Ogievetsky:1978mt}.

In the case of 3D rigid supersymmetry, $\cN=3$ and $\cN=4$ harmonic superspaces
were introduced by Zupnik  \cite{ZH,Zupnik2,Zupnik3,Zupnik4}. This approach was used, 
in particular, 
to describe ABJM models in $\cN=3$ harmonic superspace \cite{Buchbinder:2008vi}.
No harmonic superspace formulation for supergravity in three dimensions has yet been 
constructed. Three-dimensional  $\cN=3$ and $\cN=4$ projective superspace approaches 
have recently been 
developed \cite{KPT-MvU} to describe general superconformal field theories.
It should be mentioned that the 3D $\cN=4$ projective superspace 
${\mathbb R}^{3|8} \times {\mathbb C}P^1$ was introduced by Lindstr\"om and Ro\v{c}ek 
in 1988 \cite{LR-projective} as a direct generalization of their four-dimensional construction 
\cite{KLR}. It follows  from the analysis in \cite{KPT-MvU}
that two mirror copies of ${\mathbb C}P^1$ are required to provide a natural superspace setting 
for general off-shell  $\cN=4$ supermultiplets. 

In this paper, the 3D $\cN \leq 4$ supergravity-matter couplings are formulated in terms of 
superspace and superfields.  
The  issue of component reduction will be addressed in a separate  publication.

This paper is organized as follows. In section 2 we develop 
the superspace geometry of $\cN$-extended conformal supergravity
in three space-time dimensions. Matter couplings in supergravity theories with $\cN=1,2,3$ and 4 
are studied, on a case-by-case basis,  in sections 3 to 6. Our final comments and conclusions
are given in section 7. The main body of the paper is accompanied by two appendices. 
Our 3D notation and conventions are collected in Appendix A.
Appendix B is devoted to the derivation of the left projection operator.

\section{Geometry of $\cN$-extended conformal supergravity}
\setcounter{equation}{0}

In this section we develop a formalism of differential geometry 
in a curved three-dimensional  $\cN$-extended 
superspace, which is locally parametrized by real   bosonic ($x^m$) 
and real fermionic ($\q^{\mu}_{\tt I}$) coordinates
\bea
z^{M}=(x^m,\q^{\mu}_{\tt I})~,
 \qquad m=0,1,2
~,~~~
\mu=1,2
~,~~~
{\tt I}
={\bf 1},\cdots,{\bm \cN}~,~~
\eea
that is suitable to describe $\cN$-extended conformal supergravity.
A natural condition upon such a geometry is that it should reduce to that 
of $\cN$-extended Minkowski superspace ${\mathbb R}^{3|2\cN}$ in a flat limit.
We recall that
the  spinor covariant derivatives $D_\a^I$ 
associated with  Minkowski superspace
satisfy the anti-commutation relations 
\bea
\{D_\a^{I},D_\b^{J}\}=2\ri \, \d^{IJ}(\g^c)_{\a\b}\,\pa_c~.
\eea
An explicit realization of  $D_\a^I$ is 
\bea
D_\a^I=\frac{\pa}{\pa\q^\a_I}+\ri \,
 (\s^b)_{\a \b} \, \q^{\b}_I \,\pa_{b}~.
\eea
As usual, there is no need to distinguish  between upper and lower SO($\cN$) vector indices.

As compared to the 4D supersymmetry, the three-dimensional case  has an important  specific  
feature which is due to the fact that 3D spinors are real. 
This feature is the conjugations rule:
given a superfield $F$ of Grassmann parity $\e (F)$, it holds that  
\bea
(D_\a^I F)^*=-(-)^{\e(F)}D_\a^I \bar{F}~,
\label{c.c.1}
\eea
with $\bar F:=(F)^*$ the complex conjugate of $F$.

\subsection{The algebra of covariant derivatives}
\label{algebra of covariant derivatives}
We choose the structure group to be ${\rm SL}(2,{\mathbb R}) \times {\rm SO} (\cN)$, 
and denote by $\cM_{ab}=-\cM_{ba}$ and $\cN_{IJ} =-\cN_{JI}$ the corresponding generators.
The covariant derivatives have the form:
\bea
\cD_{A}&\equiv& (\cD_a, \cD^I_\a )= 
E_{A}
+\O_A
+\F_A
~.
\eea
Here $E_A=E_A{}^M(z)\pa_M$ is the supervielbein, with $\pa_M=\pa/\pa z^M$,
\bea
\O_A=\hf\O_{A}{}^{bc}\cM_{bc}=-\O_{A}{}^b\cM_b=\hf\O_{A}{}^{\b\g}\cM_{\b\g}~,
~~~~
\cM_{ab}=-\cM_{ba}~,~~\cM_{\a\b}=\cM_{\b\a}
\eea
is the Lorentz connection, and 
\bea
\F_A=\hf\F_A{}^{KL}\cN_{KL}~,~~~~\cN_{KL}=-\cN_{LK}
\eea
is the SO($\cN$)-connection.
The Lorentz generators with two vector indices ($\cM_{ab}$), with one vector index ($\cM_a$)
and with two spinor indices ($\cM_{\a\b}$) are related to each other by the rules:
$\cM_a=\hf \ve_{abc}\cM^{bc}$ and $\cM_{\a\b}=(\g^a)_{\a\b}\cM_a$ 
(for more details see Appendix \ref{3Dconventions}).
The generators of SL(2,${\mathbb R}$)$\times$SO($\cN$) act on the covariant derivatives as 
follows:\footnote{The operation of (anti)symmetrization of $n$ indices is defined to involve 
a factor of $(n!)^{-1}$.}
\bsubeq
\bea
&&{\big [}{\cal M}_{ab} ,\cD_{\a}^{I}{\big ]} =\hf \ve_{abc}(\g^c)_\a{}^\b\cD_\b^{I}~,~~
{\big [}  {\cal M}_a,\cD_{\a}^{I} {\big]} =-\hf(\g_a)_\a{}^\b\cD_\b^{I} 
~,~~ 
~~~~~~~
\\
&&
{\big [}{\cal M}_{\a\b} ,\cD_{\g}^{I}{\big ]} =\ve_{\g(\a}\cD_{\b)}^{I}~,~~~
{\big [} {\cal M}_{ab} , \cD_{c}  {\big]} =2\eta_{c[a}\cD_{b]}~,~~~
{\big [} {\cal M}_{a} , \cD_{b}  {\big]} =\ve_{abc}\cD^{c}~,~~~~~~
\label{acM}
\\
&&
{\big [} {\cal N}{}_{KL},\cD_{\a}^{I}{\Big]} =2\d^I_{[K}\cD_{\a L]} ~,~~
{\big [} {\cal N}{}_{KL},\cD_{a}{\Big]} =0 ~.
~~~~~~~~~
\eea
\esubeq

The supergravity gauge group is generated by local transformations of the form
\bea
\d_K\cD_A=[K,\cD_A]~,~~~~~~
K=K^C(z)\cD_C+\hf K^{cd}(z)\cM_{cd}+\hf K^{PQ}(z)\cN_{PQ}
~,
\label{SUGRA-gauge-group1}
\eea
with all the gauge parameters obeying natural reality conditions but otherwise arbitrary.
Given a tensor superfield $T(z)$, it transforms as follows:
\bea
\d_KT=KT ~.
\label{SUGRA-gauge-group2}
\eea

The covariant derivatives satisfy the (anti)commutation relations
\bea
{[}\cD_{{A}},\cD_{{B}}\}&=&T_{{A}{B}}{}^{{C}}\cD_{{C}}
+\hf R_{AB}{}^{KL}\cN_{KL}
+\hf R_{{A}{B}}{}^{cd}\cM_{cd}
~,
\label{algebra-4-2-N}
\eea
with  $T_{AB}{}^C$ the torsion,  $R_{AB}{}^{cd}$ the Lorentz curvature and
 $R_{AB}{}^{KL}$ the SO($\cN$) curvature.
 The torsion and the curvature are related to each other by 
the  Bianchi identities:
\bea
\sum_{[ABC)}{[}\cD_{{A}},{[}\cD_{{B}},\cD_{{C}}\}\} =0~.
\eea
To describe conformal supergravity, we impose 
conventional constraints on the torsion.
They are:
\bsubeq
\bea
T_{\a}^I{\,}_{\b}^J{\,}^c=2\ri\d^{IJ}(\g^{c})_{\a\b}~,~~~
&&~~~~~~({\rm dimension~0})
\label{constr-0-4-N}
\\
T_{\a}^I{\,}_{\b}^J{\,}^{\g}_K=0 ~, \qquad T_{\a}^I{\,}_{b}{\,}^c=0~,~~~
&&~~~~~~({\rm dimension~1/2})
\label{constr-1/2-4-N}
\\
T_{ab}{}^{c}=0~, \qquad
\ve^{\b\g}T_{a}{}_\b^{[J}{}_\g^{K]}=0
~.~~~
&&~~~~~~({\rm dimension~1})
\label{constr-1-4-N}
\eea
\esubeq
We emphasize 
that for any $\cN$ the torsion has  no dimension-1/2 components (this differs from Howe's 
formulation for 4D $\cN$-extended conformal supergravity \cite{Howe}).
The above constraints have been introduced in \cite{HIPT}. 
However, no explicit solution to the Bianchi identities has been given in \cite{HIPT}.
Solutions to some of the Bianchi identities are implicit in the results of \cite{HIPT}.

Under the conventional constraints (\ref{constr-0-4-N})--(\ref{constr-1-4-N}),
the solution to the Bianchi identities is given by the following algebra of covariant 
derivatives\footnote{The results are presented to dimension-3/2 in the torsion and curvature. 
We plan to give a complete solution to the Bianchi identities elsewhere.}:
\bsubeq
\bea
\{\cD_\a^I,\cD_\b^J\}&=&
2\ri\d^{IJ}(\g^c)_{\a\b}\cD_c
-2\ri\ve_{\a\b}C^{\g\d}{}^{IJ}\cM_{\g\d}
-4\ri S^{IJ}\cM_{\a\b}
\non\\
&&
+\Big(
\ri\ve_{\a\b}X^{IJKL}
-4\ri\ve_{\a\b}S^{K}{}^{[I}\d^{J]L}
+\ri C_{\a\b}{}^{KL}\d^{IJ}
-4\ri C_{\a\b}{}^{K(I}\d^{J)L}
\Big)
\cN_{KL}
~,~~~~~~~~~
\label{alg-1}
\\
{[}\cD_{\a\b},\cD_\g^K{]}
&=&
-\Big(
\ve_{\g(\a}C_{\b)\d}{}^{KL}
+\ve_{\d(\a}C_{\b)\g}{}^{KL}
+2\ve_{\g(\a}\ve_{\b)\d}S^{KL}
\Big)
\cD^\d_L
\non\\
&&
+\hf R_{\a\b}{}_\g^K{}_{\d\r}\cM^{\d\r}
+\hf R_{\a\b}{}_\g^K{}^{PQ}\cN_{PQ}
\label{alg-3/2}
~.
\eea
\esubeq
Here all the dimension-1 components are  real and satisfy the symmetry properties
\bea
X^{IJKL}=X^{[IJKL]}~,~~~~
S^{IJ}=S^{(IJ)}
~,~~~~
C_a{}^{IJ}=C_a{}^{[IJ]}
~.
\eea
The torsion superfield of $S^{IJ}$ can be decomposed
into its trace and traceless parts as
\bea
S^{IJ}=\cS\d^{IJ}+\cS^{IJ}~,~~~~~~
\cS=\frac{1}{\cN}\d_{IJ}S^{IJ}~,~~
\d_{IJ}\cS^{IJ}=0~.
\eea
The dimension-3/2 components of the torsion and the curvature are
\bsubeq
\bea
T_{ab}{}^\g_K &=&
\hf\ve_{abc}(\g^c)^{\a\b}\ve^{\g\d}T_{\a\b}{}_\d^K~, \non \\
T_{\a\b}{}_\g^{K}
&=&
\ri C_{\a\b\g}{}^K
-{4\ri\over 3}\ve_{\g(\a}(\cD_{\b)}^{K}\cS)
-{4(\cN-1)\ri\over 3\cN}\ve_{\g(\a}\cS_{\b)}{}^{K}
~,
\\
R_{\a\b}{}_\g^K{}_{\d\rho}&=&
4\ve_{\g(\a}C_{\b)\d\r}{}^K
+\frac{16}{3}\ve_{\g(\d}\ve_{\r)(\a}(\cD_{\b)}^K\cS)
-\frac{4}{3}\ve_{\a(\d}\ve_{\r)\b}(\cD_{\g}^K\cS)
\non\\
&&
+{16(\cN-1)\over 3\cN}\ve_{\g(\d}\ve_{\r)(\a}\cS_{\b)}{}^{K}
-{4(\cN-1)\over 3\cN}\ve_{\a(\d}\ve_{\r)\b}\cS_{\g}{}^{K}
~,
\\
R_{\a\b}{}_\g^K{}^{PQ}
&=&
{2\over 3}\ve_{\g(\a}\Big(
-2C_{\b)}{}^{KPQ}
+3\cT_{\b)}{}^{PQK}
+8(\cD_{\b)}^{[P} \cS)\d^{Q]K}
+{(5\cN-8)\over \cN}\cS_{\b)}{}^{[P}\d^{Q]K}
\Big)
\non\\
&&
+C_{\a\b\g}{}^{KPQ}
+2C_{\a\b\g}{}^{[P}\d^{Q]K} 
~.
\eea
\esubeq
The superfields $C_{\a\b\g}{}^{KPQ},C_{\a\b\g}{}^{K},\cT_\a{}^{KPQ},\cS_\a{}^K$ are defined 
through the differential constraints satisfied by the dimension-1 torsion and 
curvature superfields. At dimension-3/2 the Bianchi identities imply
\bsubeq
\bea
\cD_\a^{I} \cS^{JK}&=&
2\cT_{\a}{}^{I(JK)}
+\cS_\a{}^{(J}\d^{K)I}
-{1\over \cN}\cS_\a{}^{I}\d^{JK}~,
\label{22.18a} \\
\cD_{\a}^{I} C_{\b\g}{}^{JK}
&=&
{2\over 3}\ve_{\a(\b}\Big(
C_{\g)}{}^{IJK}
+3\cT_{\g)}{}^{JKI}
+4(\cD_{\g)}^{[J} \cS)\d^{K]I}
+{(\cN-4)\over \cN}\cS_{\g)}{}^{[J}\d^{K]I}
\Big)
\non\\
&&
+C_{\a\b\g}{}^{IJK}
-2 C_{\a\b\g}{}^{[J}\d^{K]I}
~,
\label{22.18b}\\
\cD_\a^{I}X^{JKLP}
&=&
X_\a{}^{IJKLP}-4C_{\a}{}^{[JKL}\d^{P]I}
\label{22.18c}
~.
\eea
\esubeq
The symmetry properties of the superfields
$\cT_{\a}{}^{IJK},C_{\a\b\g}{}^{IJK},C_{\a\b\g}{}^{I},X_\a{}^{IJKPQ}$
are
\bsubeq
\bea
&\cT_{\a}{}^{IJK}=\cT_{\a}{}^{[IJ]K}~, \qquad
\d_{JK}\cT_{\a}{}^{IJK}=
\cT_{\a}{}^{[IJK]}=0
~,
\\
&C_{\a\b\g}{}^{IJK}=C_{(\a\b\g)}{}^{IJK}=C_{\a\b\g}{}^{[IJK]}~,
\qquad
C_{\a\b\g}{}^{I}=C_{(\a\b\g)}{}^{I}~,
\\
&C_{\a}{}^{IJK}=C_{\a}{}^{[IJK]}~,
\qquad
X_\a{}^{IJKPQ}=X_\a{}^{[IJKPQ]}
~.
\eea
\esubeq
A remarkable result in superfield supergravity is Dragon's theorem \cite{Dragon}
which states that 
the curvature is completely determined by the torsion. 
More precisely, this result concerns the Lorentz curvature
and it does not necessarily apply to the curvature associated with the $R$-symmetry subgroup 
of the structure group. 
This is exactly what happens in three dimensions for $\cN\geq 4$.
The antisymmetric tensor $X^{IJKL}$ 
appears only in the SO($\cN$) curvature but not as a component of  the torsion.

In this paper we will often use the well-known rule for integration by parts  in superspace:
given a vector superfield $V=V^A\cD_A$,  it holds that 
\bea
\int\rd^3x\,\rd^{2\cN}\q\, E \,(-1)^{\ve_A} \Big\{ \cD_AV^A
- (-1)^{\ve_B}T_A{}_B{}^B\,V^A\Big\}=0
~,\qquad
E^{-1}= {\rm Ber}(E_A{}^M)
~.~~~~~~
\eea
In particular, the fact that the torsion has no dimension-1/2 components
implies the following useful result:
\bea
\int\rd^3x\,\rd^{2\cN}\q \,E\, \cD_\a^IV^\a_I=0
~.
\eea


\subsection{Super-Weyl transformations}
\label{Super-Weyl transformations}
The constraints (\ref{constr-0-4-N})--(\ref{constr-1-4-N}) 
can be shown to be invariant under arbitrary super-Weyl transformations of the form
\bsubeq
\bea
\d_\s\cD_\a^I&=&
\hf \s\cD_\a^I + (\cD^{\b I}\s)\cM_{\a\b}+(\cD_{\a J} \s)\cN^{IJ}
~,
\label{N-sW-1}
\\
\d_\s\cD_a&=&
\s\cD_a
+{\ri\over 2}(\g_a)^{\g\d}(\cD_{\g}^K \s)\cD_{\d K}
+\ve_{abc}(\cD^b\s)\cM^{c}
\non\\
&&
+{\ri\over 16}(\g_a)^{\g\d}([\cD_\g^{K},\cD_\d^{L}]\s)\cN_{KL}~,
\label{N-sW-2}
\eea
where $\s$ is a real unconstrained superfield. This leads to
\bea
\d_\s S^{IJ}&=&
\s S^{IJ}
-{\ri\over 8}([\cD^{\g(I},\cD_{\g}^{J)}]\s)
~,
\label{N-sW-S}
\\
\d_\s C_{a}{}^{IJ}&=&
\s C_{a}{}^{IJ}
-{\ri\over 8}(\g_a)^{\g\d}([\cD_\g^{[I},\cD_\d^{J]}]\s)
~,
\label{N-sW-C}
\\
\d_\s X^{IJKL}&=&\s X^{IJKL}~.
\label{N-sW-X}
\eea
\esubeq
This invariance is essential for the geometry under consideration to describe conformal 
supergravity.

\subsection{Coupling to a vector multiplet}

We now couple the multiplet of conformal supergravity to an Abelian $\cN$-extended 
vector multiplet
$V={\rm d}z^M V_M = E^A V_A$, with $V_A :=E_A{}^M V_M$.
For this we modify the covariant derivatives as
\bea
\cD_A~~~\longrightarrow~~~
{\bm\cD}_A:=\cD_A+V_A{\bm Z}
~, \qquad [{\bm Z}, \cD_A ]=0~,
\label{2.23}
\eea
with $V_A(z)$ the gauge connection
associated with a  generator $\bm Z$.\footnote{For $\cN>1$, one can interpret $\bm Z$ as 
a central charge.}
The gauge transformation of $V_A $ is 
\bea
\d V_A = -\cD_A \t~,
\eea
with $\t (z)$ an arbitrary scalar superfield.

The algebra of covariant derivatives is
\bea
[{\bm \cD}_A,{\bm \cD}_B\}=T_{AB}{}^C{\bm \cD}_C
+\hf R_{AB}{}^{cd}\cM_{cd}
+\hf R_{AB}{}^{KL}\cN_{KL}
+F_{AB}{\bm Z}
~.
\eea
Here $F_{AB}$ is the gauge-invariant field strength, and the torsion and curvatures are the same 
as above.
The field strength satisfies the Bianchi identities
\bea
\sum_{[ABC)}\big(\cD_AF_{BC}-T_{AB}{}^{D}F_{DC}\big)=0~.
\eea
To describe an irreducible vector multiplet, we have to impose covariant constraints on $F_{AB}$.
Their structure is different for $\cN=1$ and for $\cN >1$. 

In the $\cN=1$ case, 
one imposes the covariant constraint  \cite{Siegel}
\bea
F_{\a\b}=0~.
\eea
Then, from the Bianchi identities one gets
\bsubeq
\bea
F_{a\b}&=&-\hf (\g_a)_\b{}^\g W_\g
~,\\
F_{ab}&=&\frac{\ri}{4}\ve_{abc}(\g^c)^{\g\d}\cD_\g W_\d
~,
\eea
\esubeq
together with the dimension-2 differential constraint on the spinor field strength 
\bea
\cD^\a W_\a=0~.
\eea

For  $\cN>1$ one imposes the following dimension-1 covariant  constraint 
\cite{HitchinKLR,ZP,ZH}
\bea
F_{\a}^I{}_\b^J&=&
2\ri\ve_{\a\b}W^{IJ}~, \qquad W^{IJ}=-W^{JI}~
\eea
which is a natural generalization of the 4D $\cN>1$ constraints \cite{GSW,Sohnius}.
The Bianchi identities are solved by the following 
expressions for the field strengths 
\bsubeq
\bea
F_{a}{}_\a^I&=&
-{1\over (\cN-1)}(\g_a)_\a{}^{\b}\cD_{\b J} W^{I J}
~,
\\
F_{ab}&=&
{\ri\over 4\cN(\cN-1)}\ve_{abc}(\g^c)^{\r\t}[\cD_{\r}^{ K},\cD_{\t}^{ L}] W_{ K L}
+{2\over \cN}\ve_{abc}C^{c}{}^{ K L}W_{ K L}
~.
\eea
\esubeq

The case $\cN=2$ is special in the sense that the field strength 
$W^{IJ}$
and the torsion $C^c{}^{KL}$
are proportional to the antisymmetric tensor $\ve^{IJ}$ (normalized as 
$\ve^{{\bf 1}{\bf 2}}=1$), 
\bea
W^{IJ}= \ve^{IJ} G~, \qquad C^c{}^{KL}=\ve^{KL}C^c~.
\eea
As a result, the components of $F_{AB}$ become
\bsubeq
\bea
F_\a^I{}_\b^J&=&2\ri\ve_{\a\b}\ve^{IJ}G
~,
\\
F_a{}_\b^J&=&-\ve^{JK}(\g_a)_\b{}^\g \cD_{\g K}G
~,
\\
 F_{ab}&=&
\ve_{abc}\Big(
\frac{\ri}{4}(\g^c)^{\g\d}\ve^{KL}\cD_{\g K} \cD_{\d L} 
+2C^c
\Big)G~.
\eea
\esubeq
Further analysis of the Bianchi identities 
shows that $G$ obeys the dimension-2 constraint
\bea
\Big(\ve^{K(I}\cD^{\g J)} \cD_{\g K}
-4\ri \ve^{K(I}S_K{}^{J)}\Big)G=0 ~.
\label{2.32}
\eea

Unlike eq. (\ref{2.32}),  in the case  $\cN>2$ the field strength $W^{I J}$ is constrained by the 
dimension-3/2 Bianchi identity 
\bea
\cD_{\g}^{I} W^{ J K}&=&
\cD_{\g}^{[I} W^{ J K]}
-{1\over (\cN-1)}\big(\d^{I J}\cD_{\g L} W^{ K L}
-\d^{I K}\cD_{\g L} W^{ J L}\big)
~.
\label{2.35}
\eea
This constraint can be shown to define an on-shell multiplet for $\cN>4$.

The concept of  super-Weyl transformations introduced in subsection 
\ref{Super-Weyl transformations} can be extended   to the gauge-covariant derivatives
(\ref{2.23}).  The key observation is that the one-form $V={\rm d}z^M V_M $ is invariant under the 
super-Weyl 
transformation, and this determines the super-Weyl transformation law of $V_A$ defined by
$V= E^A V_A$.
After that one can read of the transformation law of the field strength. 
In the case  $\cN=1$ one finds  
\bea
\d_\s W_\a=\frac{3}{2}\s W_\a~,
\eea
while for $\cN>1$ the field strength $W^{IJ}$ transforms as
\bea
\d_\s W^{IJ}=\s W^{IJ}
~.
\label{2.37}
\eea


\section{Matter couplings in $\cN=1$ supergravity}
\setcounter{equation}{0}

The geometry of 3D $\cN=1$ supergravity and its matter couplings have been studied 
in the literature \cite{HT,BG,GGRS,vN,Uematsu,ZP,Becker:2003wb}.
Here the structure group coincides with the 3D Lorentz group. 
The algebra of covariant derivatives becomes
\bsubeq
\bea
\{\cD_\a,\cD_\b\}&=&
2\ri\cD_{\a\b}
-4\ri \cS\cM_{\a\b}
~,~~~~~~~~~
\label{N=1alg-1}
\\
{[}\cD_{\a\b},\cD_\g{]}
&=&
-2\cS\ve_{\g(\a}\cD_{\b)}
+2\ve_{\g(\a}C_{\b)\d\r}\cM^{\d\r}
+{2\over 3}\big((\cD_{\g}\cS)\cM_{\a\b}
-4(\cD_{(\a}\cS)\cM_{\b)\g}\big)
~,~~~~~~
\label{N=1alg-3/2}
\\
{[}\cD_{a},\cD_b{]}
&=&
\hf\ve_{abc}(\g^c)^{\a\b}\Big{\{}
-\ri C_{\a\b\g}\cD^\g
-{4\ri\over 3}(\cD_{\a}\cS)\cD_\b
+\ri\cD_{(\a}C_{\b\g\d)}\cM^{\g\d}
\non\\
&&~~~~~~~~~~~~~~~~
-\Big(
\frac{2\ri}{3}(\cD^2\cS)
+4\cS^2\Big)\cM_{\a\b}
\Big{\}}
~.
~~~~~~~~~~~~
\label{N=1alg-2}
\eea
\esubeq
The torsion and the curvature are expressed in terms of 
a dimension-1 scalar $\cS$ and a dimension-3/2 totally symmetric spinor 
$C_{\a\b\g}=C_{(\a\b\g)}$, in complete agreement with, e.g., \cite{GGRS}.
They obey the constraint
\bea
\cD_{\a}C_{\b\g\d}&=&
\cD_{(\a}C_{\b\g\d)}
-\ri\ve_{\a(\b}\cD_{\g\d)}\cS
~.
\eea
As is seen from (\ref{N=1alg-1}), the vector derivative can be re-defined 
by $\cD_{\a\b} \to 
\cD_{\a\b}-2 \cS\cM_{\a\b}$ such that the relation (\ref{N=1alg-1}) 
takes the same form as in flat superspace  \cite{GGRS}.

The super-Weyl transformation of the covariant derivatives, given e.g. in \cite{ZP,LR-brane}, 
is
\bsubeq
\bea
\d_\s\cD_\a&=&
\hf \s\cD_\a + (\cD^{\b}\s)\cM_{\a\b}
~,
\\
\d_\s\cD_a&=&
\s\cD_a
+{\ri\over 2}(\g_a)^{\g\d}(\cD_{\g} \s)\cD_{\d}
+\ve_{abc}(\cD^b\s)\cM^{c}
~.
\eea
\esubeq
The induced transformation of the torsion is:
\bsubeq
\bea
\d_\s \cS&=&
\s\cS
-{\ri\over 4} \cD^{\g}\cD_{\g}\s ~,
\label{s-Weyl-S}
\\
\d_\s C_{\a\b\g}&=&
\frac{3}{2}\s C_{\a\b\g}
-\frac{1}{2} \cD_{(\a\b}\cD_{\g)}\s ~.
\eea
\esubeq

With the technical tools presented,
it is easy to derive  a locally supersymmetric and super-Weyl
invariant action principle.
It is constructed 
in terms of a {\it purely imaginary}
 Lagrangian  $\cL$ whose super-Weyl 
transformation is
\bea
\d_\s\cL=2\s\cL~,
\label{N=1sWL}
\eea
modulo  total derivatives. 
The action is
\bea
S&=&\int\rd^3x\rd^2\q E \,\cL
~,~~~
\qquad E^{-1}= {\rm Ber}(E_A{}^M)
~, \qquad (\cL)^* =-\cL~.
\label{N=1Ac}
\eea
It is super-Weyl invariant since $\d_\s E=-2\s E$.

Let us construct a nonlinear sigma-model coupled to $\cN=1$ conformal supergravity.
Its dynamics will be described by  real scalar superfields $\varphi^\mu$  taking values in a 
Riemannian manifold $\cM$.
Consider the kinetic term
\bea
\cL_0=
-\frac{1}{2}g_{\mu\nu}(\varphi)(\cD^\a\varphi^\mu)\cD_\a\varphi^\nu
~,
\label{N=1La-0}
\eea
where $g_{\mu\nu}(\varphi)$ is the metric on the target space.
We are looking for a  Lagrangian of the form $\cL= \cL_0 +\dots$ that  
transforms homogeneously as (\ref{N=1sWL}) modulo  total derivatives.
Postulating  the super-Weyl transformation of $\vf^\m$
\bea
\d_\s\varphi^{\mu}=\hf \s\chi^\mu(\varphi)~,
\eea
for some vector field $\c= \c^\m( \vf ) \pa_\m$ on $\cM$, we find
\bea
\d_\s \cL_0 =-\hf g_{\m \n }(\vf) (\cD^\a\vf^\m ) (\cD_\a \vf^\l)
\Big( \nabla_\l \c^\n (\vf) +\d_\l^\n \Big)\s
-\hf g_{\m \n }(\vf) (\cD^\a \vf^\m ) \c^\n (\cD_\a \s)~.~~~
\label{N=1sWL0-1}
\eea
In the case that   $\s ={\rm const}$, the action $S_0=\int\rd^3x\rd^2\q E \,\cL_0$ 
is invariant only if
\bea
\nabla_\mu\chi^\nu=\d_\mu^\nu \quad \Longrightarrow \quad
\chi_{\mu}(\varphi)=\pa_\mu f(\varphi)~,~~~
f(\varphi):=\hf g_{\mu\nu}(\varphi)\chi^{\mu}(\varphi)\chi^{\nu}(\varphi)
~.
\label{3.10}
\eea
We see that $\chi=\chi^\mu(\vf)\pa_\mu$ is  a  homothetic conformal Killing vector field
such that $\chi^{\mu}$ is the gradient of a function over the target space.
Therefore, the sigma-model target space $\cM$ is  a Riemannian cone \cite{GR}, 
as in the rigid superconformal case \cite{KPT-MvU,BCSS}.
Now, the variation (\ref{N=1sWL0-1}) becomes 
\bea
\d_\s \cL_0 =2 \s \cL_0
-\hf (\cD^\a f) (\cD_\a \s)
=2 \s \cL_0
+\hf f (\cD^\a\cD_\a \s)
-\hf \cD^\a (f\cD_\a \s)
~.
\label{N=1sWL0-2}
\eea
It remains to recall the transformation law (\ref{s-Weyl-S}) as well as to notice that eq. 
(\ref{3.10}) implies the homogeneity condition 
\bea 
\c^\m \pa_\m f = 2f~.
\eea
We then observe that 
\bea
\d_\s \Big(\ri\cS f(\varphi)\Big)=
2 \s \Big( \ri \cS f(\varphi) \Big)
+\frac{1}{4}(\cD^\a\cD_\a \s)f(\varphi)
~,
\eea
and therefore 
\bea
\cL:= -\frac{1}{2}g_{\mu\nu}(\varphi)(\cD^\a\varphi^\mu)(\cD_\a\varphi^\nu)- 2\ri\cS f(\varphi)
\eea
is the required Lagrangian.

The above Lagrangian can be modified by adding a potential term
\bea
\cL=
-\frac{1}{2}\Big(g_{\mu\nu}(\varphi)(\cD^\a\varphi^\mu)(\cD_\a\varphi^\nu)+4\ri\cS f(\varphi)\Big)
+\ri V(\varphi)
~.
\label{N=1La}
\eea
For the action to be super-Weyl invariant, the potential  
should satisfy the homogeneity condition
\bea
\chi^{\mu}(\varphi)V_\mu(\varphi)=4V(\varphi)~.
\eea

In the rigid supersymmetric case, the Lagrangian (\ref{N=1La})  reduces to that 
corresponding to the general $\cN=1$ superconformal sigma-model 
\cite{KPT-MvU}.

In the case of a single scalar superfield $\vf$, 
the general form for   (\ref{N=1La}) is 
\bea
\cL=
-\frac{1}{2}\Big( (\cD^\a\varphi)(\cD_\a\varphi )+2\ri\cS \varphi^2 \Big)
+ \l \ri  \varphi^4~,\qquad \l={\rm const}~.
\eea
We can choose $\vf $ to be a superconformal compensator,  if we think of Poincar\'e  supergravity 
as conformal supergravity coupled to the compensator. Then we should use $(-\cL)$ 
as a supergravity Lagrangian, 
and interpreteÊ $\l$ as a cosmological constant.

\section{Matter couplings in $\cN=2$ supergravity}
\setcounter{equation}{0}

Three-dimensional $\cN=2$ supergravity and its matter couplings have not been studied
as thoroughly as in the $\cN=1$ case. This is partly due to  the fact  that 
3D $\cN=2$ supergravity can be obtained by dimensional reduction from 
that with $\cN=1$ supersymmetry in four dimensions, and therefore 
much is known about the component structure of 3D $\cN=2$ supergravity-matter systems.
However, there are several reasons to achieve a better understanding 
of the superspace geometry of 
$\cN=2$ supergravity, for instance,  in the context of
 massive 3D supergravity 
\cite{Andringa:2009yc,Bergshoeff:2010mf,Bergshoeff:2010ui} 

\subsection{Complex basis for spinor covariant derivatives}
The $R$-symmetry subgroup of the $\cN=2$ superspace structure group is 
${\rm SO}(2) \cong {\rm U}(1)$.
Instead of dealing with the anti-Hermitian generator $\cN_{KL} = - \cN_{LK}$ of SO(2),  
as defined in subsection \ref{algebra of covariant derivatives},
it is convenient to introduce a scalar Hermitian generator $\cJ$ 
defined by\footnote{The antisymmetric tensors 
$\ve^{IJ}=\ve_{IJ}$ 
are normalized as $\ve^{\bf{1}\bf{2}}=\ve_{\bf{1}\bf{2}}=1$.
The normalization of $\ve_{IJ}$ is nonstandard as compared with the 
definitions given in Appendix A.}
\bea
\cN_{KL}=\ri\ve_{KL}\cJ~,~~~
\cJ=-\frac{\ri}{2}\ve^{PQ}\cN_{PQ}~,
\eea
which acts on the covariant derivatives as
\bea
[\cJ,\cD_\a^I]&=& -\ri\ve^{IJ}\cD_{\a J} ~.
\eea
It is also convenient to switch to a complex basis for the spinor covariant derivatives, 
$\cD^I_\a \to (\cD_\a, \cDB_\a) $, in which $\cD_\a$ and $\cDB_\a$ have definite U(1) charges.
 We define
\bea
&&\cD_\a={1\over \sqrt{2}}(\cD_\a^1-\ri\cD_\a^2)~,~~~
\cDB_\a=-{1\over \sqrt{2}}(\cD_\a^1+\ri\cD_\a^2)~,~~~
\eea
such that
\bea
{[}\cJ,\cD_\a{]}&=&\cD_\a
~,~~~~~~
{[}\cJ,\cDB_\a{]}=-\cDB_\a
~.
\eea
 The SO(2) connection and the corresponding curvature 
take the form
\bea
\hf \Phi_{A}{}^{KL}\cN_{KL}=\ri\Phi_{A}\cJ
~,~~~~~~
\hf R_{AB}{}^{KL}\cN_{KL}=\ri R_{AB}\cJ
~.
\eea
Given a complex superfield $F$ and its complex conjugate $\bar{F}:=(F)^*$,
the following rule for complex conjugation holds
\bea
(\cD_\a F)^*=(-1)^{\ve(F)}\cDB_\a\bar{F}
~, 
\eea
which can be compared with (\ref{c.c.1}).

In the  $(\cD,\cDB )$ basis introduced, the supergravity algebra 
(\ref{alg-1}) and (\ref{alg-3/2}) takes the form
\bsubeq
\bea
\{\cD_\a,\cD_\b\}
&=&
-4\bar{R}\cM_{\a\b}
~,~~~~~~
\{\cDB_\a,\cDB_\b\}
=
4{R}\cM_{\a\b}~,
\label{N=2-alg-1}
\\
\{\cD_\a,\cDB_\b\}
&=&
-2\ri\cD_{\a\b}
-2\cC_{\a\b}\cJ
-4\ri\ve_{\a\b}\cS\cJ
+4\ri\cS\cM_{\a\b}
-2\ve_{\a\b}\cC^{\g\d}\cM_{\g\d}
~,
\\
{[}\cD_{\a\b},\cD_\g{]}
&=&
-\ri\ve_{\g(\a}\cC_{\b)\d}\cD^{\d}
+\ri\cC_{\g(\a}\cD_{\b)}
-2\ve_{\g(\a}\cS\cD_{\b)}
-2\ri\ve_{\g(\a}\bar{R}\cDB_{\b)}
\non\\
&&
+2\ve_{\g(\a}C_{\b)\d\r}\cM^{\d\r}
-\frac{4}{3}\Big(
2\cD_{(\a}\cS
+\ri\cDB_{(\a}\bar{R}
\Big)\cM_{\b)\g}
+\frac{1}{3}\Big(
2\cD_{\g}\cS
+\ri\cDB_{\g}\bar{R}
\Big)\cM_{\a\b}
\non\\
&&
+\Big(
C_{\a\b\g}
+\frac{1}{3}\ve_{\g(\a}\Big(
8(\cD_{\b)}\cS)
+\ri\cDB_{\b)}\bar{R}
\Big)
\Big)\cJ
~.
\eea
\esubeq
Here we have accounted for the fact that in the $\cN=2$ case
\bea
X^{IJKL}=0~,~~~
C_a{}^{IJ}=\ve^{IJ}\cC_a~,
\eea
as well as we  have defined the scalar torsion superfields
\bea
R&:=&-\frac{\ri}{2}(S^{{\bf 1}{\bf 1}}-S^{{\bf 2}{\bf 2}}+2\ri S^{{\bf 1}{\bf 2}})~,~~~
\bar{R}:=\frac{\ri}{2}(S^{{\bf 1}{\bf 1}}-S^{{\bf 2}{\bf 2}}-2\ri S^{{\bf 1}{\bf 2}})~,
\\
\cS&:=&\hf\d_{IJ}S^{IJ}=\hf(S^{{\bf 1}{\bf 1}}+S^{{\bf 2}{\bf 2}})~.
\eea
The U(1) charges  of $R$ and its conjugate are 
\bea
\cJ \bar{R}=2 \bar{R}~,\qquad
\cJ {R}=-2{R}~,~~~
\eea
while the real fields  $\cS$ and $\cC_a$ are obviously neutral.
The dimension-3/2 differential constraints on the dimension-1 torsion superfields
are 
\bea
\cD_\a\bar{R}&=&0~,~~~~~~
\cDB_\a R=0~,
\\
\cD_{\a}\cC_{\b\g}
&=&
\ri C_{\a\b\g}
-{1\over 3}\ve_{\a(\b}\Big(
\cDB_{\g)}\bar{R}
+4\ri \cD_{\g)}\cS
\Big)
\label{4.13}
\eea
where we have defined the completely symmetric complex spinors
\bea
C_{\a\b\g}&:=&\frac{1}{\sqrt{2}}(C_{\a\b\g}{}^1-\ri C_{\a\b\g}{}^{2})
~,\qquad
{\bar{C}}_{\a\b\g} :=-\frac{1}{\sqrt{2}}(C_{\a\b\g}{}^1+\ri C_{\a\b\g}{}^{2})~,
\eea
which are charged under the U(1)-group
\bea 
&&\cJ C_{\a\b\g}= C_{\a\b\g}~, \qquad
\cJ \bar{C}_{\a\b\g}= - \bar{C}_{\a\b\g}
~.
\eea
It follows from (\ref{4.13}) that ${\cal S}$ is a real linear superfield, 
\bea
(\cDB^2-4R) {\cal S} =(\cD^2 - 4\bar R ) {\cal S} =0~.
\label{4.16--}
\eea

The super-Weyl transformation of the covariant derivatives becomes
\bsubeq
\bea
\d_\s\cD_\a&=&
\hf \s\cD_\a + (\cD^{\b}\s)\cM_{\a\b}-(\cD_{\a} \s)\cJ
~,
\\
\d_\s\cDB_\a&=&
\hf \s\cDB_\a + (\cDB^{\b}\s)\cM_{\a\b}+(\cDB_{\a} \s)\cJ
~,
\\
\d_\s\cD_a&=&
\s\cD_a
-{\ri\over 2}(\g_a)^{\g\d}(\cD_{\g}\s)\cDB_{\d}
-{\ri\over 2}(\g_a)^{\g\d}(\cDB_{\g} \s)\cD_{\d}
+\ve_{abc}(\cD^b\s)\cM^{c}
\non\\
&&
-{\ri\over 8}(\g_a)^{\g\d}([\cD_{\g},\cDB_{\d}]\s)\cJ
~.
\eea
${}$From here we can read off the transformation  of the torsion 
\bea
\d_\s \cS&=&
\s\cS
+{\ri\over 8} [\cD^{\g},\cDB_{\g}]\s
~,~~~~~~\\
\d_\s \cC_{a}
&=& \s\cC_{a}
+{1\over 8}(\g_a)^{\g\d} [\cD_{\g},\cDB_{\d}]\s
~,
\\
\d_\s R&=&
\s R
+\frac{1}{4}\cDB^2\s
~,~~~~~~
\d_\s \bar{R}=
\s\bar{R}
+\frac{1}{4} \cD^2\s
~,
\eea
\esubeq
where we have defined
\bea
\cD^2:=\cD^\a\cD_\a~,\qquad
\cDB^2:=\cDB_\a\cDB^\a~.
\eea

\subsection{Scalar and vector multiplets}
\label{subsection4.2}

We now wish to study in some detail the properties of scalar and vector multiplets.
Consider a covariantly chiral scalar $\F$, ${\bar \cD}_\a \F=0$, which is a primary 
field under the super-Weyl group, $\d_\s\F=w\s\F$. Then its super-Weyl weight $w$ 
and its U(1) charge have the same value and opposite signs,
\bea
{\bar \cD}_\a \F=0 ~, \qquad
\cJ\F=-w\F~,\qquad 
\d_\s\F=w \s\F~.
\label{4.18}
\eea
Consider now a complex scalar $\Psi$ with the properties
\bea
\cJ\Psi=(2-w) \Psi~,\qquad
\d_\s \Psi=(w-1)\s\Psi~,
\label{N=2psi}
\eea
for some constant super-Weyl weight $w$. Then, the superfield 
\bea
\F=\bar{\D}\Psi~, \qquad \bar{\D}:=-\frac{1}{4}(\cDB^2-4R)
\label{N=2chiralPsi}
\eea
is characterized by the properties (\ref{4.18}). The operator $\bar \D$ 
is the   $\cN=2$ chiral projection operator.
The fact that the explicit form of $\bar \D$  is identical to that for the chiral 
projection operator in 4D $\cN=1$  supergravity  \cite{WZ,Zumino78},
is not surprising since the anticommutator $\{\cD_\a,\cD_\b\}$ 
in (\ref{N=2-alg-1})  is algebraically identical to that in the 4D $\cN=1$ case.

We next turn to a complex linear superfield $\S$. It is defined to obey the  constraint 
\bea 
(\cDB^2-4R)\S=0
\label{complex-linear}
\eea
and no reality condition. If $\S$ is chosen to transform homogeneously under the 
super-Weyl transformations, 
then its  U(1) charge is determined by the super-Weyl weight, 
\bea
\d_\s \S=w\s\S \quad \Longrightarrow \quad \cJ\S=(1-w) \S
~.
\label{complex-linear2}
\eea
This follows from the identity
\bea
\d_\s(\cDB^2-4R)&=&
\s(\cDB^2-4R)
 +2(\cDB^{\a}\s)\cDB_\a
 -2(\cDB^{\a} \s)\cDB_\a\cJ
 +2 (\cDB^{\a}\s)\cDB^\b\cM_{\a\b}
 \non\\
&&
-(\cDB^2\s)
+(\cDB^2 \s)\cJ
~,
\eea
Indeed, using the relations (\ref{complex-linear2})
allows us to prove that
\bea
\d_\s \Big((\cDB^2-4R)\S\Big)=(1+w) \s(\cDB^2-4R)\S=0~.
\eea

Unlike the chiral and the complex linear superfields, the superconformal transformation law 
of the vector multiplet is uniquely fixed. 
Consider an Abelian vector multiplet described by its gauge-invariant field strength $G$. 
The latter is real, $(G)^* =G$,  and obeys the constraint\footnote{Eq. (\ref{N=2realLinear})
 is a 3D version of the constraint 
defining the 4D $\cN=1$ tensor multiplet \cite{Siegel-spinor}.} 
\bea
(\cDB^2-4R)G=0 \quad  \stackrel{\bar G =G}{\Longrightarrow} \quad (\cD^2-4\bar{R})G=0~,
\label{N=2realLinear}
\eea
which is equivalent to eq. (\ref{2.32}).
Since $G$ is neutral under the group U(1), $\cJ G =0$, eq. (\ref{complex-linear2}) tells us 
that the super-Weyl transformation of $G$ is 
\bea
\d_\s G=\s G~.
\label{N=2sWrealLinear}
\eea
The constraint (\ref{N=2realLinear}) can be solved in terms of 
a real unconstrained prepotential $V$, 
\bea
G= \ri {\bar \cD}^\a \cD_\a V~, 
\label{G-prep}
\eea
which is defined modulo arbitrary gauge transformations of the form:
\be 
\d V = \l + \bar \l~, \qquad  \cJ \l =0~, \quad {\bar \cD}_\a \l =0~.
\ee
It is consistent to consider the gauge prepotential $V$ to be inert under the super-Weyl 
transformations, 
\be
\d_\s V =0~.
\ee

\subsection{Matter couplings}
We are now prepared to introduce interesting matter couplings in $\cN=2$ supergravity.
Let us first elaborate on locally supersymmetric and super-Weyl invariant actions.
Given a real Lagrangian $\cL$ with the super-Weyl transformation law
\bea
\d_\s\cL=\s\cL~,
\label{4.21}
\eea
the action
\bea
S&=&\int\rd^3x\rd^2\q\rd^2\qb E \,\cL
~,~~~
\qquad E^{-1}= {\rm Ber}(E_A{}^M)
~,
\label{N=2Ac}
\eea
is invariant under the supergravity gauge group. 
It is also  super-Weyl invariant since the corresponding transformation law of $E$ is
$\d_\s E=-\s E$.

The existence of covariantly chiral superfields in $\cN=2$  conformal supergravity  implies 
that the action (\ref{N=2Ac})  can also be rewritten as an integral over a chiral subspace, 
for instance,  using the approach 
developed in \cite{KT-M2}.\footnote{There should exist a prepotential formulation 
for 3D $\cN=2$ supergravity that is similar to that developed many years ago 
for 4D $\cN=1$ supergravity \cite{SG}. 
In such a formulation eq. (\ref{N=2Ac-2}) could be derived using a covariant chiral representation.}
\bea
S&=&\int\rd^3x\rd^2\q\rd^2\qb\, E \,\cL
=\int\rd^3x\rd^2\q\, \cE \,\bar{\D}\cL
~.
\label{N=2Ac-2}
\eea
Here $\cE$ denotes the chiral density, ${\bar \cD}_\a \cE=0$, and $\bar \D$ the chiral
projection operator (\ref{N=2chiralPsi}).
As follows from (\ref{N=2psi}) and (\ref{4.21}), the chiral superfield $\bar{\D}\cL$
has super-Weyl weight +2 and U(1) charge $-2$. 
Thus the chiral density has the properties 
\bea
\cJ  \cE = 2 \cE~, \qquad 
\d_\s\cE=-2\s\cE~.
\eea

The construction (\ref{N=2Ac-2}) allows us to introduce a different action principle.
Given a chiral Lagrangian $\cL_c$ of super-Weyl weight two
\bea
\cDB_\a\cL_c=0~,~~~~~~
\d_\s\cL_c=2\s\cL_c~,
\eea
the following {\rm chiral} action
\bea
S_c&=&\int\rd^3x\rd^2\q\rd^2\qb\, E \,\frac{\cL_c}{R}
=\int\rd^3x\rd^2\q\, \cE \,\cL_c 
\label{4.26}
\eea
is locally supersymmetric and super-Weyl invariant. The first representation in  (\ref{4.26})
is analogous to that derived by Zumino \cite{Zumino78} in 4D $\cN=1$ supergravity.

We now wish to uncover conditions on the target space geometry under which a 3D $\cN=2$
rigid supersymmetric sigma-model \cite{Zumino} can be coupled to conformal supergravity.
Consider the  $\cN=2$ locally supersymmetric sigma-model  action
\bea
S= \int {\rm d}^3x {\rm d}^2 \q {\rm d}^2 {\bar \q}\, E\,K(\F^I, {\bar \F}^{\bar J})~, \qquad 
{\bar {\mathbb \cD}}_\a \F^I=0~,
\label{7.7}
\eea
where the dynamical variables $\F^I$ are covariantly chiral scalar superfields,  
$K(\F , \bar\F) $  is the K\"ahler potential of a K\"ahler manifold $\cM$. As usual, we denote by
 $g_{I \bar J} (\F , \bar \F) $ the K\"ahler metric on $\cM$.
 We postulate the super-Weyl transformation of the chiral superfields 
 \bea
\d_\s \F^I = \hf \s \c^I (\F)~, 
\label{genN=2sctr}
\eea
where $\chi^\mu:=(\chi^I,\bar{\chi}^{\bar{J}})$ is a holomorphic vector field on the target space.
 The action (\ref{7.7}) can be seen to be invariant 
 provided the K\"ahler  potential satisfies the condition
\bea
\c^I (\F) K_I (\F , \bar \F ) = K(\F , \bar \F)~.
\label{4.47}
\eea
This condition turns out to imply that $\chi^I$ is a homothetic holomorphic Killing vector on
the target space. It has the properties\begin{subequations}
\bea
\nabla_I \c^J &=& \d_I^J~, \qquad {\bar \nabla}_{\bar I} \c^J ={\bar \pa}_{\bar I} \c^J =0
~,
 \label{7.8a} \\
 \c_I := g_{I\bar J}  {\bar \c}^{\bar J}&=& \pa_I K~, \qquad g_{I\bar J} 
=\pa_I {\bar \pa}_{\bar J} K
~, \label{7.8b}
\eea
\end{subequations}
where $K$ can be chosen to be 
\bea
 K =g_{I\bar J} \c^I  {\bar \c}^{\bar J}~.
 \label{7.9}
 \eea
These properties mean that the target space $\cM$ is a K\"ahlerian cone \cite{GR}. 

There is an important consistency condition. For the chirality condition 
${\bar \cD}_\a \F^I =0 $ to be super-Weyl invariant, $\d_\s ( {\bar \cD}_\a \F^I )=0$, 
the U(1) charge of $\F^I$ is uniquely fixed as 
\bea
\cJ \F^I = -\frac{1}{2}  \c^I (\F)~.
\eea
The sigma-model action (\ref{7.7}) is invariant under local U(1) transformations, 
as a consequence of (\ref{4.47}).

The sigma-model (\ref{7.7}) can be generalized  to include a superpotential.
\bea
S= \int {\rm d}^3x {\rm d}^2 \q {\rm d}^2 {\bar \q}\,E 
 K(\F^I, {\bar \F}^{\bar J})
 +\int {\rm d}^3x {\rm d}^2 \q {\rm d}^2 {\bar \q}\,E \left\{
\frac{W(\F^I)}{R}   +{\rm c.c.}
\right\}
~,
\label{4.33}
\eea
with $W(\F)$  a holomorphic scalar field on the target space. It should obey 
the homogeneity condition
\bea
\c^I (\F) W_I (\F  ) = 4 W(\F )~
\eea
for the second term in (\ref{4.33}) to be locally supersymmetric and super-Weyl invariant.
The theory (\ref{4.33}) is a locally supersymmetric extension  of the general 3D $\cN=2$ 
superconformal sigma-model presented in \cite{KPT-MvU}.

Local complex coordinates, $\F^I$, on $\cM$ can be chosen in such a way that $\c^I =\F^I$.
Then $K(\F^I, {\bar \F}^{\bar J}) $ obeys the following homogeneity condition:
\bea
\F^I \frac{\pa}{\pa \F^I} K(\F, \bar \F) =  K( \F,   \bar \F)~.
\label{Kkahler2}
\eea

Locally supersymmetric nonlinear sigma-models can also be generated from self couplings
of vector multiplets. Consider a system of Abelian vector multiplets 
described by  real field strengths $G^i $, with $i=1,\dots, n$,
constrained by
\be
(\cDB^2-4R) G^i = (\cD^2-4\bar{R}) G^i =0~, \qquad i=1, \dots, n~.
\label{vmc}
\ee
Their dynamics  can be described by an action of the form
\bea
S=\int {\rm d}^3x {\rm d}^2 \q {\rm d}^2 {\bar \q}\,E\, L(G^i)~.
\label{vm-action}
\eea
We know that the constraints (\ref{vmc}) require 
the super-Weyl transformation law $\d_\s G^i= \s G^i$.  The action is therefore 
super-Weyl invariant 
provided the Lagrangian is a homogeneous function of $G^i$ of first degree, 
\be
G^i L_i (G) =L (G)~.
\ee
This theory is a local supersymmetric extension of the $\cN=2$ superconformal model presented 
in \cite{KPT-MvU}.
In the case of a single vector multiplet, 
there is a super-Weyl invariant action generated by  the Lagrangian 
$L(G) \propto (- G \ln G + 4V{\cal S} ) $.
Such an action describes an improved vector multiplet  \cite{HitchinKLR} 
which is a 3D version of the 
4D $\cN=1$ improved tensor multiplet \cite{deWR}.

The vector multiplet model (\ref{vm-action}) can be generalized to include 
gauge-invariant Chern-Simons couplings 
\bea
S_{\rm CS}=\int {\rm d}^3x {\rm d}^2 \q {\rm d}^2 {\bar \q}\,E\,\Big\{  L(G^i) + \hf m_{ij} V^i G^j \Big\}~, 
\qquad m_{ij} =m_{ji} =(m_{ij})^*={\rm const}~.~~~~~~
\label{vm-CS-action}
\eea
Here $V^i$ is the gauge prepotential for $G^i$ defined as in (\ref{G-prep}).

\subsection{Conformal compensators}

As is well-known, Poincar\'e supergravity can be realised as conformal supergravity coupled 
to a compensating supermultiplet 
(or compensator) \cite{KT}.
Different choices of compensator lead, in general, to different off-shell formulations 
for Poincar\'e supergravity, 
as has been shown in detail in the case of 4D $\cN=1$ supergravity \cite{Howe,GGRS}. 
In complete analogy with 4D $\cN=1$  supergravity, there are three different types 
of compensator for $\cN=2$ supergravity in three dimensions: (i) a chiral scalar $\F$ and 
its conjugate $\bar \F$;
(ii) a real linear scalar $G$; (iii) a complex linear scalar $\S$ and its conjugate $\bar \S$.
Here we briefly discuss these choices.

In the case  (i), the compensator $\F$ can be chosen to to have  super-Weyl weight 1/2, 
\bea
\d_\s\F=\hf \s\F~.
\eea
The freedom to perform the super-Weyl and local 
U(1) transformations can be used to impose the gauge
\bea
\F=1
~.
\eea
Such a gauge fixing is accompanied by  the consistency conditions 
\bea
0=\cDB_\a\F=-\frac{\ri}{2}\F_\a~,\qquad 
0=\{\cD_\a,\cDB_\b\}\F
=
-\F_{\a\b}
+\cC_{\a\b}
-2\ri\ve_{\a\b}\cS
~,
\eea
and therefore
\bea
\F_\a= \cS=0~,\qquad
\F_{\a\b}= \cC_{\a\b}~.
\eea
The formulation is the analogue of old-minimal 4D $\cN=1$ supergravity (see \cite{GGRS,Ideas}
for reviews).

Another choice for compensator is the field strength of a vector multiplet, $G=\bar G$, 
which is subject to the linear constraint (\ref{N=2realLinear})
and has the super-Weyl transformation (\ref{N=2sWrealLinear}).
The super-Weyl gauge freedom can be used to impose the condition
\bea
G=1
~.
\eea
The local U(1) group remains unbroken. 
The resulting geometry is characterized by the properties
\bea
R=\bar{R}=0 \quad \Longleftrightarrow  \quad \{\cD_\a,\cD_\b\}=\{\cDB_\a,\cDB_\b\}=0
~.
\eea
This is clearly the 3D analogue of 4D $\cN=1$ new minimal supergravity 
(see \cite{GGRS} for a review).

A conformal compensator can be chosen to be 
a complex linear superfield $\S$ which obeys  the constrain
(\ref{complex-linear}) and is characterized by the 
local U(1) and super-Weyl transformation properties (\ref{complex-linear2}).
These local symmetries can be used to impose the gauge condition
\bea
\S=1
\label{4.50}
\eea
which implies some restrictions on the geometry. To describe such restrictions, it is useful to 
 split the covariant derivatives as 
\bea
\cD_\a=\nabla_\a+\ri T_\a\cJ
~,~~~
\cDB_\a=\bar{\nabla}_\a+\ri \bar{T}_\a\cJ
\eea
where we have renamed the 
original U(1) connection $\F_\a$ as $T_\a$.
The operators $\nabla_\a$ and ${\bar \nabla}_\a$
have no U(1) connection.
In the gauge (\ref{4.50}), the  constraint $({\bar \cD}^2 -4R)\S=0$ 
turns into 
\bea
R&=&-\frac{\ri(w-1)}{4}\Big(
\bar{\nabla}_\a\bar{T}^\a
-\ri w\bar{T}_\a\bar{T}^\a
\Big)~.
\label{4.52}
\eea
We see that $R$ becomes a descendant of $T_\a$ and its complex conjugate. 
Eq. (\ref{4.52})  is not the only constraint which is induced by the gauge fixing (\ref{4.50}).
By evaluating $\{ \cD_\a, \cD_\b\} \S $ and  $\{ \cD_\a, {\bar \cD}_\b\} \S $
and then setting $\S=1$
gives
\bsubeq
\bea
&&\de_{(\a}T_{\b)}=0~,~~~~~~
\cS=\frac{1}{8}\Big(
\deb^\a T_\a
-\de^\a\bar{T}_\a
+2\ri T^\a\bar{T}_\a
\Big)
~,
\\
&&~~~~~~
\F_{\a\b}
=
\cC_{\a\b}
+\frac{\ri}{2}\de_{(\a}\bar{T}_{\b)}
+\frac{\ri}{2}\deb_{(\a} T_{\b)}
+T_{(\a}\bar{T}_{\b)}
~.
\eea
\esubeq
If we define a new vector covariant derivative $\nabla _a$ by $\cD_a = \nabla_a +\ri \F_a$, 
then the algebra of the covariant derivatives $\nabla_A =(\nabla_a, \nabla_\a , {\bar \nabla}_\a)$
proves to be 
\bsubeq
\bea
\{\de_\a,\de_\b\}&=&
-2\ri T_{(\a}\de_{\b)}
-\ri(w-1)\Big(
{\nabla}^\g{T}_\g
+\ri w{T}^\g{T}_\g
\Big)
\cM_{\a\b}
~,
\\
\{\de_\a,\deb_\b\}&=&
-2\ri\de_{\a\b}
-\ri\bar{T}_\b\de_\a
+\ri T_\a\deb_\b
-2\ve_{\a\b}\cC^{\g\d}\cM_{\g\d}
\non\\
&&
+\frac{\ri}{2}\Big(
\deb^\g T_\g
-\de^\g\bar{T}_\g
+2\ri T^\g\bar{T}_\g
\Big)
\cM_{\a\b}
~.
\eea
\esubeq
The emerging formulation for 3D $\cN=2$ supergravity
is analogous  to 4D $\cN=1$ non-minimal  supergravity (see \cite{GGRS} for a review).

The procedure of de-gauging described in this subsection is completely similar to that presented 
in the book \cite{GGRS} which in turn closely followed Howe's approach  \cite{Howe}.


\section{Matter couplings in $\cN=3$ supergravity}
\setcounter{equation}{0}

To the best of our knowledge, three-dimensional $\cN=3$ supergravity in superspace 
is terra incognita.
Here we set out to explore this continent.

In this and the following sections, we build on the projective-superspace formulations 
for general 5D $\cN=1$  and 4D $\cN=2$ supergravity-matter theories which were 
developed in \cite{KT-M5D-1,KT-M5D-2,KLRT-M1,K-08,KLRT-M2}, as well as 
on the recent results obtained in \cite{KPT-MvU} concerning the off-shell $\cN=3$ 
and $\cN = 4$ rigid superconformal  sigma-models in three dimensions.

\subsection{Elaborating on the $\cN=3$ superspace geometry}

In accordance with the geometric formulation developed  in
section 2, the structure group of 
 $\cN=3$ conformal supergravity 
is ${\rm SL}(2,\mathbb{R})\times {\rm SO}(3)$, with  the spinor derivatives $\cD^I_\a$ 
transforming in the
defining (vector) representation of SO(3). 
In order to define a large class of matter multiplets coupled to supergravity, however, 
it is convenient to  switch to
an isospinor notation using the isomorphism ${\rm SO}(3) \cong {\rm SU}(2)/{\mathbb Z}_2$.
As usual, this is achieved by replacing 
any SO(3) vector  index by a  symmetric pair of 
SU$(2)$ spinor indices, $\cD^I_\a \to \cD^{ij}_\a = \cD^{ji}_\a$.
In this subsection, our isospinor notation is defined and the supergravity algebra
is rewritten using this notation.

Isospinor indices are raised and lowered with the aid of  the  SU(2) invariant  antisymmetric
tensors $\ve^{ij}$ and $\ve_{ij}$ ($\ve^{12}=\ve_{21}=1$) according to the rule
\bea
\psi^{i}=\ve^{ij}\psi_j~,~~~~~~
\psi_{i}=\ve_{ij}\psi^j
~.
\eea
Given a real isovector $V_I$, we associate with it  
the symmetric isospinor $V_{ij}$ defined by 
\bea
V_I~~\to~~V_{ij}:=(\t^I)_{ij}V_I=V_{ji}
~,\qquad
V_I=(\t_I)^{ij}V_{ij}~,~~(V_{ij})^*=V^{ij}
~,
\eea
see Appendix A for  the  definition of the $\t$-matrices.
The normalization of the $\t$-matrices is such that 
\bea
A^IB_I=A^{ij}B_{ij}~, 
\eea
for any isovectors $A_I$ and $B_I$ and the associated symmetric
isospinors $A_{ij}$ and $B_{ij}$. 
 Consider now  an antisymmetric second-rank isotensor, $A_{IJ}=-A_{JI}$. 
 Its counterpart with isospinor 
indices, $A_{ijkl}=-A_{klij}=A_{IJ}(\t^I)_{ij}(\t^J)_{kl}$ can be decomposed as
\bea
A_{ijkl}=\hf\ve_{jl}A_{ik}+\hf\ve_{ik}A_{jl}~,\qquad
A^{ijkl}=-\hf\ve^{jl}A^{ik}-\hf\ve^{ik}A^{jl}~,\qquad
A_{ij}=A_{ji}
~.
\eea
In particular,  if  $A_{IJ}=-A_{JI}$ and  $B_{IJ}=-B_{JI}$  are  two antisymmetric isotensors, 
and  $A_{ij}$ and $B_{ij}$ are their isospinor  counterparts, then it holds that 
\bea
\hf A^{IJ}B_{IJ}
&=&
\hf A^{kl}B_{kl}~.
\eea
Finally, let us derive the isospinor analogue of 
the completely antisymmetric third-rank 
tensor $\ve_{IJK}$  ($\ve_{{\bf 1}{\bf 2}{\bf 3}}=1$). 
The definition $\ve_{ijklpq}=\ve_{IJK}(\t^I)_{ij}(\t^J)_{kl}(\t^K)_{pq}$ 
leads to 
\bea
\ve_{ijklpq}&=&
-\frac{1}{\sqrt{2}}\Big(
\ve_{p(k}\ve_{l)(i}\ve_{j)q}
+\ve_{q(k}\ve_{l)(i}\ve_{j)p}
\Big)
~.
\eea

We are now ready to rewrite the results obtained in section 2 
for the case $\cN=3$  in the isospinor notation introduced.
The covariant derivatives are
\bea
\cD_{A}&\equiv& (\cD_a , \cD^{ij}_\a ) =
E_{A} +\O_{A} +\F_A
~,
\eea
where  the SO(3) connection $\F_A$ 
takes the form
\bea
\F_A=\hf\Phi_{A}{}^{KL}\cN_{KL}=\hf\Phi_{A}{}^{kl}\cJ_{kl}~.
\eea
Here we have introduced the SU(2) generator $\cJ_{kl}$ which is obtained from  $\cN_{KL}$ as
\bea
\cN_{KL}~\to~\cN_{ijkl}=\hf\ve_{jl}\cJ_{ik}+\hf\ve_{ik}\cJ_{jl}~,~~~
\cN^{ijkl}=-\hf\ve^{jl}\cJ^{ik}-\hf\ve^{ik}\cJ^{jl}
~.
\eea
It acts on the spinor covariant derivatives $\cD_\a^{ij}:=\cD_\a^I(\t_I)^{ij}$ as follows
\bea
&&
{\big [} {\cJ}{}^{kl},\cD_{\a}^{ij}{\Big]} =\ve^{i(k} \cD_{\a}^{ l)j}+\ve^{j(k} \cD_{\a}^{ l)i} 
~.
\eea

In the $\cN=3$ case under consideration,  the dimension-1 components of 
the torsion and the curvature
can be  rewritten as 
\bea
C_a{}^{IJ}~&\to&~C_a{}^{ijkl}=-\hf\ve^{ik}C_a{}^{jl}-\hf\ve^{jl}C_a{}^{ik}
~,~~~~~~
C_a{}^{ij}=C_a{}^{ji}~,~~
\\
S^{IJ}~&\to&~S^{ijkl}=\cS^{ijkl}-\ve^{i(k}\ve^{l)j}\cS~,~~~
\cS^{ijkl}=\cS^{(ijkl)}~.
\eea
The algebra of spinor covariant derivatives 
becomes
\bea
\{\cD_\a^{ij},\cD_\b^{kl}\}&=&
-2\ri\ve^{i(k}\ve^{l)j}(\g^c)_{\a\b}\cD_c
-\ri\ve_{\a\b}(\ve^{jl}\cS^{ikpq}
+\ve^{ik}\cS^{jlpq})\cJ_{pq}
+2\ri\ve_{\a\b}\cS\Big(
\ve^{jl}\cJ^{ik}
+\ve^{ik}\cJ^{jl}
\Big)
\non\\
&&
-\ri\ve^{i(k}\ve^{l)j}C_{\a\b}{}^{pq}\cJ_{pq}
+\ri C_{\a\b}{}^{i(k}\cJ^{l)j}
+\ri C_{\a\b}{}^{j(k}\cJ^{l)i}
+\ri C_{\a\b}{}^{k(i}\cJ^{j)l}
+\ri C_{\a\b}{}^{l(i}\cJ^{j)k}
\non\\
&&
-\ri C_{\a\b}{}^{ij}\cJ^{kl}
-\ri C_{\a\b}{}^{kl}\cJ^{ij}
+\ri\ve_{\a\b}(\ve^{ik}C^{\g\d}{}^{jl}+\ve^{jl}C^{\g\d}{}^{ik})\cM_{\g\d}
\non\\
&&
-4\ri(\cS^{ijkl}-\ve^{i(k}\ve^{l)j}\cS)\cM_{\a\b}
~,
\label{algebraN=3}
\eea
where we have taken into account the fact that
$X^{IJKL}=0$ for $\cN=3$.

The dimension-3/2 Bianchi identities 
become
\bsubeq
\bea
\cD_\a^{ij} \cS^{klpq}&=&
-\frac{1}{2}\ve^{jl}\cT_{\a}{}^{ikpq}
-\frac{1}{2}\ve^{ik}\cT_{\a}{}^{jlpq}
-\frac{1}{2}\ve^{jq}\cT_{\a}{}^{iklp}
-\frac{1}{2}\ve^{ip}\cT_{\a}{}^{jklq}
\non\\
&&
-\hf\cS_\a{}^{kl}\ve^{p(i}\ve^{j)q}
-\hf\cS_\a{}^{pq}\ve^{k(i}\ve^{j)l}
+\frac{1}{3}\cS_\a{}^{ij}\ve^{k(p}\ve^{q)l}
~,
\label{N=3-3/2-S}
\\
\cD_{\a}^{ij}C_{\b\g}{}^{kl}
&=&
\sqrt{2}\ve^{i(k}\ve^{l)j}C_{\a\b\g}
+C_{\a\b\g}{}^{k(i}\ve^{j)l}
+C_{\a\b\g}{}^{l(i}\ve^{j)k}
+\frac{2\sqrt{2}}{3}\ve^{i(k}\ve^{l)j}\ve_{\a(\b}C_{\g)}
+2\ve_{\a(\b}\cT_{\g)}{}^{ijkl}
\non\\
&&
-{4\over 3}\ve_{\a(\b}\Big((\cD_{\g)}^{k(i} \cS)\ve^{j)l}
+(\cD_{\g)}^{l(i} \cS)\ve^{j)k}\Big)
+{1\over 9}\ve_{\a(\b}\Big(\cS_{\g)}{}^{k(i}\ve^{j)l}
+\cS_{\g)}{}^{l(i}\ve^{j)k}\Big)
~.~~~~~~~~~
\label{N=3-3/2-C}
\eea
\esubeq
Here the dimension-3/2 component superfields possess the symmetry properties
\bsubeq
\bea
&\cT_{\a}{}^{ijkl}=\cT_{\a}{}^{(ijkl)}~,~~~
\cS_\a{}^{ij}=\cS_\a{}^{ji}
~,
\\
&C_{\a\b\g}=C_{(\a\b\g)}~,~~~
C_{\a\b\g}{}^{ij}=C_{\a\b\g}{}^{ji}=C_{(\a\b\g)}{}^{ij}
~.
\eea
\esubeq

We conclude by giving the super-Weyl transformation in isospinor notation. 
It holds\bsubeq
\bea
\d_\s \cD_\a^{ij}&=&
\hf \s \cD_\a^{ij} + (\cD^{\b {ij}}\s )\cM_{\a\b}
-(\cD_{\a k}{}^{(i} \s )\cJ^{j)k}~,
\label{sWsD}
\\
\d_\s \cD_a&=&
\s \cD_a
+{\ri\over 2}(\g_a)^{\g\d}(\cD_{\g}^{kl} \s )\cD_{\d kl}
+\ve_{abc}(\cD^b\s )\cM^{c}
+{\ri\over 16}(\g_a)^{\g\d}([\cD_\g^{p(k},\cD_{\d p}^{l)}]\s )\cJ_{kl}
~.~~~~~~~~~~~~
\eea
The super-Weyl transformation laws of the torsion superfields  are
\bea
\d_\s  \cS^{ijkl}&=&
\s \cS^{ijkl}
-{\ri\over 8}  [\cD^{\g(ij},\cD_{\g}^{kl)}]\s 
~,~~~~~~
\label{dUcSijkl}
\d_\s  \cS=
\s \cS
-{\ri\over 24} [ \cD^{\g kl},\cD_{\g kl}]\s 
~,~~~~~~
\\
\d_\s  C_{a}{}^{ij}&=&
\s C_{a}{}^{ij}
-{\ri\over 8}( \g_a)^{\g\d} [\cD_\g^{k(i},\cD_{\d k}^{j)}]\s ~.
\eea
\esubeq

\subsection{Covariant projective multiplets}
\label{subsection5.2}

In this section we introduce a large family of $\cN=3$ (matter) supermultiplets  
coupled to conformal supergravity -- covariant projective multiplets. 
One of the simplest projective multiplets, the so-called $\cO(2)$ multiplet, 
is naturally associated with the field strength of a $\cN=3$  vector  multiplet. 
Although being the simplest in the family, it displays many properties of the general 
projective multiplets. 
We therefore start by  considering this particular multiplet,
and then turn to the general case.

The antisymmetric field strength of the 
vector multiplet,  $W^{IJ}$,  is equivalently described by the symmetric isospinor
$W^{ij}$ which originates as
\bea
W^{IJ}~\to~W^{ijkl}=-\hf\ve^{jl}W^{ik}-\hf\ve^{ik}W^{jl}~.
\eea
In terms of $W^{ij}$ the 
Bianchi identity
(\ref{2.35}) 
turns into
the analyticity constraint
\bea
\cD_\a^{(ij}W^{kl)}=0
~.
\label{a0}
\eea

Let us
introduce a complex commuting isospinor, $v^{i} \in {\mathbb C}^2 \setminus  \{0\}$,
and use it to define the  derivative\footnote{Our conventions for isospinor bosonic variables and
projective multiplets differ slightly from \cite{KLRT-M1,KLRT-M2}, but agree with 
those adopted in  \cite{Kuzenko:2010bd}.}
\bea
\cD_\a^{(2)}:=v_iv_j\cD_\a^{ij}~,
\eea
as well as  the superfield
\bea
W^{(2)}:=v_iv_jW^{ij}~.
\eea
Then, the constraint (\ref{a0}) is equivalent to 
\bea
\cD_\a^{(2)}W^{(2)}=0~.
\label{5.21}
\eea
The superscripts, which are attached to $W^{(2)}$ and $\cD^{(2)}_\a$,
indicate the degree of homogeneity in $v^i$.
Similarly to the local superspace coordinates $z^M$, the isospinor $v^{i}$ is defined
to be inert under the local structure-group transformations. Its sole role is to package 
the field strength  $W^{ij}$ 
into an index-free object. 
This  interpretation of $v^i$ as a book-keeping device
is discussed in detail in \cite{KLRT-M1}.

In accordance with  (\ref{algebraN=3}), 
the spinor covariant derivatives $\cD_\a^{(2)}$  satisfy the algebra
\bsubeq
\bea
&\{\cD_\a^{(2)},\cD_\b^{(2)}\}=
-4\ri \cS^{(4)}\cM_{\a\b}
+2\ri C_{\a\b}^{(2)}\cJ^{(2)}
~,
\label{N=3D2D2}
\eea
where we have defined
\bea
&
C_{\a\b}^{(2)}:=v_iv_jC_{\a\b}{}^{ij}~,\qquad
\cS^{(4)}:=v_iv_jv_kv_l\cS^{ijkl}
~,\qquad
\cJ^{(2)}:=v_iv_j\cJ^{ij}~.
\label{N=3Ta-1}
\eea
\esubeq
It follows from (\ref{N=3D2D2}) that the constraint (\ref{5.21}) is consistent.
Indeed, the  SU(2) transformation 
\bea
\cJ_{ij}W_{kl}=-\ve_{k(i}W_{j)l}-\ve_{l(i}W_{j)k}
\eea
implies $\cJ^{(2)} W^{(2)}=0$.

Under the infinitesimal supergravity gauge transformation,  
\bea
\d_K\cD_A=[K,\cD_A]~,\qquad
K=K^C(z)\cD_C+\hf K^{cd}(z)\cM_{cd}+\hf K^{kl}(z)\cJ_{kl}~,
\eea
the field strength $W^{ij}$ changes as 
\bea
\d_K W^{ij} = K^C \cD_C W^{ij} + W^{l(i}K^{j)}{}_l~.
\eea
In terms of 
 $W^{(2)}$, this transformation law can be rewritten in the form:
\bsubeq
\bea
\d_K W^{(2)} 
&=& \Big( K^{{C}} \cD_{{C}} + \hf K^{ij} \cJ_{ij} \Big)W^{(2)} ~,  
\label{W2t1}
\\ 
K^{ij} \cJ_{ij}  W^{(2)}&=& -\Big(K^{(2)} {\bm \pa}^{(-2)} 
-2 \, K^{(0)}\Big) W^{(2)} ~. 
\label{W2t2}
\eea
\esubeq
Here we have denoted
\bea
K^{(2)} :=K^{ij}\, v_i v_j 
~,\qquad
K^{(0)} :=\frac{v_i u_j }{(v,u)}K^{ij}~,
\qquad (v,u):=v^iu_i
\label{W2t3}
\eea
and also introduced the differential operator 
\bea
{\bm \pa}^{(-2)} :=\frac{1}{(v,u)}u^{i}\frac{\pa}{\pa v^{i}}~.
\label{5.28}
\eea
The expressions in (\ref{W2t3}) and (\ref{5.28})
involve a new isospinor  $u_{i}$ 
which is subject to
the condition $(v,u)\ne0$, but otherwise completely arbitrary.
By definition, $W^{(2)}$ is independent of $u_i$. The variation  $\d_K W^{(2)} $
can be seen to be independent of $u_i$ as well, in spite 
of the fact that each of the two terms on the right-hand side of (\ref{W2t2}) involves $u_i$.

In accordance with (\ref{2.37}),  the super-Weyl transformation of $W^{(2)}$ is
\bea
\d_\s W^{(2)}=\s W^{(2)}~.
\eea
It may be seen that the analyticity constraint  (\ref{5.21}) and the functional form of $W^{(2)}$
uniquely determine the super-Weyl transformation law of $W^{(2)}$.
This is  similar to the properties of the $\cN=1$ and $\cN=2$ vector multiplets. 

The condition  $(v,u)\ne0$ means that  $v^{i}$ and $u^{i}$ form a basis for ${\mathbb C}^2$. 
Therefore the isospinors $v^{i}$ and $u^{i}$ can be used to 
define a new  basis for the isospinor indices, with the aid of the completeness relation
\bea
\d_j^i=\frac{1}{ (v,u)}\big(v^{i}u_j- v_j u^{i} \big) ~.
\label{completeness}
\eea
Specifically, associated with a symmetric valence-$n$ isospinor
$T^{i_1 \dots i_n} = T^{(i_1 \dots i_n)}$
is a set of $(n+1)$ index-free objects
\bea
T^{(n-2m)} := T^{i_1 \cdots i_{n-m} i_{n-m+1} \cdots  i_n} v_{i_1} \cdots 
v_{i_{n-m}} \frac{u_{i_{n-m+1}}}{(v,u)} \cdots \frac{u_{i_n}}{(v,u)}~, \quad 
m=0,1,\dots, n~~~
\label{ChangingBasis}
\eea
which are homogeneous in  $v$  and $u$ of  degrees $n-2m$ and 0, 
respectively.\footnote{in some situations, 
in order to avoid possible misunderstanding, 
it would be more precise to use the notation $T^{(n-m,m)} $ instead of $T^{(n-2m)} $.
Such a notation is not used in this paper.}
For example, starting from the spinor covariant derivatives $\cD_\a^{ij}$, we generate
\bea
\cD_\a^{(2)}:=v_iv_j\cD_\a^{ij}~,\qquad
\cD^{ (0)}_\a:=\frac{v_iu_j}{(v,u)}\cD^{ij}_\a~,
\qquad
\cD^{ (-2)}_\a:=\frac{u_iu_j}{(v,u)^2}\cD^{ij}_\a~.
\eea
Applying this rule to the SU(2) generators $\cJ^{ij}$ gives
\bea
\cJ^{(2)}:=v_iv_j\cJ^{ij}~,\qquad
\cJ^{ (0)}:=\frac{v_iu_j}{(v,u)}\cJ^{ij}~,
\qquad
\cJ^{ (-2)}:=\frac{u_iu_j}{(v,u)^2}\cJ^{ij}~.
\eea

We are now prepared to define general projective multiplets.
A {\em covariant projective supermultiplet} of weight $n$,
$Q^{(n)}(z,v)$, is defined to be 
a Lorentz-scalar superfield that 
lives on the curved  $\cN=3$ superspace ${\cM}^{3|6}$, 
is holomorphic with respect to 
the isospinor variables $v^i $ on an open domain of 
${\mathbb C}^2 \setminus  \{0\}$, 
and is characterized by the following conditions:\\
(i) it obeys the covariant analyticity constraint
\be
\cD^{(2)}_{\a} Q^{(n)}  =0~;
\label{ana}
\ee  
(ii) it is  a homogeneous function of $v$ 
of degree $n$, that is,  
\be
Q^{(n)}(z,c\,v)\,=\,c^n\,Q^{(n)}(z,v)~, \qquad c\in \mathbb{C}^* \equiv {\mathbb C} \setminus  \{0\}~;
\label{weight}
\ee
(iii)  supergravity gauge transformations act on $Q^{(n)}$ 
as follows:
\bea
\d_K Q^{(n)} 
&=& \Big( K^{{C}} \cD_{{C}} + \hf K^{ij} \cJ_{ij} \Big) Q^{(n)} ~,  
\non \\ 
K^{ij} \cJ_{ij}  Q^{(n)}&=& -\Big(K^{(2)} {\bm \pa}^{(-2)} 
-n \, K^{(0)}\Big) Q^{(n)} ~.
\label{harmult1}   
\eea 
Note that by construction, $Q^{(n)}$ is independent of $u$, 
i.e. $\pa  Q^{(n)} / \pa u^{i} =0$.
One can see that $\d_K Q^{(n)} $ 
is also independent of the isotwistor $u$, $\pa (\d_K Q^{(n)})/\pa u^{i} =0$,
due to (\ref{weight}). 
It is also important to note that eq. (\ref{harmult1}) implies that
\bea
\cJ^{(2)} Q^{(n)}=0~,
\label{J++}
\eea
and hence the covariant analyticity constraint (\ref{ana}) is indeed consistent.

The analyticity constraint (\ref{ana}) and the homogeneity condition (\ref{weight}) 
are consistent with the interpretation that 
the isospinor
$ v^{i} \in {\mathbb C}^2 \setminus\{0\}$ is   defined modulo the equivalence relation
$ v^{i} \sim c\,v^{i}$,  with $c\in {\mathbb C}^*$, {hence it parametrizes ${\mathbb C}P^1$}.
Therefore, the projective multiplets live in ${\cM}^{3|6} \times {\mathbb C}P^1$.

There exists 
a real structure on the space of projective multiplets.
Given a  weight-$n$ projective multiplet $ Q^{(n)} (v^{i})$, 
its {\it smile conjugate},
$ \breve{Q}^{(n)} (v^{i})$, is defined by 
\bea
 Q^{(n)}(v^{i}) \longrightarrow  {\bar Q}^{(n)} ({\bar v}_i) 
  \longrightarrow  {\bar Q}^{(n)} \big({\bar v}_i \to -v_i  \big) =:\breve{Q}^{(n)}(v^{i})~,
\label{smile-iso}
\eea
with ${\bar Q}^{(n)} ({\bar v}_i)  :=\overline{ Q^{(n)}(v^{i} )}$
the complex conjugate of  $ Q^{(n)} (v^{i})$, and ${\bar v}_i$ the complex conjugate of 
$v^{i}$. One can show that $ \breve{Q}^{(n)} (v)$ is a weight-$n$ projective multiplet.
In particular,   $ \breve{Q}^{(n)} (v)$
obeys the analyticity constraint $\cD_\a^{(2)}\breve{Q}^{(n)} =0$,
unlike the complex conjugate of $Q^{(n)}(v) $.
One can also check that 
\bea
\breve{ \breve{Q}}^{(n)}(v) =(-1)^n {Q}^{(n)}(v)~.
\label{smile-iso2}
\eea
Therefore, if  $n$ is even, one can define real projective multiplets, 
 $\breve{Q}^{(2n)} = {Q}^{(2n)}$.
Note that geometrically, the smile-conjugation is complex conjugation composed
with the antipodal map on the projective space ${\mathbb C}P^1$.

Let $Q^{(n)} (z,v) $ be a projective supermultiplet of weight $n$.
Assuming that it transforms homogeneously under the super-Weyl transformations,
the analyticity constraints uniquely fix its transformation law:
\be
\d_{\s} Q^{(n)} =\frac{n}{2}\s Q^{(n)} ~.
\label{Q(n)super-Weyl}
\ee

Our definition of the 3D $\cN=3$ projective multiplets given above is  reminiscent of 
the covariant projective multiplets in 4D $\cN=2$ conformal supergravity \cite{KLRT-M2} 
or 5D $\cN=1$ conformal supergravity \cite{KT-M5D-2}. However, the three-dimensional 
case has two 
specific features as compared to four and five dimensions. 
First of all, the analyticity constraint (\ref{ana}) is formulated 
in terms of two spinor operators, $\cD^{(2)}_\a$, while the 4D projective multiplets 
are annihilated by four 
 derivatives $\cD^{(1)}_\a:= \cD^i_\a v_i$ and ${\bar \cD}^{(1)}_\ad := {\bar \cD}^i_\ad v_i$.
Secondly, the operators $\cD^{(2)}_\a$ are quadratic in the isotwistor variables $v_i$, while 
their four-dimensional analogues, $\cD^{(1)}_\a$ and ${\bar \cD}^{(1)}_\ad $, are linear in $v_i$.

We now list several projective multiplets that can be  used to describe superfield 
dynamical variables.
A natural generalization of the field strength $W^{(2)} (v)$ is a real $\cO(2n) $ multiplet, 
with $n=1,2,\dots$. It  is described by a real weight-$2n$ projective superfield $H^{(2n)} (v)$ 
of the form:
\bea
H^{(2n)} (v) &=& H^{i_1 \dots i_{2n}} v_{i_1} \dots v_{i_{2n}} 
=\breve{H}^{(2n)} (v) ~.
\eea
The analyticity constraint (\ref{ana}) is equivalent to 
\bea
\cD_\a^{(ij} H^{k_1 \dots k_{2n} )} =0~.
\eea
The reality condition $\breve{H}^{(2n)}  = {H}^{(2n)} $ is equivalent to 
\bea
\overline{ H^{i_1 \dots i_{2n}} } &=& H_{i_1 \dots i_{2n}}
=\ve_{i_1 j_1} \cdots \ve_{i_{2n} j_{2n} } H^{j_1 \dots j_{2n}} ~.
\eea
The field strength of the vector multiplet is a real $\cO(2) $ multiplet. For $n>1$, 
the real $\cO(2n) $ multiplet can be used to describe an off-shell (neutral) hypermultiplet. 

An off-shell (charged) hypermultiplet can be described in term of the so-called {\it arctic} 
weight-$n$ multiplet $\U^{(n)} (v)$ which is defined to be 
holomorphic  in the north chart  $\mathbb C$, 
of the projective space ${\mathbb C}P^1 ={\mathbb C} \cup \{\infty \}$: 
\bea
\U^{(n)} ( v) &=&  (v^{1})^n\, \U^{[n]} ( \z) ~, \qquad 
\U^{ [n] } ( \z) = \sum_{k=0}^{\infty} \U_k  \z^k 
~, 
\label{arctic1}
\eea
and  its smile-conjugate {\it antarctic} multiplet $\breve{\U}^{(n)} (v) $,
 \bea
\breve{\U}^{(n)} (v) &=& 
(v^{2}  \big)^{n}\, \breve{\U}^{[n]}(\z) =
(v^{1} \,\z \big)^{n}\, \breve{\U}^{[n]}(\z) ~, \qquad
\breve{\U}^{[n]}( \z) = \sum_{k=0}^{\infty}  {\bar \U}_k \,
\frac{(-1)^k}{\z^k}~.~~~
\label{antarctic1}
\eea
Here we have introduced the inhomogeneous complex coordinate 
$\z= v^2/v^1$ on the north chart of  ${\mathbb C}P^1$.
The pair consisting of $\U^{[n]} ( \z)$ and $\breve{\U}^{[n]}(\z) $ 
constitutes the so-called polar weight-$n$ multiplet.
The spinor covariant derivative $\cD^{(2)}_\a$ can be represented as 
\bea
\cD^{(2)}_{\a} = (v^1)^2 \cD^{[2]}_{\a}~, \qquad \cD^{[2]}_{\a} (\z)
&=& \cD^{22}_\a -2 \z \cD^{12}_\a +\z^2 \cD^{11}_\a  ~.
\eea
It follows from this representation that the analyticity  condition (\ref{ana}) 
relates,  in a nontrivial way, the superfield coefficients $\U_k$ in the series (\ref{arctic1}).

Our last example is  the real {\it tropical} multiplet $\cU^{(2n)} (v) $ of weight $2n$ defined by 
\bea
\cU^{(2n)} (v) &=&\big({\rm i}\, v^{1} v^{2}\big)^n \cU^{[2n]}(\z) =
\big(v^{1}\big)^{2n} \big({\rm i}\, \z\big)^n \cU^{[2n]}(\z)~,  \non \\
\cU^{[2n]}(\z) &=& 
\sum_{k=-\infty}^{\infty} \cU_k  \z^k~,
\qquad  {\bar \cU}_k = (-1)^k \cU_{-k} ~.
\label{2n-tropica1}
\eea
As will be shown below, the case $n=0$ can be used to describe a 
gauge prepotential of the vector multiplet.

\subsection{Analytic projection operator}

In this subsection we show how to engineer covariant projective multiplets.

The torsion superfield $\cS^{(4)}$  was defined in subsection \ref{subsection5.2}, 
 eq. (\ref{N=3Ta-1}).
It proves to be a real $\cO(4)$ multiplet.
Indeed, the equation (\ref{N=3-3/2-S}) implies the relation $\cD_\a^{(ij}\cS^{klpq)}=0$
which is equivalent to the analyticity constraint $\cD_\a^{(2)}\cS^{(4)}=0$. 
It is easy to see that $\cS^{(4)}$  does not enjoy the super-Weyl transformation law
(\ref{Q(n)super-Weyl}).
As follows from  eq. (\ref{dUcSijkl}),
its super-Weyl transformation is inhomogeneous, 
\bea
\d_\s\cS^{(4)}
=
2\s\cS^{(4)}
-\frac{\ri}{4}\Big(\cD^{(4)}-4\ri\cS^{(4)}\Big)\s
~,
\label{dUcS4}
\eea
where $\cD^{(4)}$ is defined by
\bea
\cD^{(4)}:=\cD^{(2)\g}\cD^{(2)}_\g
~.
\eea
The appearance in (\ref{dUcS4}) of the following differential operator 
\bea
\D^{(4)}:=\frac{\ri}{4}\Big(\cD^{(4)}-4\ri\cS^{(4)}\Big)
\eea
is not accidental. 
This operator turns out to be a $\cN=3$ {\it analytic projection operator}.
In particular,    $\D^{(4)}$ is 
 such that the constraint $\cD_\a^{(2)}\cS^{(4)}=0$ is preserved under the super-Weyl 
 transformations, $\d_\s(\cD_\a^{(2)}\cS^{(4)})=0$. This follows from the
transformation rule
\bea
\d_\s\cD_\a^{(2)}&=&
\hf \s\cD_\a^{(2)} 
+ (\cD^{(2)\b}\s)\cM_{\a\b}
+(\cD_{\a}^{(2)} \s)\cJ^{(0)}
-(\cD_{\a}^{(0)} \s)\cJ^{(2)}
~,
\label{N=3sWD2}
\eea
the identity
\bea
\cD_\a^{(2)}\D^{(4)}
&=&
\frac{1}{2} C_{\a\b}^{(2)}\cD^{(2)\b}\cJ^{(2)}
+\frac{1}{6} \Big(\cD^{(2)\b}C_{\a\b}^{(2)}\Big)\cJ^{(2)}
-\cS^{(4)}\cD^{\b(2)}\cM_{\a\b}
~,
\label{N=3D2D4}
\eea
and the obvious relation $\cJ^{(0)}\cS^{(4)}=-2\cS^{(4)}$.
Note that the above super-Weyl transformations of $\cD_\a^{(2)}$ follows from (\ref{sWsD}).

Let us formulate more precisely what we mean by `analytic projection operator.'  First of all, 
we have to introduce the concept 
of isotwistor superfields, following \cite{KLRT-M1}.
By definition, a weight-$n$ isotwistor superfield  $U^{(n)}$ is a {\it tensor} superfield 
(with suppressed Lorentz indices)
that  lives on  ${\cM}^{3|6}$, 
is holomorphic with respect to 
the isospinor variables $v^i $ on an open domain of 
${\mathbb C}^2 \setminus  \{0\}$, 
is a homogeneous function of $v^i$ of degree $n$,
\bsubeq
\bea
&&U^{(n)}(c\,v)\,=\,c^n\,U^{(n)}(v)~, \qquad c\in \mathbb{C}^*,
\eea
and is  characterized by the supergravity gauge transformation
\bea
\d_K U^{(n)} 
&=& \Big( K^{{C}} \cD_{{C}} +  \hf K^{ab} \cM_{ab}+\hf K^{ij} \cJ_{ij} \Big) 
U^{(n)} ~,  ~~~\non \\
\cJ_{ij}  U^{(n)}&=& -\Big(v_{(i}v_{j)}{\bm \pa}^{(-2)} 
-\frac{n}{(v,u)}v_{(i}u_{j)}\Big) U^{(n)} \quad \Longrightarrow \quad
\cJ^{(2)}U^{(n)}=0~.
\label{iso2}
\eea 
\esubeq
It is clear that any weight-$n$ projective multiplet  is an isotwistor superfield, but not vice versa.
If $U^{(n-4)}$ is a Lorentz scalar, it follows from  (\ref{N=3D2D4}) that 
   the weight-$n$ isotwistor superfield $Q^{(n)}:=\D^{(4)}U^{(n-4)}$
 obeys  the analyticity constraint  $\cD_\a^{(2)}Q^{(n)}=0$, and therefore it is a projective multiplet.
 One can also prove that if under the super-Weyl  transformations $U^{(n-4)}$ 
 varies as a primary field of special weight,
 \bea
 \d_\s U^{(n-4)}=\frac{(n-2)}{2}\s U^{(n-4)}
 ~,
\label{super-Weyl-U}
 \eea
then $\D^{(4)}U^{(n-4)}$ also transforms homogeneously according to 
eq. (\ref{Q(n)super-Weyl}).
The derivation of 
this property requires 
some straightforward algebra making use of  eq. (\ref{N=3sWD2})
and the relations
\bea
{[}\cJ^{(2)},\cD_\a^{(2)}{]}=0~,~~~
{[}\cJ^{(0)},\cD_\a^{(2)}{]}=-\cD_\a^{(2)}~,~~~
\cJ^{(0)}U^{(n-4)}=-\frac{(n-4)}{2}U^{(n-4)}
~.
\eea
As a simple application of the construction described, we
note that one can build a weight-$4$ projective superfield, 
$\D^{(4)}P$, from 
a $v$-independent scalar superfield $P$. 
The only condition that $P$ has to satisfy is to have weight one under
super-Weyl transformations $\d_\s P=\s P$.

The careful reader should have noticed that the explicit form of the analytic projector operator
is formally equivalent to that of  the antichiral projection operator 
\bea
\D=-\frac{1}{4} (\cD^2 -4 \bar R )
\eea
in $\cN=2$ conformal supergravity, see subsection \ref{subsection4.2}.
This is not surprising if one notes that the anti-commutation relation  (\ref{N=3D2D2})
reduces to 
\bea
\{\cD_\a^{(2)},\cD_\b^{(2)}\} U^{(n)}=
-4\ri \cS^{(4)}\cM_{\a\b} U^{(n)}
\label{5.57}
\eea
when acting on an arbitrary {\it isotwistor} superfield  $U^{(n)}$.
This result  is analogous to the first anti-commutation relation in (\ref{N=2-alg-1}),
\bea
\{\cD_\a,\cD_\b\} U &=&
-4\bar{R}\cM_{\a\b} U~,
\label{5.58}
\eea
for any $\cN=2$ tensor superfield. 
The relations (\ref{5.57}) and (\ref{5.58}) show an analogy between 
$\cN=3$ projective multiplets, $\cD^{(2)}_\a Q^{(n)}=0$, and 
$\cN=2$ antichiral  superfields, $\cD_\a \bar \J =0$.
In particular, both $Q^{(n)}$ and $\bar \J$ must be scalar with respect to the Lorentz group.

\subsection{Vector multiplet prepotential}
In this subsection we show that the constraints obeyed 
by the $\cN=3$  vector-multiplet field strength $W^{ij}$
 can be solved in terms of a real weight-zero tropical prepotential
$V(v^i)$ defined modulo arbitrary gauge transformations of the form
\bea
\d V = \l + \breve{\l}~, 
\label{5.59}
\eea
where $\l (v^i)$ is an arctic weight-zero multiplet.\footnote{In 4D $\cN=2$ rigid supersymmetry, 
the idea to describe the massless vector multiplet in terms of a  tropical multiplet appeared for the 
first time 
in \cite{LR-projective2}. The transformation law (\ref{5.59}) is a locally supersymmetric version  of 
that given in \cite{LR-projective2}.}
Conceptually, this is similar to the situation in 4D $\cN=2$ conformal supergravity 
in which the covariantly chiral field strength of a vector multiplet is also given 
in terms of a weight-zero real tropical prepotential \cite{KT-M_4DconfFlat,K-08}, as an extension 
of the rigid-supersymmetric constructions given in \cite{LR-projective2,K-compactified}. 
Technically, the 3D solution which we are going to present turns out to differ significantly from its 
four-dimensional counterpart.

We start from the real weight-zero tropical multiplet $V(v^i)$ and associate with it a 
weight-two isotwistor superfield $W^{(2)} (w^i)$ defined by 
\bea
W^{(2)}(w)&:=&\frac{1}{8\pi\ri}\oint_\g (v,\rd v)
\Big\{
(w,v)^2\cD^{\a (-2)}\cD_\a^{(-2)}
-4\frac{(w,v)(w,u)}{(v,u)}\cD^{\a (-2)}\cD_\a^{(0)}
\non\\
&&~~~
+4\frac{(w,u)^2}{(v,u)^2}\cD^{\a (0)}\cD_\a^{(0)}
-4\ri(w,v)^2\cS^{(-4)}
+8\ri\frac{(w,v)(w,u)}{(v,u)}\cS^{(-2)}
\non\\
&&~~~
-16\ri\frac{(w,u)^2}{(v,u)^2}\cS^{(0)}
+8\ri \frac{(w,u)^2}{(v,u)^2}\cS
\Big\}
V(v)
~,
\label{N=3_Vector_prepot}
\eea
for some closed integration contour $\g$.
Here the integrand involves the  superfields
\bea
&&\cS^{(-4)}:=\frac{u_iu_ju_ku_l}{(v,u)^4}\cS^{ijkl}~,~~~
\cS^{(-2)}:=\frac{v_iu_ju_ku_l}{(v,u)^3}\cS^{ijkl}~,~~~
\cS^{(0)}:=\frac{v_iv_ju_ku_l}{(v,u)^2}\cS^{ijkl}~
~~~~~~
\eea
which are defined in accordance with our general conventions introduced earlier.
It follows from (\ref{N=3_Vector_prepot}) that $W^{(2)}(w)$ has the following functional form:
$W^{(2)} (w)= W^{ij} w_i w_j $, for some real SU(2) triplet $W^{ij}$.
The field strength (\ref{N=3_Vector_prepot}) is indeed invariant under the gauge 
transformations (\ref{5.59}).

A crucial property of (\ref{N=3_Vector_prepot}) is that it does not depend on the auxiliary 
isospinor $u^i$. This property can be proved  considering an infinitesimal transformation 
$\d u^i = a v^i$
and then making use of the analyticity condition $\cD^{(2)}_\a V=0$ in conjunction with 
the anticommutation 
relations for the spinor covariant derivatives.

 The fact that (\ref{N=3_Vector_prepot}) 
 is independent  of $u^i$ can be used to derive two important implications.
 First of all, it allows us to prove the invariance of $W^{(2)}$ under
the gauge transformation (\ref{5.59}).
Secondly, it makes it possible to prove that  $W^{(2)} (w)$ is a projective multiplet.
Indeed, let us 
choose $u^i=w^i$ in (\ref{N=3_Vector_prepot}) 
and also re-label $w^i \to v^i$ and $v^i \to \hat{v}{}^i$. Then (\ref{N=3_Vector_prepot}) 
turns into\footnote{To prove this
the reader should use eq. (\ref{N=3D2D4}) and the fact that 
$w_iw_j\cJ^{ij}\oint\frac{(v,\rd v)}{(v,w)^2}V(z,v)=0$.}
\bea
W^{(2)}(v)=
 \D^{(4)} \oint_\g \frac{ ({\hat v},\rd {\hat v})}{  2\p(v, {\hat v})^2}\,V( {\hat v})
~.
\label{5.622}
\eea
This representation makes it 
manifest that $W^{(2)}(v)$ is a projective multiplet.

We postulate the prepotential $V$ to be inert under the super-Weyl transformations,
\bea
\d_\s V=0~
\eea
This  leads to the correct transformation law for $W^{(2)}$.

\subsection{Supersymmetric  action principle}

With the results obtained in the previous subsections, we are now prepared
to formulate a locally supersymmetric and super-Weyl invariant action principle.

Similarly to the off-shell supergravity-matter systems with eight supercharges
in four and five dimensions \cite{KT-M5D-2,KLRT-M2},
our Lagrangian $\cL^{(2)}$ is chosen  
to be a real weight-2  projective multiplet, with the following
super-Weyl transformation law 
\bea
\d_\s\cL^{(2)}=\s \cL^{(2)}~.
\label{5.64}
\eea
Associated with $\cL^{(2)}$ is the action
\bea
S(\cL^{(2)})&=&
\frac{1}{2\pi\ri} \oint_\g (v, \rd v)
\int \rd^3 x \,{\rm d}^6\q\,E\, C^{(-4)}\cL^{(2)}~, 
\qquad E^{-1}= {\rm Ber}(E_A{}^M)~.
\label{InvarAc}
\eea
Here the superfield $C^{(-4)}$ is required to be a Lorentz-scalar 
isotwistor superfield of weight $-4$ 
such that 
the following two conditions hold:
\begin{subequations}
\bea
\d_\s C^{(-4)}&=&-\s C^{(-4)}~,
\label{AcComp-a}\\
\D^{(4)}C^{(-4)}&=&1~.
\label{AcComp-b}
\eea
\end{subequations}
These conditions prove to guarantee 
that the action (\ref{InvarAc}) is invariant under the  supergravity gauge
and the super-Weyl transformations.
The invariance of $S$ under the supergravity gauge transformations can be proven 
in complete analogy 
to the 5D and 4D cases \cite{KT-M5D-1,KT-M5D-2,KLRT-M1,KLRT-M2}. 
To show that the action (\ref{InvarAc}) is 
super-Weyl invariant, 
it is necessary to make use of the super-Weyl transformation laws
(\ref{5.64}) and  (\ref{AcComp-a}), 
as well as to use the observation that 
\bea
\d_\s E=0~,
\eea
which is similar to the 4D $\cN=2$ case.

All information about a concrete dynamical system is encoded in its Lagrangian $\cL^{(2)}$.
It may look somewhat odd that the action  (\ref{InvarAc}) also involves 
the `compensating' superfield $C^{(-4)}$, in principle one and the same for all dynamical systems.
The important point, however, is that
the action (\ref{InvarAc}) does not 
depend on  $C^{(-4)}$ if the Lagrangian $\cL^{(2)} $ is independent of $C^{(-4)}$.
To prove this statement, let us represent the Lagrangian as
$\cL^{(2)}=\D^{(4)} U^{(-2)}$, for some
isotwistor superfield  $U^{(-2)}$ of weight $-2$.
Then, making use of eq. (\ref{AcComp-b}) allows us 
to rewrite the action  in the form
\bea
S&=&
\frac{1}{2\pi\ri} \oint_\g (v ,\rd v)
\int \rd^3 x \,{\rm d}^6\q\,E\, U^{(-2)}~.
\label{InvarAc2}
\eea
This representation makes manifest the fact that  the action does not depend on $C^{(-4)}$.

A natural choice for  $C^{(-4)}$ is available  if the theory under consideration 
possesses an Abelian  vector 
multiplet such that its field strength $W^{ij}$ is nowhere vanishing, 
that is $W:=\sqrt{W^{ij}W_{ij}}\neq 0$.
Such a vector multiplet may be a conformal compensator. 
Since the super-Weyl transformation of $W$ is 
\bea
\d_\s W=\s W~,
\eea
we immediately observe that 
$C^{(-4)}$ can be chosen as
\bea
C^{(-4)}:=\frac{W}{\S^{(4)}}~,~~~~~~
\S^{(4)}:=\D^{(4)}W
~.
\eea
Indeed, the condition (\ref{AcComp-a}) holds since the super-Weyl transformation 
of $\S^{(4)}$ is 
\bea
\d_\s\S^{(4)}=2\s\S^{(4)}~.
\eea
The condition (\ref{AcComp-b}) holds, since $\S^{(4)}$ is an $\cO(4)$ multiplet.

More generally, given a real weight-$n$ isotwistor superfield $\cU^{(n)}$,
with the super-Weyl transformation law (\ref{super-Weyl-U}),
it is possible to define 
$C^{(-4)}$ as 
\bea
C^{(-4)}=\frac{\cU^{(n)}}{\D^{(4)}\cU^{(n)}}
~,
\eea
provided $\big( \D^{(4) } \cU^{(n)} \big)^{-1}$ is well defined.

The action (\ref{InvarAc}) has the following important property:
\be
S\Big(W^{(2)} (\l +  \breve{\l} )\Big) =0~, 
\ee 
with 
$W^{(2)} $ a real $\cO(2)$ multiplet and $\l$ an arctic weight-zero multiplet.

\subsection{Conformal compensators}
\label{subsection5.6}

As is well known, conformal supergravity is a useful starting point to construct 
Poincar\'e supergravity
theories \cite{KT}. This is achieved by coupling 
the ({\it Weyl multiplet}) (i.e. the multiplet of conformal supergravity)
to a compensating matter multiplet ({\it compensator}). The latter
allows one to gauge away part of the local symmetries by imposing appropriate gauge conditions.  
In 4D $\cN=2$ supergravity, two compensators are required of which one is a vector multiplet
(see \cite{deWPV} and references therein). In the case of $\cN=3$ supergravity 
in three dimensions,
 the vector multiplet can be chosen as  a compensator.
Its field strength $W^{ij}$ must be  nowhere vanishing, 
that is $W:=\sqrt{W^{ij}W_{ij}}\neq 0$. Then, the super-Weyl gauge freedom can be used 
to impose 
the gauge condition $W=1$. After that, the local SU(2) symmetry allows one to set 
$W^{ij} = w^{ij}$,
for some constant SU(2) triplet $w^{ij}$ of unit length.

The supergravity Lagrangian is
\bea
\cL^{(2)}_{\rm SUGRA} = \frac{1}{\k^2} \,
W^{(2)} \ln \frac{W^{(2)} }{\ri \Upsilon^{(1)} \breve\Upsilon^{(1)}} 
+\frac{\x}{\k^2}\, V W^{(2)}~,
\label{5.74}
\eea
with $\k^2$ and $\x$ the gravitational and cosmological  constants, respectively. 
The cosmological term is described by a U(1) Chern-Simons term.
The action is invariant under the gauge transformations (\ref{5.59}). 
The first term in (\ref{5.74}) is (minus) the Lagrangian for a massless
 improved vector multiplet coupled to conformal supergravity. Its 4D $\cN=2$ counterpart
was given in \cite{K-08}  as a locally supersymmetric extension of the projective-superspace 
formulation \cite{KLR} for the 4D $\cN=2$ improved tensor multiplet \cite{deWPV,LR83}.

The supergravity action can  equivalently be described by the following Lagrangian
\bea
\tilde{\cL}{}^{(2)}_{\rm SUGRA} = \frac{1}{\k^2} \,
V \Big\{ 
{\mathbb W}^{(2)} 
+
\x W^{(2)}\Big\}~,
\eea
where 
\bea
{\mathbb W}^{(2)} := {\mathbb W}^{ij}v_iv_j
=\D^{(4)}
\oint\frac{({\hat v},\rd {\hat v})}{ 2\pi (v, {\hat v} )^2}\ln \frac{W^{(2)} ( {\hat v}) }
{\ri \Upsilon^{(1) } ( {\hat v}) \breve\Upsilon^{(1)}  ( {\hat v})} ~, \qquad
\cD^{(2)}_\a {\mathbb W}^{(2)} =0
\label{5.62}
\eea
is a composite real $\cO (2) $ multiplet. The contour integral in (\ref{5.62}) can be evaluated 
using the technique developed in \cite{BK}.

In the case $\x =0$, we can construct a dual supergravity formulation by considering 
the first-oder model 
\bea
\cL^{(2)}_{\mbox{first-order}} = \frac{1}{\k^2} \,\cU^{(2)} \Big(
 \ln \frac{\cU^{(2)} }{\ri \Upsilon^{(1)} \breve\Upsilon^{(1)}} -1 \Big)~,
\eea
where $\cU^{(2)}$ is a real weight-two tropical multiplet.
Varying the first-order action with respect to $\U^{(1)} $ and its conjugate gives 
$\cU^{(2)} =W^{(2)} $, 
and then we return to the original formulation. On the other hand, varying the first-order action 
with respect to $\cU^{(2)}$ gives $\cU^{(2)}  =\ri \Upsilon^{(1)} \breve\Upsilon^{(1)}$, and 
we arrive at the dual formulation
\bea
\cL^{(2)}_{\rm SUGRA,dual} = - \frac{\ri}{\k^2} \,
\Upsilon^{(1)} \breve\Upsilon^{(1)}~
\eea
in which the compensator is an off-shell hypermultiplet.

If the cosmological constant is non-zero, $\x \neq 0$, then the theory (\ref{5.74}) proves to be
 self-dual under a different type of duality transformation that is similar to the one considered in 
 \cite{K-08}.

\subsection{Locally supersymmetric sigma-models}
\label{subsection5.7}

The Lagrangian in (\ref{InvarAc}) is required to be a real weight-two covariant projective multiplet 
with the super-Weyl transformation law (\ref{5.64}). Otherwise $\cL^{(2)}$ may be 
completely arbitrary. 
This freedom in the choice of  $\cL^{(2)}$ means that practically any off-shell  $\cN=3$ 
rigid superconformal theory
\cite{KPT-MvU} can be coupled to $\cN=3$ conformal supergravity.

We consider a system of interacting weight-one
 arctic  multiplets, 
$\U^{(1) I} (v) $, and their smile-conjugates,
$ \breve{\U}^{(1)\bar I }(v)$, described by a Lagrangian\footnote{The action generated by 
the Lagrangian (\ref{conformal-sm}) is real due to (\ref{smile-iso2}).}
of the form \cite{K-hyper}:
\bea
\cL^{(2)} 
= {\rm i} \, K (\U^{(1)I}, \breve{\U}^{(1) \bar J})~.
\label{conformal-sm}
\eea
Here $K(\F^I, {\bar \F}^{\bar J}) $ is a real function
of $n$ complex variables $\F^I$, with $I=1,\dots, n$, 
satisfying the homogeneity condition
\bea
\F^I \frac{\pa}{\pa \F^I} K(\F, \bar \F) =  K( \F,   \bar \F)~.
\label{Kkahler22}
\eea
The function  $K(\F^I, {\bar \F}^{\bar J}) $ can be interpreted as the K\"ahler potential 
of a {\it K\"ahlerian cone} $\cM$ written in special complex coordinates in which 
the homothetic conformal Killing vector field $\c^I (\F)$ has the form $\c^I (\F)=\F^I$. 

There exists a more geometric formulation of the theory (\ref{conformal-sm}) described in detail 
in \cite{KLvU}.
It is realized in terms of a single weight-one arctic multiplet $\U^{(1)}$
and $n-1$ weight-zero arctic multiplets $\X^{ i} $. The corresponding Lagrangian is
\bea
K (\U^{(1)I}, \breve{\U}^{(1) \bar J}) = \U^{(1)} \breve{\U}^{(1)} \,
\exp \Big\{ \cK (\X^{ i}, \breve{\X}^{\bar j}) \Big\}~,
\eea
where the original variables $\U^{(1)I}$ are related to the new ones by a 
holomorphic reparametrization.
The arctic variables $\U^{(1)}$ and $\X^{ i} $ parametrize a holomorphic line bundle over a 
K\"ahler-Hodge 
manifold with K\"ahler potential $\cK (\vf^i, {\bar \vf}^{\bar j})$, see  \cite{KLvU} for more details.

Consider a system of $n$ Abelian vector multiplets, and let $W^{(2)}_I$
be  their field strengths,  $I=1,\dots, n$. 
Its dynamics  can be described by  a Lagrangian of the form
\bea
\cL^{(2)} = \cL (W^{(2)}_I)~,
\label{5.82}
\eea
where $\cL$ is a real homogeneous function of degree  $+1$, 
\bea
W^{(2)}_I \frac{\pa }{\pa W^{(2)}_I} \cL =\cL~.
\eea
The vector multiplet model (\ref{5.82}) can be generalized to include a Chern-Simons term
\bea
\cL^{(2)}_{\rm CS} = \cL (W^{(2)}_I) +\hf m^{IJ} V_I W^{(2)}_J~, \qquad
m^{IJ}=m^{JI} =(m^{IJ})^*={\rm const}~.
\eea
Here $V_I$ is the weight-zero tropical prepotential for the field strengths $W_I$, eq. (\ref{5.622}). 
The action associated with $\cL^{(2)}_{\rm CS}$ is invariant under gauge transformations 
$\d V_I = \l_I + \breve{\l}_I$, with 
with $\l_I $  arctic weight-zero multiplets.


\section{Matter couplings in $\cN=4$ supergravity}
\setcounter{equation}{0}

The structure of multiplets in  3D $\cN=4 $ supersymmetry is largely determined 
by the fact that the Lie algebra of the $R$-symmetry group is reducible, ${\frak so}(4) \cong
{\frak su}(2)  \oplus {\frak su}(2) $.

\subsection{Elaborating on the $\cN=4$ superspace geometry}
\label{subsection6.1}
Within  the geometric formulation developed  in section 2, the structure group of 
 $\cN=4$ conformal supergravity 
is ${\rm SL}(2,\mathbb{R})\times {\rm SO}(4)$, with  the spinor derivatives $\cD^I_\a$ 
transforming in the
defining (vector) representation of SO(4).
In order to define a large class of matter multiplets coupled to supergravity,
it is advantageous to make use of the isomorphism 
 ${\rm SO}(4) \cong  \big( {\rm SU}(2)_{\rL}\times {\rm SU}(2)_{\rR}\big)/{\mathbb Z}_2$
and  switch to an isospinor notation, $\cD^I_\a \to \cD^{i\bar i}_\a$, by replacing each SO(4) vector 
index  by a pair of isospinor ones. We use the notation $\j_i$ and  $\chi_{\bar i}$
to denote the isospinors which transform under the defining representations of 
SU$(2)_{\rm L}$ and  SU(2)$_{\rm R}$, respectively. 
The rules for  raising and lowering isospinor indices are spelled out in Appendix A.
The algebraic structure underlying 
 the correspondence $\cD^I_\a \to \cD^{i\bar i}_\a$ is also explained in Appendix A.
For completeness, here we only repeat the definition.
Associated with a real SO(4) vector $V_I$ is a second-rank  isospinor $V_{i\bai}$ 
defined as
\bea
V_I~\to~V_{i\bai}:=(\t^I)_{i\bai}V_I ~,\qquad
V_I=(\t_I)^{i\bai}V_{i\bai}~,\qquad(V_{i\bai})^*=V^{i\bai}~,
\label{6.1}
\eea
see Appendix A for the definition of the $\t$-matrices.
If  $V_I$ and $U_I$ are two SO(4) vectors, and  
$V_{i\bai}$ and $U_{i\bai}$ the associated second-rank isospinors, then
\bea
V^I U_I=V^{i\bai} U_{i\bai}~.
\label{6.2}
\eea

Along with the relations (\ref{6.1}) and (\ref{6.2}), we need a few more general results.
Given an antisymmetric second-rank SO(4) tensor, $A_{IJ}=-A_{JI}$, its counterpart with isospinor 
indices, $A_{i\bai j\baj}=-A_{j\baj i\bai}=A_{IJ}(\t^I)_{i\bai}(\t^J)_{j\baj}$ can be decomposed as
\bea
A_{i\bai j\baj}=\ve_{ij}A_{\bai\baj}+\ve_{\bai\baj}A_{ij}~ \longrightarrow ~
A^{i\bai j\baj}=-\ve^{ij}A^{\bai\baj}-\ve^{\bai\baj}A^{ij}~,\qquad
A_{ij}=A_{ji}~,~~A_{\bai\baj}=A_{\baj\bai}
~.~~~
\eea
Here the two independent symmetric isospinors  $A_{ij}$ and $A_{\bai\baj}$ represent the
self-dual and anti-self-dual 
parts of the antisymmetric tensor $A_{IJ}$.
Given another antisymmetric second-rank SO(4) tensor, 
$B_{IJ}=-B_{JI}$,
and the corresponding isospinor counterparts 
$B_{ij} $ and $B_{\bai\baj}$,
one can check that 
\bea
\hf A^{IJ}B_{IJ}
&=& A^{ij}B_{ij}+A^{\bai\baj}B_{\bai\baj}
~.
\eea
Finally, consider the completely antisymmetric fourth-rank 
tensor $\ve_{IJKL}$ normalized by 
$\ve_{{\bf 1}{\bf 2}{\bf 3}{\bf 4}}=1$.
Its isospinor counterpart is
\bea
\ve_{i\bai j\baj k\bak l\bal}&:=&
\ve_{IJKL}(\t^I)_{i\bai}(\t^J)_{j\baj}(\t^K)_{k\bak}(\t^L)_{l\bal} 
=\Big(\ve_{ij}\ve_{kl}\ve_{\bai\bal}\ve_{\baj\bak}
-\ve_{il}\ve_{jk}\ve_{\bai\baj}\ve_{\bak\bal}\Big)
~.
\eea

We are now prepared to specify the $\cN$-extended supergravity algebra, which was
derived in section 2, to the case $\cN=4$ and rewrite it 
using the isospinor notation introduced.
The covariant derivatives are
\bea
\cD_{A}&=& (\cD_a, \cD^{i\bar i}_\a )=E_{A}
+\O_{A}
+\F_A
~,
\eea
where the original SO(4) connection $\F_A$ now turns into a sum of  two SU(2) connections,
the left $(\F_{\rm L})_A$ and  the right $(\F_{\rm R})_A$ ones,
\bea
\F_A=(\F_{\rL})_A+(\F_{\rR})_A~,~~~~
(\F_{\rm L})_A=\Phi_{A}{}^{kl}\bL_{kl}~,~~
(\F_{\rm R})_A=\Phi_{A}{}^{\bak\bal}\bR_{\bak\bal}
~.
\eea
Here $\bL_{kl}$ are the  generators of $ {\rm SU}(2)_{\rL}$
and  $\bR_{\bak\bal}$ the generators of $ {\rm SU}(2)_{\rR}$.
They are related to the SO(4) generators 
$\cN_{KL}$ as
\bea
\cN_{KL}~\to~\cN_{k\bak l\bal}=\ve_{\bak\bal}\bL_{kl}+\ve_{kl}\bR_{\bak\bal}~.
\eea
The same decomposition into left and right sectors 
takes place  for the SO(4) curvature
and for the SO(4) gauge parameters.
The two sets of SU(2) generators act on 
the spinor covariant derivatives $\cD_\a^{i\bai}:=\cD_\a^I(\t_I)^{i\bai}$ as follows:
\bea
&&
{\big [} {\bL}{}^{kl},\cD_{\a}^{i\bai}{\Big]} =\ve^{i(k} \cD_{\a}^{ l)\bai}
~,~~~
{\big [} {\bR}{}^{\bak\bal},\cD_{\a}^{i\bai}{\Big]} =\ve^{\bai(\bak} \cD_{\a}^{i \bal)}
~.
\label{acL-R}
\eea

As shown in section 2, in $\cN$-extended curved superspace the torsion and the curvature 
of dimension 1
are given in terms of the three tensor superfields:  $X^{IJKL}$, $C_a{}^{IJ}$ and $S^{IJ}$.
We recall that the completely antisymmetric curvature $X^{IJKL}$ does not occur for $\cN<4$.
In the $\cN=4$ case, these superfields
take the form:
\bea
X^{IJKL}~&\to &~
X^{i\bai j\baj k\bak l\bal}
=
\ve^{i\bai j\baj k\bak l\bal}X
=\Big(\ve^{ij}\ve^{kl}\ve^{\bai\bal}\ve^{\baj\bak}
-\ve^{il}\ve^{jk}\ve^{\bai\baj}\ve^{\bak\bal}\Big)
X
~,
\\
C_a{}^{IJ}~ & \to &~C_a{}^{i\bai j\baj}=-\ve^{\bai\baj}B_a{}^{ij}-\ve^{ij}C_a{}^{\bai\baj}
~,~~~
B_a{}^{ij}=B_a{}^{ji}~,~~
C_a{}^{\bai\baj}=C_a{}^{\baj\bai}~,
\\
S^{IJ}~ &\to &~ \cS^{ij}{}^{\bai\baj}+\ve^{ij}\ve^{\bai\baj}\cS~,~~~
\cS^{ij}{}^{\bai\baj}=\cS^{ji}{}^{\bai\baj}=\cS^{ij}{}^{\baj\bai}~.
\eea
The algebra of spinor covariant derivatives becomes
\bea
\{\cD_\a^{i\bai},\cD_\b^{j\baj }\}&=& \phantom{+}
2\ri\ve^{ij}\ve^{\bai \baj }(\g^c)_{\a\b}\cD_c
+{2\ri}\ve_{\a\b}\ve^{\bai \baj }(2\cS+X)\bL^{ij}
-2\ri\ve_{\a\b}\ve^{ij}\cS^{kl}{}^{\bai \baj }\bL_{kl}
+4\ri C_{\a\b}{}^{\bai \baj }\bL^{ij}
\non\\
&&
+2\ri\ve_{\a\b}\ve^{ij}(2\cS-X)\bR^{\bai \baj }
-2\ri\ve_{\a\b}\ve^{\bai \baj }\cS^{ij}{}^{\bak\bal}\bR_{\bak\bal}
+4\ri B_{\a\b}{}^{ij}\bR^{\bai \baj }
\non\\
&&
+2\ri\ve_{\a\b}(\ve^{\bai \baj }B^{\g\d}{}^{ij}+\ve^{ij}C^{\g\d}{}^{\bai \baj })\cM_{\g\d}
-4\ri(\cS^{ij}{}^{\bai \baj }+\ve^{ij}\ve^{\bai \baj }\cS)\cM_{\a\b}
~.
\label{N=4alg}
\eea

It can be shown that the dimension-3/2 Bianchi identities 
take the form:
\bsubeq
\bea
\cD_\a^{i\bai} \cS^{jk\baj\bak}&=&
2\cT_\a^{ijk(\baj}\ve^{\bak)\bai}
-2\ve^{i(j}\cT_\a^{k)\bai\baj\bak}
-\ve^{i(j}\cS_\a{}^{k)(\baj}\ve^{\bak)\bai}
~,
\\
\cD_{\a}^{i\bai}B_{\b\g}{}^{jk}
&=&
-\ve^{i(j}\Big(A_{\a\b\g}{}^{k)\bai}-C_{\a\b\g}{}^{k)\bai}\Big)
-{2\over 3}\ve_{\a(\b}\ve^{i(j}\Big(\cD_{\g)}^{k)\bai}(2 \cS-X)\Big)
+2\ve_{\a(\b}\cT_{\g)}{}^{ijk\bai}
~,~~~~~~~~~
\\
\cD_{\a}^{i\bai}C_{\b\g}{}^{\baj\bak}
&=&
-\Big(A_{\a\b\g}{}^{i(\baj}+ C_{\a\b\g}{}^{i(\baj}\Big)\ve^{\bak)\bai}
+{2\over 3}\ve_{\a(\b}\Big(\cD_{\g)}^{i(\baj} (2\cS+X)\Big)\ve^{\bak)\bai}
+2\ve_{\a(\b}\cT_{\g)}{}^{i\bai\baj\bak}
~.~~~~~~~~~
\eea
\esubeq
Here the superfields appearing in the right-hand sides have the following algebraic properties:
\bea
\cT_\a^{k\bai\baj\bak}=\cT_\a^{k(\bai\baj\bak)}~,~~
\cT_\a^{ijk\bak}=\cT_\a^{(ijk)\bak}~,~~
C_{\a\b\g}{}^{i\bai}=C_{(\a\b\g)}{}^{i\bai}
~,~~
A_{\a\b\g}{}^{i\bai}=A_{(\a\b\g)}{}^{i\bai}
~.
\eea
These superfields are related to those introduced in 
eqs. (\ref{22.18a})--(\ref{22.18c}) as follows:
\bsubeq
\bea
\cT_\a^{IJK}&\to&\cT_{\a}{}^{i\bai j\baj k\bak}=
-\ve^{\bai\baj}\cT_\a^{ijk\bak}-\ve^{ij}\cT_\a^{k\bai\baj\bak}~,~~~~~~
\\
C_{\a\b\g}{}^{IJK}&\to&
C_{\a\b\g}{}^{i\bai j\baj k\bak}=
\ve^{ij}A_{\a\b\g}{}^{k\bai}\ve^{\baj\bak}
-\ve^{jk}A_{\a\b\g}{}^{i\bak}\ve^{\bai\baj}
~,
\\
C_{\a}{}^{IJK}&\to&
C_{\a}{}^{i\bai j\baj k\bak}=
-\ve^{ij}(\cD_{\a}^{k\bai}X)\ve^{\baj\bak}
+\ve^{jk}(\cD_{\a}^{i\bak}X)\ve^{\bai\baj}~.
\eea 
\esubeq

An important property of the $\cN=4$ curved superspace geometry is its invariance under 
the discrete transformation
\bea
{\frak M} : ~ {\rm SU}(2)_{\rL}~ \longleftrightarrow ~ {\rm SU}(2)_{\rR}
\eea
which  changes the tensor types of superfields 
as 
${\rm D}^{(p/2)}_{\rL} \otimes {\rm D}^{(q/2)}_{\rR} 
\to {\rm D}^{(q/2)}_{\rL} \otimes {\rm D}^{(p/2)}_{\rR}$, 
where ${\rm D}^{(p/2)}$ denotes the spin-$p$ 
representation of SU(2). In the rigid supersymmetric case, this transformation is 
an outer automorphism
of the $\cN = 4$ super-Poincar\'e algebra, which underlies mirror symmetry in  
 3D $\cN=4$ Abelian gauge theories \cite{IS}.
It  has been  studied by Zupnik \cite{Zupnik3,Zupnik4} within the 3D $\cN=4$ rigid 
harmonic superspace 
\cite{Zupnik2}. Following \cite{Zupnik3,Zupnik4},
we  call $\frak M$ the {\it mirror map}.

The various geometric objects behave differently under the mirror map: 
\begin{subequations}
\bea
&{\frak M}\cdot  \cS =\cS ~,   \qquad &{\frak M}\cdot     \cS^{ij\bai\baj} = \cS^{ij\bai\baj} ~,
\qquad {\frak M}\cdot X=-X  ~, \\
& {\frak M}\cdot C_a^{\bai\baj}=B_a^{ij} ~, \qquad  &{\frak M}\cdot B_a^{ij} =C_a^{\bai\baj}~;\\
 & {\frak M}\cdot \cS_\a{}^{i \,\bai}=\cS_\a{}^{i \,\bai}    ~,
\qquad   
&{\frak M}\cdot  A_{\a\b\g}{}^{i \,\bai} =-A_{\a\b\g}{}^{i \,\bai}~, \quad
 {\frak M}\cdot C_{\a\b\g}{}^{i \,\bai} =C_{\a\b\g}{}^{i \,\bai}~,~~~~~~\\
& {\frak M}\cdot  \cT_\a{}^{ijk \,\bai}  =\cT_\a{}^{i \,\bai\baj\bak} ~, \qquad       
& {\frak M}\cdot   \cT_\a{}^{i \,\bai\baj\bak} =\cT_\a{}^{ijk \,\bai}~.
\eea
\end{subequations}

We conclude by giving the super-Weyl transformation in the isospinor notation: 
\bsubeq
\bea
\d_\s\cD_\a^{i\bai}&=&
\hf \s\cD_\a^{i\bai} + (\cD^{\b i\bai}\s )\cM_{\a\b}
-(\cD_{\a j}^\bai \s )\bL^{ij}
-(\cD_{\a \baj}^i \s )\bR^{\bai\baj}
~,
\\
\d_\s\cD_a&=&
\s \cD_a
+{\ri\over 2}(\g_a)^{\g\d}(\cD_{\g}^{k\bak} \s )\cD_{\d k\bak}
+\ve_{abc}(\cD^b\s )\cM^{c}
\non\\
&&
+{\ri\over 16}(\g_a)^{\g\d}([\cD_\g^{(k\bak},\cD_{\d \bak}^{l)}]\s )\bL_{kl}
+{\ri\over 16}(\g_a)^{\g\d}([\cD_\g^{k(\bak},\cD_{\d k}^{\bal)}]\s )\bR_{\bak\bal}
~.
\eea
The dimension-1 torsion and curvature superfields transform as follows:
\bea
\d_\s \cS^{ij}{}_{\bai\baj}&=&
\s \cS^{ij}{}_{\bai\baj}
-{\ri\over 8} [\cD^{\g (i}_{(\bai},\cD_{\g \baj)}^{j)}]\s 
~,~~~
\d_\s \cS=
\s \cS
-{\ri\over 32}[\cD^{\g k\bak},\cD_{\g k\bak}]\s 
~,
\\
\d_\s B_{a}{}^{ij}&=&
\s B_{a}{}^{ij}
-{\ri\over 16}(\g_a)^{\g\d} [\cD_\g^{(i\bak},\cD_{\d \bak}^{j)}]\s 
~,
\\
\d_\s C_{a}{}^{\bai\baj}&=&
\s C_{a}{}^{\bai\baj}
-{\ri\over 16}(\g_a)^{\g\d} [\cD_\g^{k(\bai},\cD_{\d k}^{\baj)}]\s 
~,
\\
\d_\s X&=&\s X~.
\eea
\esubeq

\subsection{Covariant projective multiplets}
\label{subsection6.2}
In this section we introduce a curved-superspace extension of the  $\cN=4$ 
superconformal projective multiplets \cite{KPT-MvU}. As in the $\cN=3$ case, it is 
natural to start our analysis with a more detailed look at the properties of the $\cN=4$ 
vector multiplet 
in conformal supergravity, and then turn to more general supermultiplets.

In accordance with the consideration of subsection \ref{subsection6.1},
the vector-multiplet field strength $W^{IJ} =-W^{JI}$ 
is equivalently described by two symmetric second-rank isospinors, $W^{ij}$ and $W^{\bai\baj}$,
which are defined as 
\bea
W^{IJ}~\to~W^{i\bai j\baj}=-\ve^{\bai\baj}W^{ij}-\ve^{ij}W^{\bai\baj}
~,~~~~
W^{ij}=W^{ji}~,~~
W^{\bai\baj}=W^{\baj\bai}
\eea
and transform under the local groups ${\rm SU}(2)_{\rL} $ and ${\rm SU}(2)_{\rR}$, respectively.
The dimension-3/2 Bianchi identity (\ref{2.35}) 
turns into the two {\it independent} analyticity
constraints
\bsubeq
\bea
\cD_\a^{(i\bai}W^{kl)}&=&0
~,
\label{a0-L}
\\
\cD_\a^{i(\bai}W^{\bak\bal)}&=&0
~.
\label{a0-R}
\eea
\esubeq
As a result, the field strengths  
$W^{ij}$ and $W^{\bai\baj}$ 
are completely independent of each other. Therefore,  the $\cN=4$ supermultiplet described by 
$W^{IJ}$ is reducible 
and is, in fact,  a superposition of two inequivalent off-shell $\cN=4$ vector multiplets.
One of them is 
characterized by the condition $W^{\bai\baj}=0$, 
while for the other vector multiplet $W^{ij}=0$.
The existence of two inequivalent off-shell $\cN=4$ vector multiplets in three dimensions 
was discovered by Brooks and Gates  \cite{BrooksG} (see also \cite{KS} where the results
of \cite{BrooksG} were recast in terms of $\cN=2$ superfields).

A superfield $W^{ij}$ under the constraint (\ref{a0-L}) will be called a {\it left linear multiplet}.
Similarly, eq. (\ref{a0-R}) defines a {\it right linear multiplet}. 
These multiplets are 3D analogues of the 4D $\cN=2$ linear multiplet \cite{BS,SSW}.

The constraints (\ref{a0-L}) and (\ref{a0-R}) can be rewritten as generalized chirality  conditions. 
This can be achieved, as in the $\cN=3$ case studied earlier, 
by allowing for auxiliary bosonic dimensions.
Specifically, let us introduce left and right isospinor variables, 
$v_\rL:=v^i \in {\mathbb C}^2 \setminus  \{0\}$ and 
$v_\rR:=v^\bai \in {\mathbb C}^2 \setminus  \{0\}$, 
and use them to define two different subsets,  $\cD_\a^{(1)\bai}$ and $\cD_\a^{(\bau)i}$,
in the set of spinor covariant  derivatives $\cD^{i\bar i}_\a$,  
\bea
\cD_\a^{(1)\bai}:=v_i\cD_\a^{i\bai}~,~~~~~~
\cD_\a^{(\bau)i}:=v_\bai\cD_\a^{i\bai}~,
\eea
as well as the index-free superfields
\bea
W_\rL^{(2)}:=v_iv_jW^{ij}\equiv W^{(2)}~,~~~~~~
W_\rR^{(2)}:=v_\bai v_\baj W^{\bai\baj}\equiv W^{(\bad)}
\eea
associated with the left and the right linear multiplets, respectively.
Now, the constraints (\ref{a0-L}) and (\ref{a0-R}) become
\begin{subequations}
\bea
\cD_\a^{(1)\bai}W_\rL^{(2)}&=&0~, 
\label{6.24a} \\
\cD_\a^{(\bau)i}W_\rR^{(2)}&=&0~.
\label{6.24b} 
\eea
\end{subequations}
The parenthesized superscripts attached to $\cD_\a^{(1)\bai}$ and $W_\rL^{(2)}$
indicate the degree of homogeneity in the left isospinor $v_i$. 
The same convention is used for  
the right objects   $\cD_\a^{(\bau)i}$ and $W_\rR^{(2)}$, 
but  with $v_i \to  v_{\bar i}$.
In complete analogy with the $\cN=3$ case, both  $v_i $ and $  v_{\bar i}$ are chosen
to be inert under the local $ {\rm SU}(2)_{\rL}$ and $ {\rm SU}(2)_{\rR}$  transformations.

Since the right  linear multiplet, $W^{\bar i \bar j}$, can be obtained from the left one, $W^{ij}$, 
by applying the mirror map, it  suffices to restrict our  analysis to the latter. 
Consider an infinitesimal supergravity gauge transformation
\bea
\d_K\cD_A=[K,\cD_A]~,\qquad
K=K^C \cD_C+\hf K^{cd} \cM_{cd}+ K^{kl} \bL_{kl} +K^{\bar k \bar l} \bR_{\bar k \bar l} ~.
\eea
It acts on 
$W^{ij}$  as 
\bea
\d_K W^{ij} = K^C \cD_C W^{ij} + 2 W^{l(i}K^{j)}{}_l~.
\eea
In terms of $W_\rL^{(2)}$ this transformation law  takes the form
\bsubeq
\bea
\d_K W_\rL^{(2)} 
&=& \Big( K^{{C}} \cD_{{C}} + K^{ij} \bL_{ij} \Big)W_\rL^{(2)} ~,  
\label{W2t1-L}
\\ 
&&K^{ij} \bL_{ij}  W_\rL^{(2)}= -\Big(K_\rL^{(2)} {\bm \pa}_\rL^{(-2)} 
-2 \, K_\rL^{(0)}\Big) W_\rL^{(2)} ~, 
\label{W2t2-L}
\eea
\esubeq
where we have used the notations:
\bea
K_\rL^{(2)} =K^{ij}\, v_i v_j 
~,\qquad
K_\rL^{(0)} =\frac{v_i u_j }{(v_\rL,u_\rL)}K^{ij}~,\qquad \quad
(v_\rL,u_\rL):=v^{i}u_i~.
\label{W2t3-L}
\eea
The differential operator ${\bm \pa}_\rL^{(-2)} $ is defined as
\bea
{\bm \pa}_\rL^{(-2)} :=\frac{1}{(v_\rL,u_\rL)}u^{i}\frac{\pa}{\pa v^{i}}~.
\eea
Here we have introduced a second left isospinor variable $u_\rL:=u^{i}$ 
which is restricted to be linearly independent of $v_\rL$,
that is
$(v_\rL,u_\rL)\ne0$. 
Thus $v^{i}$ and $u^{i}$ can be used  
to define a new  basis for the left isospinor indices, with the aid of the completeness relation
\bea
\d_j^i=\frac{1}{ (v_\rL,u_\rL)}\big(v^{i}u_j- v_j u^{i} \big)~,
\label{completeness-L}
\eea
in complete analogy with our previous consideration for $\cN=3$ supergravity, 
see eq. (\ref{ChangingBasis}). 
For example, the generators $\bL^{ij}$ 
of the group ${\rm SU}(2)_\rL$
turn into 
\bea
\bL^{(2)} := v_iv_j\bL^{ij} ~, \qquad 
\bL^{(0)} :=\frac{1}{(v_\rL,u_\rL)}v_iu_j\bL^{ij} ~, \qquad
\bL^{(-2)} :=\frac{1}{(v_\rL,u_\rL)^2}u_iu_j\bL^{ij}~.~~~
\eea
Then it follows from (\ref{W2t2-L}) that 
\bea
\bL^{(2)} W_\rL^{(2)} =0~.
\label{6.31}
\eea
This identity is crucial for the consistency of the constraints (\ref{6.24a}). 
Indeed,  
the spinor covariant derivatives  $\cD_\a^{(1)\bai}$ 
obey the anticommutation relations
\bea
\{\cD_\a^{(1)\bai},\cD_\b^{(1)\baj}\}&=&
{2\ri}\ve_{\a\b}\ve^{\bai\baj}(2\cS+X)\bL^{(2)}
+4\ri C_{\a\b}{}^{\bai\baj}\bL^{(2)}
-2\ri\ve_{\a\b}\ve^{\bai\baj}\cS^{(2)}{}^{\bak\bal}\bR_{\bak\bal}
+4\ri B_{\a\b}^{(2)}\bR^{\bai\baj}
\non\\
&&
+2\ri\ve_{\a\b}\ve^{\bai\baj}B^{\g\d}{}^{(2)}\cM_{\g\d}
-4\ri \cS^{(2)}{}^{\bai\baj}\cM_{\a\b}~,
\label{N=4D1D1}
\eea
where we have defined 
\bea
B_{\a\b}^{(2)}&:=&B_{\a\b}{}^{ij}v_iv_j
~,\qquad
\cS^{(2)}{}^{\bai\baj}:=\cS^{ij}{}^{\bai\baj}v_iv_j 
~.
\label{N=4Ta-1}
\eea
Since $ W_\rL^{(2)} $ is inert under both the Lorentz and ${\rm SU} (2)_\rR$ transformations, 
eq. (\ref{6.31}) guarantees that the requirement 
$\{\cD_\a^{(1)\bai},\cD_\b^{(1)\baj}\} W_\rL^{(2)} =0$
holds.

The properties of  $ W_\rL^{(2)} $, which we have just described, 
are analogous to those of the $\cO (2)$ multiplet in 4D $\cN=2$ supergravity  
\cite{KLRT-M1,KLRT-M2}. 
To comply with the four-dimensional terminology,
$W_\rL^{(2)}$ and $W_\rR^{(2)}$ will be called  left and right $\cO(2)$ multiplets, 
respectively.

The super-Weyl transformation of $W_\rL^{(2)}$ is 
\bea
\d_\s W_\rL^{(2)}=\s  W_\rL^{(2)}~.
\eea

We are now 
prepared to introduce a large family of off-shell supermultiplets with properties 
similar to those of $W_\rL^{(2)}$.
A {\it covariant left projective multiplet} of weight $n$,
$Q_\rL^{(n)}(z,v_\rL)$, is defined to be 
a Lorentz and ${\rm SU}(2)_\rR$ scalar superfield that 
lives on the curved $\cN=4$ superspace  ${\cM}^{3|8}$, 
is holomorphic with respect to 
the isospinor variables $v^i $ on an open domain of 
${\mathbb C}^2 \setminus  \{0\}$, 
and is characterized by the following conditions:\\
(i) it obeys the covariant analyticity constraints 
\be
\cD^{(1)\bai}_{\a} Q_\rL^{(n)}  =0~;
\label{ana-L}
\ee  
(ii) it is  a homogeneous function of $v_\rL$ 
of degree $n$, that is,  
\be
Q_\rL^{(n)}(c\,v_\rL)\,=\,c^n\,Q_\rL^{(n)}(v_\rL)~, \qquad c\in \mathbb{C}^*~;
\label{weight-L}
\ee
(iii)  the supergravity gauge transformations act on $Q_\rL^{(n)}$ 
as follows:
\bea
\d_K Q_\rL^{(n)} 
&=& \Big( K^{{C}} \cD_{{C}} + K^{ij} \bL_{ij} \Big)Q_\rL^{(n)} ~,  
\non \\ 
K^{ij} \bL_{ij}  Q_\rL^{(n)}&=& -\Big(K^{(2)} {\bm \pa}_\rL^{(-2)} 
-n \, K^{(0)}\Big) Q_\rL^{(n)} ~.
\label{harmult1-L}   
\eea 
By construction, $Q_\rL^{(n)}$ is independent of $u_\rL$. 
One can see that $\d_K Q_\rL^{(n)} $ 
is also independent of the isospinor $u_\rL$, 
due to (\ref{weight-L}). 

It is  important to note that 
\bea
\bL^{(2)} Q_\rL^{(n)}=0~, \qquad \bL^{(0)}Q_\rL^{(n)}=-\frac{n}{2}Q_\rL^{(n)} ~,
\label{J++-L}
\eea
as a consequence of (\ref{harmult1-L}). 
Since $Q_\rL^{(n)}$ is invariant under the Lorentz and SU(2)$_\rR$ 
transformations,  the first relation in  (\ref{J++-L})
guarantees that
the covariant analyticity constraints (\ref{ana-L}) are indeed consistent.

As is clear from the above consideration, the isospinor 
$ v^{i} \in {\mathbb C}^2 \setminus\{0\}$ is   defined modulo the equivalence relation
$ v^{i} \sim c\,v^{i}$,  with $c\in {\mathbb C}^*$, {hence it parametrizes ${\mathbb C}P^1$}.
Therefore, the covariant left projective multiplets live in 
{\it curved projective superspace}, ${\cM}^{3|8} \times {\mathbb C}P^1$.

Let $Q_\rL^{(n)} (v_\rL) $ be a left
 projective supermultiplet of weight $n$.
Assuming that it varies homogeneously under the super-Weyl transformations,
the analyticity constraints (\ref{ana-L}) uniquely fix its transformation law to be
\be
\d_{\s} Q_\rL^{(n)} =\frac{n}{2}\s Q_\rL^{(n)} ~.
\label{Q(n)super-Weyl-L}
\ee
This relation can be derived by noticing that the transformation rules of the $\cD_\a^{(1)\bai}$ 
derivatives under super-Weyl transformations are 
\bea
\d_\s\cD_\a^{(1)\bai}&=&
\hf \s\cD_\a^{(1)\bai} + (\cD^{(1)\b \bai}\s)\cM_{\a\b}
-(\cD_{\a \baj}^{(1)} \s)\bR^{\bai\baj}
+(\cD_{\a}^{(1)\bai} \s)\bL^{(0)}
-(\cD_{\a}^{(-1)\bai} \s)\bL^{(2)}
~,~~~~~~~~~
\label{sWD1-L}
\eea
where 
\bea
\cD_{\a}^{(-1)\bai} :=\frac{1}{(v_\rL,u_\rL)}u_i\cD_\a^{i\bai}~.~~~
\eea

We conclude this subsection with two comments. 
Firstly, for any integer $n$, the space of left weight-$n$ projective superfields 
can be endowed with a real structure.  Associated with $Q_\rL^{(n)} (v_\rL )$
is its smile-conjugate $\breve{Q}_\rL^{(n)} (v_\rL )$ which is 
defined according to eq. (\ref{smile-iso}) with obvious modifications.
The important property  (\ref{smile-iso2}) also extends to the $\cN=4$ left projective multiplets.
Thus, if $n$ is even, we can consistently  define real left projective superfields.

Our second comment is that  applying the mirror map to 
$Q_\rL^{(n)} (v_\rL )$  gives a {\it covariant right projective multiplet} of weight $n$, 
$Q_\rR^{(n)} (v_\rR )$.
The entire consideration of this section naturally extends  to   the right projective multiplets.
In what follows, for the left and right projective multiplets we often use two alternative types of
notation, specifically
\bea
Q_\rL^{(n)} \equiv Q^{(n)} ~, \qquad 
Q_\rR^{(n)} \equiv Q^{(\bar n)} ~.
\eea

\subsection{Hybrid projective multiplets} 
The definitions and properties of the left projective multiplets, which we presented
 in subsection \ref{subsection6.2},
are completely analogous to those given in \cite{KLRT-M1,KLRT-M2}
for the 4D $\cN=2$ covariant projective multiplets.
A nontrivial new aspect of the 3D case is that there exist two types of $\cN=4$ 
covariant projective multiplets, the left and the right ones.
Moreover, in three dimensions we can define 
{\it hybrid projective multiplets} of the form
\bea
Q^{(n,m)} (v_\rL, v_\rR ):=\sum_{Q_\rL \, ,\, Q_\rR} Q_\rL^{(n)} (v_\rL ) Q_\rR^{(m)} (v_\rR )~.
\label{6.43}
\eea
They obey the following analyticity constraint 
\bea
\cD^{(1,{1})}_\a Q^{(n,m)} (v_\rL, v_\rR ) = 0~, \qquad 
\cD^{ (1,{1})}_\a := \cD^{i \bar i}_\a \,v_i v_{\bar i}
\label{6.44}
\eea
and are characterized by the algebraic properties 
\bea
\bL^{(2)}Q^{(n,m)}= \bR^{(\bar{2})} Q^{(n,m)}  =0~.
\eea
The analyticity constraint is consistent, 
since the operators $\cD_\a^{(1,{1})}$ satisfy  the anticommutation relations:
\bea
\{\cD_\a^{(1,{1})},\cD_\b^{(1,{1}) }\}&=&
4\ri C_{\a\b}^{ (\bar{2}) } \bL^{(2)}
+4\ri B_{\a\b}^{ (2) }\bR^{(\bar{2})}
-4\ri \cS^{(2,{2}) } \cM_{\a\b}~.
\label{hybrid-algebra}
\eea
It should be remarked that $\cS^{(2,2)}=v_iv_jv_\bai v_\baj\cS^{ij\bai\baj}$
 is a hybrid projective multiplet, 
\bea
\cD_\a^{(1,{1})} \cS^{(2,{2}) } =0~.
\eea

The explicit representation (\ref{6.43}) can be formalized.
A  {\it hybrid projective multiplets} of weight $(n,{m})$, $Q^{(n,{m})} (v_\rL, v_\rR )$, 
 is defined 
 to be a scalar superfield that 
lives on the curved $\cN=4$ superspace  ${\cM}^{3|8}$, 
is holomorphic with respect to 
the isospinor variables $v^i,v^{\bai} $ on an open domain of 
${\mathbb C}^2 \setminus  \{0\}\times {\mathbb C}^2 \setminus  \{0\}$, 
and is characterized by the following conditions:\\
(i) it obeys the covariant analyticity constraint (\ref{6.44});\\
(ii) it is  a homogeneous function of degree $n$ in $v_\rL$ and of degree ${m}$ in
$v_\rR$, that is,  
\be
Q^{ (n,{m})} ( c_\rL v_\rL, c_\rR v_\rR)\,=\,c_\rL^n \,c_\rR^{m}
\,Q^{(n,{m})}(v_\rL,v_\rR)~,~~~
 c_\rL , c_\rR  \in \mathbb{C}^*~;
\label{6.48}
\ee
(iii)  under the supergravity gauge group, $Q^{(n,{m})}$ transforms
as follows:
\begin{subequations}
\bea
\d_K Q^{(n,{m})} 
&=& \Big( K^{{C}} \cD_{{C}} + K^{ij} \bL_{ij}+ K^{\bai\baj} \bR_{\bai\baj} \Big)Q^{(n,{m})} ~,   
\label{6.49a}\\ 
K^{ij} \bL_{ij} Q^{(n,{m})}&=& -\Big(K_\rL^{(2)} {\bm \pa}_\rL^{(-2)} 
-n \, K_\rL^{(0)}\Big)Q^{(n,{m})} ~,
 \label{6.49b} \\ 
K^{\bai\baj} \bR_{\bai\baj}  Q^{(n,{m})}&=& -\Big(K_\rR^{({2})} {\bm \pa}_\rR^{(-{2})} 
-m \, K_\rR^{({0})}\Big)Q^{(n,{m})} ~.
\label{6.49c}
\eea 
\end{subequations}

If  $Q^{(n,{m})}$  has a  homogeneous  super-Weyl 
transformation law, $\d_\s Q^{(n,{m})} \propto \s Q^{(n,{m})}$,
then it proves to have the unique form:
\bea
\d_\s Q^{(n,{m})}=\hf (n+m) \s Q^{(n,{m})}~.
\label{hybrid-super}
\eea
There exist hybrid projective multiplets with  inhomogeneous 
super-Weyl  transformation laws. For example, the torsion $\cS^{(2,{2})}$ transforms as
\bea
\d_\s \cS^{(2,{2})}=\s \cS^{(2,{2})}
-\frac{\ri}{4}\cD^{(2,{2})}\s~,~~~~~~
\cD^{(2,{2})}:=\cD^{\a(1,{1})}\cD_\a^{(1,{1})}
~.
\eea

The results of this subsection are consistent with the interpretation that the isospinors
$ v^{i} ,
v^{\bai} \in {\mathbb C}^2 \setminus\{0\}$ 
are  defined modulo the equivalence relations
$ v^{i} \sim c_\rL\,v^{i}$, $ v^{\bai} \sim c_\rR\,v^{\bai}$,  
with $c_\rL,c_\rR\in {\mathbb C}^*$, {hence $(v^i,v^\bai)$ parametrizes 
${\mathbb C}P^1\times {\mathbb C}P^1$}.
Therefore, the hybrid projective multiplets live in
{\it curved bi-projective superspace} 
${\cM}^{3|8} \times {\mathbb C}P^1\times {\mathbb C}P^1$.

Hybrid projective multiplets 
can naturally be defined in rigid $\cN=4$ bi-projective superspace 
${\mathbb R}^{3|8} \times {\mathbb C}P^1\times {\mathbb C}P^1$, but this possibility 
has not been considered in \cite{KPT-MvU}.
Let us dimensionally reduce this superspace
to two dimensions. The result is 
the 2D $\cN=(4,4)$ bi-projective superspace 
${\mathbb R}^{2|8} \times {\mathbb C}P^1\times {\mathbb C}P^1$
which was introduced more than twenty years ago 
by Buscher, Lindstr\"om and Ro\v{c}ek
\cite{bi-proj-rigid1}
and further studied in \cite{bi-proj-rigid2,bi-proj-rigid3}.
Its local version has been developed in 
 \cite{TartaglinoMazzucchelli:2009ip}.

\subsection{Covariant  projection operators}
\label{CovProjOp}
In this subsection we develop techniques to engineer covariant left/right and 
hybrid projective multiplets.
For this we have to introduce a new superfield type  --  {\it isotwistor  multiplets} 
of  arbitrary weight  $(n,{m})$,
with $n,m$ integers.  
Such a superfield $T^{(n,{m})} (v_\rL, v_\rR )$ has the same properties as 
the hybrid projective multiplet 
 $Q^{(n,{m})} (v_\rL, v_\rR )$ except for the analyticity condition (\ref{6.44}).
More specifically, the properties (\ref{6.48}) and (\ref{6.49a})--(\ref{6.49c}) 
are required to hold for   $T^{(n,{m})} (v_\rL, v_\rR )$. However, no analyticity constraint
is imposed on $T^{(n,{m})} (v_\rL, v_\rR )$. As a result, $T^{(n,{m})} (v_\rL, v_\rR )$ 
may transform as a tensor field with respect to the local Lorentz group
(its Lorentz indices are suppressed). 
{\it Left} and {\it right isotwistor multiplets} correspond to special 
cases of isotwistor superfields:
\begin{subequations}
\bea
T^{(n )}_\rL (v_\rL )&:=& T^{(n,{0})} (v_\rL, v_\rR )~, \qquad \frac{\pa}{\pa v_\rR} T^{(n,{0})}=0~;\\
T^{(m )}_\rR (v_\rR )&:=& T^{(0,{m})} (v_\rL, v_\rR )~, \qquad \frac{\pa}{\pa v_\rL} T^{(0,{m})}=0~.
\eea
\end{subequations}

Consider a covariant left projective multiplet $Q_\rL^{(n)} (v_\rL) $ of weight $n$.
It can be proved that  there exists
a  left isotwistor superfield $T_\rL^{(n-4)} (v_\rL )$ such that 
\bea
Q_\rL^{(n)}=\D_\rL^{(4)} T_\rL^{(n-4)}~,
\label{6.53}
\eea
where  $\D_\rL^{(4)}$ denotes the following fourth-order operator:
\bsubeq
\bea
\D_\rL^{(4)}&=&\frac{1}{96}\Big(
(\cD^{(2)\bak\bal}-16\ri \cS^{(2)\bak\bal})\cD^{(2)}_{\bak\bal}
-(\cD^{(2)\a\b}-16\ri B^{(2)\a\b})\cD^{(2)}_{\a\b}
\Big)
~~~~~~~~~~~~~~~~~
\label{6.54a} \\
&=&\frac{1}{96}\Big(
\cD^{(2)\bak\bal}(\cD^{(2)}_{\bak\bal}-16\ri \cS^{(2)}_{\bak\bal})
-\cD^{(2)\a\b}(\cD^{(2)}_{\a\b}-16\ri B^{(2)}_{\a\b})
\Big)
~,~~~~~~~~~
\label{6.54b} 
\eea
\esubeq
with
\bea
\cD^{(2)}_{\bai\baj}:=\cD^{(1)\g}_{(\bai}\cD^{(1)}_{\g \baj)}~,\qquad
\cD^{(2)}_{\a\b}:=\cD^{(1)\bak}_{(\a}\cD^{(1)}_{\b) \bak}
~.
\eea
The opposite  statement also holds. Given an arbitrary  left isotwistor superfield 
$T_\rL^{(n-4)} (v_\rL )$, 
the superfield $Q_\rL^{(n)}$ defined by eq. (\ref{6.53}) can be shown to satisfy the constraint
\bea
\cD_\a^{(1) \bar i}Q_\rL^{(n)}=0~.
\label{6.56}
\eea
We will call  $\D_\rL^{(4)}$ the left projection operator.
The derivation of   $\D_\rL^{(4)}$ and the proof of (\ref{6.56}) are given in Appendix B.

It should be pointed out that the fourth-order operators that appear in the right-hand sides of 
(\ref{6.54a}) and (\ref{6.54b}) are related to each other as follows:
\bea
\cD^{(2)\bak\bal}\cD^{(2)}_{\bak\bal}
&=&
-\cD^{(2)\a\b}\cD^{(2)}_{\a\b}
-8\ri \cS^{(2)\bak\bal}\cD^{(2)}_{\bak\bal}
-8\ri B^{(2)\a\b}\cD^{(2)}_{\a\b}
-16\ri(\cD^{(1)\a}_\bak \cS^{(2)\bak\bal})\cD^{(1)}_{\a \bal}~.~~~~~
\label{four-four-L}
\eea
This relation may be rewritten in a slightly different form using the identity
\bea
\cD^{(1)\a}_\bal \cS^{(2)\bak\bal}&=&\cD^{(1)\bak}_\b B^{(2)\a\b}
~.
\eea

Suppose that  the left isotwistor superfield  $T_\rL^{(n-4)}$ in (\ref{6.53}) 
has the super-Weyl 
transformation law 
\bea
\d_\s T_\rL^{(n-4)}=\frac{n-4}{2}\,\s\,T_\rL^{(n-4)}
~.
\label{sWU-L}
\eea
Then it can be shown that $Q_\rL^{(n)} =\D_\rL^{(4)} T_\rL^{(n-4)}$ also transforms 
homogeneously as
\bea
\d_\s Q_\rL^{(n)}=\frac{n}{2}\,\s\,Q_\rL^{(n)}
~,
\label{sWQ-L-2}
\eea
which is the unique homogeneous transformation law compatible 
with the analyticity of $Q_\rL^{(n)}$  (in  accordance with
our discussion in the previous subsection).

A simple application of the construction (\ref{6.53})
is to  choose an ordinary ($v_\rL$-independent)  superfield $P$ 
in the role of $T^{(0)}_\rL$. 
Then, 
$\S_\rL^{(4)}:=\D_\rL^{(4)}P$ is a covariant $\cO(4)$ multiplet.
If $P$ is invariant under the super-Weyl transformations, $\d_\s P =0$, 
then $\S_\rL^{(4)}$  transforms
as $\d_\s\S_\rL^{(4)}=2\s\S_\rL^{(4)}$.

The above consideration can be extended to the space of right 
projective multiplets by making use of the mirror map.
The right projection operator  $\D^{(4)}_\rR$ proves to be
\bsubeq
\bea
\D^{(4)}_\rR&=&\frac{1}{96}\Big(
(\cD^{(\bad)kl}-16\ri \cS^{(\bad)kl})\cD^{(\bad)}_{kl}
-(\cD^{(\bad)\a\b}-16\ri C^{(\bad)\a\b})\cD^{(\bad)}_{\a\b}
\Big)
~~~~~~~~~~~~~~~~~
\label{RightProj1}
\\
&=&\frac{1}{96}\Big(
\cD^{(\bad)kl}(\cD^{(\bad)}_{kl}-16\ri \cS^{(\bad)}_{kl})
-\cD^{(\bad)\a\b}(\cD^{(\bad)}_{\a\b}-16\ri C^{(\bad)}_{\a\b})
\Big)
~,~~~~~~~~~
\label{LeftProj2}
\eea
\esubeq
with
\bea
\cD^{(\bad)}_{ij}:=\cD^{(\bau)\g}_{(i}\cD^{(\bau)}_{\g j)}~,~~~
\cD^{(\bad)}_{\a\b}:=\cD^{(\bau)k}_{(\a}\cD^{(\bau)}_{\b) k}
~.
\eea
It can be shown that the fourth-order operators which appear in the right-hand sides of 
(\ref{RightProj1}) and (\ref{LeftProj2}) are related to each other as follows:
\bea
\cD^{(\bad)kl}\cD^{(\bad)}_{kl}
&=&
-\cD^{(\bad)\a\b}\cD^{(\bad)}_{\a\b}
-8\ri \cS^{(\bad)kl}\cD^{(\bad)}_{kl}
-8\ri C^{(\bad)\a\b}\cD^{(\bad)}_{\a\b}
-16\ri(\cD^{(\bau)\a}_k \cS^{(\bad)kl})\cD^{(\bau)}_{\a l}~.~~~
\eea
This relation may be rewritten in a slightly different form using the identity
\bea
\cD^{(\bau)\a}_l \cS^{(\bad)kl}&=&\cD^{(\bau)k}_\b C^{(\bad)\a\b}
~.
\eea

${}$Finally, we can construct a hybrid projection operator. 
Let $T^{(n-2,{m-2})} (v_\rL , v_\rR)$ be a Lorentz-scalar 
isotwistor superfield of weight $(n-2, m-2)$.
We introduce the second-order differential operator
\bea
\D^{(2,{2})}:=\frac{\ri}{4}\Big(\cD^{(2,{2})}-4\ri\cS^{(2,{2})}\Big)
~.
\label{6.65}
\eea
It is not difficult to verify that
\bea
Q^{(n,{m})}:=\D^{(2,{2})}T^{(n-2,{m-2})}
\label{6.66}
\eea
satisfies (\ref{6.44}), and thus $\D^{(2,{2})}$ maps any  isotwistor superfield
into a hybrid projective one. 
Therefore $\D^{(2,{2})}$ is the hybrid projection operator.

Suppose that  the  isotwistor superfield  $T^{(n-2,{m-2})}$  in (\ref{6.66}) 
has the super-Weyl  transformation law 
\bea
\d_\s T^{(n-2,{m -2})}=\hf (n+m-2) \s \,T^{(n-2,{m-2})}~.
\eea
Then, it can be shown that the super-Weyl transformation  of 
the hybrid projective multiplet 
$Q^{(n,{m})}:=\D^{(2,{2})}T^{(n-2,{m-2})}$
is given by eq. (\ref{hybrid-super}).

A simple application of the construction (\ref{6.66})
is to  choose an ordinary (i.e., independent of $v_\rL $ and $ v_\rR$)  superfield $P$, 
with the super-Weyl transformation $\d_\s P= \s P$,  in the role of $T^{(0,{0})}$.
Then, $Q^{(2,{2})}=\D^{(2,{2})} P =Q^{ i j \, \bar i \bar j} v_iv_j v_{\bar i } v_{\bar j}$
 is hybrid projective.

The careful reader could have noticed that the left and right projection operators $\D^{(4)}_\rL$
and $\D^{(4)}_\rR$ have a structure which is formally equivalent to the chiral projector
of 4D $\cN=2$ supergravity \cite{Muller}. This property is not accidental and will be used
in appendix B. Recently, in the projective superspace approach to 4D $\cN=2$ supergravity,
a new powerful representation of the chiral projector has been derived \cite{KT-M2}.
It is interesting that this recent result similarly holds for $\D^{(4)}_\rL$
and $\D^{(4)}_\rR$. 
In particular, it turns out that 
in terms of 
isotwistor superfields one can obtain alternative 
representations for $\D^{(4)}_\rL$ and $\D^{(4)}_\rR$.
These are
\bsubeq
\bea
\D_\rL^{(4)}\oint(v_\rR,\rd v_\rR)\, T^{(n,-2)}
&=&
-\frac{\ri}{4}
\oint(v_\rR,\rd v_\rR)
\Big( \cD^{(2,-2)} -4\ri\cS^{(2,-2)}\Big) 
\D^{(2,2)}
T^{(n,-2)}
~,~~~
\label{proj2L}
\\
{\D}_\rR^{(4)} \oint (v_\rL, \rd v_\rL) \,T^{(-2,m)}&=&
-\frac{\ri}{4}  \oint (v_\rL ,\rd v_\rL)
\Big( \cD^{(-2,2)} -4\ri\cS^{(-2,2)}\Big) 
\D^{(2,2)}
T^{(-2,m)}
~,~~~~~~~~~
\label{proj2R}
\eea
\esubeq
with $T^{(n,-2)}$ and $T^{(-2,m)}$ 
isotwistor superfields of weight $(n,-2)$ and weight $(-2,m)$ respectively, and
\bsubeq
\bea
\cD^{(2,-2)}:=\cD^{\a(1,-1)}\cD_\a^{(1,-1)}~&,&~~~
\cD^{(-2,2)}:=\cD^{\a(-1,1)}\cD_\a^{(-1,1)}~,
\\
\cD_\a^{(1,-1)}:=\frac{1}{(v_\rR,u_\rR)}v_iu_\bai\cD_\a^{i\bai}~&,&~~~
\cD_\a^{(-1,1)}:=\frac{1}{(v_\rL,u_\rL)}u_iv_\bai\cD_\a^{i\bai}~,
\\
\cS^{(2,-2)}:=\frac{1}{(v_\rR,u_\rR)^2}v_iv_ju_\bai u_\baj\cS^{ij\bai\baj}~&,&~~~
\cS^{(-2,2)}:=\frac{1}{(v_\rL,u_\rL)^2}u_iu_jv_\bai v_\baj\cS^{ij\bai\baj}~.
\eea
\esubeq
Note that the right-hand side of (\ref{proj2L}) has the following properties:
(i) it is independent of the constant  isospinors $u_\rR = u^\bai$
 constrained by the only conditions $(v_\rR,  u_\rR) \neq 0$;
and (ii) it obeys the left analyticity constraint (\ref{ana-L}).
The proof of these statement are given in appendix B.
The mirrored  results hold for the right-hand side of (\ref{proj2R}).

Let us conclude by 
pointing out 
that the representations (\ref{proj2L}) and (\ref{proj2R}) are 
useful for applications. The point is that 
any weight-$n$ left $T_{\rL}^{(n)}(v_\rL)$
and weight-$m$ right $T_{\rR}^{(m)}(v_\rR)$  isotwistor superfields
can be represented in the following integral form:
\bea
T_{\rL}^{(n)}(v_\rL)=\oint\frac{(v_\rR,\rd v_\rR)}{2\pi}\,T_{\rL}^{(n,-2)}(v_\rL,v_\rR)
~,~~~
T_{\rR}^{(m)}(v_\rR)=\oint\frac{(v_\rL,\rd v_\rL)}{2\pi}\,T_{\rR}^{(-2,m)}(v_\rL,v_\rR)
~,~~~~~~
\label{hybrid-integral}
\eea
for some isotwistor superfields $T_{\rL}^{(n,-2)}(v_\rL,v_\rR)$
and $T_{\rR}^{(-2,m)}(v_\rL,v_\rR)$ 
of weights $(n,-2)$  and $(-2,m)$ respectively.

\subsection{Locally supersymmetric actions}

A remarkable feature of $\cN=4$ supergravity is that it allows  three types of locally 
supersymmetric 
and super-Weyl invariant actions, for which the measure involves integration over four or six
Grassmann variables only.

We introduce  three types of {\it real} Lagrangians:
(i)  a left  projective superfield $\cL_\rL^{(2)}(z,v_\rL)$;
(ii) a  right projective superfield  $\cL_\rR^{(2)}(z,v_\rR)$; 
and (iii) a hybrid multiplet  $\cL^{(0, 0)}(z,v_\rL, v_\rR)$.
All the Lagrangians are required to be real with respect to the smile-conjugation.
With the standard notation  $E^{-1}= {\rm Ber}(E_A{}^M)$,  our
locally supersymmetric and super-Weyl invariant action principle is given by
\begin{subequations}
\bea
S&=& S_{\rm left} + S_{\rm right} + S_{\rm hybrid}~,   
\label{N=4InvarAc} \\
S_{\rm left} ( \cL_\rL^{(2)}) &=& \frac{1}{2\pi} \oint (v_\rL, \rd v_\rL)
\int \rd^3 x \,{\rm d}^8\q\,E\, C_\rL^{({-4})} \cL_\rL^{(2)}~, 
\label{Action-left} \\
S_{\rm right}(\cL_\rR^{(2)} ) &=& 
\frac{1}{2\pi} \oint (v_\rR, \rd v_\rR)
\int \rd^3 x \,{\rm d}^8\q\,E\, C_\rR^{({-4})}\cL_\rR^{(2)} ~,
\label{Action-right} \\
S_{\rm hybrid} (\cL^{(0,{0})})&=& \frac{1}{(2\p)^2}
\oint {(v_\rL, \rd v_\rL)} 
\oint  {(v_\rR, \rd v_\rR)}
\int \rd^3 x \,{\rm d}^8\q\,E\, C^{(-2,-{2})}\cL^{(0,{0})}
~.~~~
\label{hybrid-Ac1}
\eea
\end{subequations}
The action involves some {\it model-independent}
Lorentz-scalar isotwistor superfields  $C_\rL^{(-4)}$, $C_\rR^{(-4)}$ and 
$C^{(-2,-{2})}$, of which  $C_\rL^{(-4)}$ and $C_\rR^{(-4)}$ are left and right respectively.
These superfields are required to be real with respect the smile-conjugation, 
to have definite super-Weyl transformation laws and obey special differential equations:
\bsubeq
\bea
\d_\s C_\rL^{(-4)}&=&-2\s C_\rL^{(-4)}~,\qquad
\D_\rL^{(4)}C_\rL^{(-4)}=1~;
\label{N=4AcComp-L}
\\ 
\d_\s C_\rR^{(-4)}&=&-2\s C_\rR^{(-4)}~, \qquad
\D_\rR^{(4)}C_\rR^{(-4)}=1~;
\label{N=4AcComp-R} \\
\d_\s C^{(-2,-{2})}&=&-\s C^{(-2,-{2})}~,~~~\,\,
\D^{(2,{2})}C^{(-2,-{2})}=1~.
\label{hybrid-AcComp}
\eea
\esubeq
All the Lagrangians are required to possess {\it uniquely defined}
homogeneous super-Weyl transformations
\bea
\d_\s\cL_\rL^{(2)}=\s\cL_\rL^{(2)}~,\qquad
\d_\s\cL_\rR^{(2)}=\s\cL_\rR^{(2)}~, \qquad \d_\s\cL^{(0,{0})}=0~.
\eea

The super-Weyl invariance of the action follows from the above transformation 
laws in conjunction with 
\bea
\d_\s E=\s E~.
\eea
The invariance of the action under the supergravity gauge transformations can be proved 
using the same considerations as in the 4D $\cN=2$ case \cite{KLRT-M1,KLRT-M2}.

It turns out that the action does not depend on the kinematic isotwistor 
superfields  $C_\rL^{(-4)}$, $C_\rR^{(-4)}$ and 
$C^{(-2,-{2})}$, provided the corresponding Lagrangians are independent. 
To prove this claim, it suffices to consider the left sector of the action, eq. (\ref{Action-left}). 
Let us represent the corresponding Lagrangian in the form 
$\cL_\rL^{(2)}=\D_\rL^{(4)}\cT_\rL^{(-2)}$, for some left isotwistor superfield $\cT_\rL^{(-2)}$.
We can now use the fact that $\D_\rL^{(4)}$ is symmetric, that is for any left isotwistor superfields
$ \J^{(-n)}$ and $\F^{(n-6)}$ it holds that 
 \bea
\int \rd^3 x \,{\rm d}^8\q\,E  \oint (v_\rL, \rd v_\rL)\, \Big\{ \J^{(-n)} \D_\rL^{(4)}\F^{(n-6)}
- 
\F^{(n-6)}  \D_\rL^{(4)}\J^{(-n)} \Big\}=0~, 
\eea
as a consequence of the representations (\ref{6.54a}) and (\ref{6.54b}).
Using this observation and the representation $\cL_\rL^{(2)}=\D_\rL^{(4)}\cT_\rL^{(-2)}$
introduced above,
the action (\ref{Action-left}) can be brought to the form
\bea
S_{\rm left}&=&
\frac{1}{2\pi} \oint (v_\rL, \rd v_\rL)
\int \rd^3 x \,{\rm d}^8\q\,E\, \cT_\rL^{(-2)}
~,
\label{N=4InvarAc2}
\eea
which makes  manifest the fact that $S_{\rm left}$ 
does not depend on 
$C_\rL^{(-4)}$. 

There is  a freedom in the choice of 
 $C_\rL^{(-4)}$, $C_\rR^{(-4)}$ and $C^{(-2,-{2})}$.
For instance,  given a real left weight-$m$ isotwistor superfield $\G_\rL^{(m)}$,
a real right weight-$n$ isotwistor superfield $\G_\rR^{(n)}$ and
a real hybrid weight-$(p,q)$ isotwistor superfield $\G^{(p,{q})}$,
we may  define $C_\rL^{(-4)}$, $C_\rR^{(-4)}$ and $C^{(-2,{-2})}$ as
\bea
C_\rL^{(-4)}=\frac{\G_\rL^{(m)}}{\D_\rL^{(4)}\G_\rL^{(m)}}
~,~~~~~~
C_\rR^{(-4)}=\frac{\G_\rR^{(n)}}{\D_\rR^{(4)}\G_\rR^{(n)}}
~,~~~~~~
C^{(-2,{-2})}=\frac{\G^{(p,{q})}}{\D^{(2,2)}\G^{(p,{q})}}
~.
\label{6.74}
\eea
Then the 
differential equations in (\ref{N=4AcComp-L})--(\ref{N=4AcComp-R})
are satisfied. To respect the super-Weyl transformation laws in 
 (\ref{N=4AcComp-L})--(\ref{hybrid-AcComp}), the superfields $\G_\rL^{(m)}$, 
 $\G_\rR^{(n)}$ and $\G^{(p,{q})}$ 
 should transform as $\d_\s \G_\rL^{(m)} = (m/2) \s \G_\rL^{(m)}$,
  $\d_\s \G_\rR^{(n)} = (n/2) \s \G_\rR^{(n)}$ and 
    $\d_\s \G^{(p,{q})} = [(p+q+2)/2)] \s \G^{(p,{q})}$.

It is natural to put forward an additional requirement 
that  the action be invariant under the mirror transformation. 
It is satisfied under the following conditions:
(i)  the Lagrangians $ \cL_\rL^{(2)}$ and  $ \cL_\rR^{(2)}$ are the mirror images of each other;
(ii)  the Lagrangian $ \cL^{(0,0)}$ is mirror invariant; 
(iii)  $C_\rL^{(-4)}$ and  $C_\rR^{(-4)}$ are the mirror images of each other;
(iv)  $C^{(-2,-{2})}$ is mirror invariant. 
If the kinematic factors are chosen as in (\ref{6.74}), 
the conditions (iii) and (iv) imply $m=n$ and $p=q$.

The simplest way to generate  $C_\rL^{(-4)}$ and $C_\rR^{(-4)}$ is 
to use  ordinary real scalar superfields $P_\rL (z) $, $P_\rR (z)$ and $P(z)$ and choose
\bea
C_\rL^{(-4)}=\frac{P_\rL}{\D_\rL^{(4)}P_\rL}~, \qquad
C_\rR^{(-4)}=\frac{P_\rR}{ \D_\rR^{(4)}P_\rR }~, \qquad
C^{(-2,-{2})}=\frac{P}{ \D^{(2,{2})} P }~.
\eea
In order to guarantee the fulfillment of
the super-Weyl transformation laws in 
(\ref{N=4AcComp-L})-- (\ref{hybrid-AcComp}), 
the superfields $P_\rL$,  $P_\rR$ and $P$ must transform as
\bea
\d_\s  P_\rL = \d_\s P_\rR =0~, \qquad \d_\s P=\s P~. 
\eea
The transformation of $P$ is similar to that appearing in the $\cN=3$ case.
If the action is chosen to be mirror invariant, then $P_\rL =P_\rR$.

In complete analogy with our four-dimensional analysis given in \cite{KLRT-M1},
it is of interest to give flat superspace versions of the actions 
(\ref{Action-left})--(\ref{hybrid-Ac1}).
In the flat superspace limit, 
 the dependence on the compensating superfields $C_\rL^{(-4)},C_\rR^{(-4)}$ and
$C^{(-2,{-2)}}$ can be seen to drop out.
The actions (\ref{Action-left}) and (\ref{Action-right}) reduce to
\begin{subequations}
\bea
S_{\rm left} (L_{\rm L}^{(2)}  )&=&\frac{1}{2\p} \oint  {(v_\rL, \rd v_\rL)} 
\int {\rm d}^3x \, D_{\rm L}^{(-4)} L_{\rm L}^{(2)}  \Big|_{\q =0}~,
\label{Action-left-flat} \\
S_{\rm right}(L_{\rm R}^{(2)} ) &=& 
\frac{1}{2\p} \oint  {(v_\rR, \rd v_\rR)} 
\int {\rm d}^3x \, D_{\rm R}^{(-4)} L_{\rm R}^{(2)} \Big|_{\q =0}~,
\label{Action-right-flat} 
\eea
\end{subequations}
with $ L_{\rm L}^{(2)}$, $ L_{\rm R}^{(2)}$ and $L^{(0,{0}) }$ the flat-superspace versions 
of the Lagrangians in (\ref{Action-left})--(\ref{hybrid-Ac1}).
Here we have introduced two  fourth-order operators, $D_{\rm L}^{(-4)} $ and $D_{\rm R}^{(-4)} $,
defined in terms of the flat covariant derivatives $D_\a^{i\bai}$, specifically
\bsubeq
\bea
D_{\rm L}^{(-4)} &:=&
\frac{1}{48}D^{(-2){\bar k}{\bar l}}D^{(-2)}_{{\bar k}{\bar l}}~,~~~
D^{(-2)}_{{\bar k}{\bar l}}:=D^{(-1)\g}_{{\bar k}}D_{\g{\bar l}}^{(-1)}~,
~~~
D_\a^{(-1)\bai}:=\frac{u_i}{(v_\rL,u_\rL)}D_\a^{i\bai}~;~~~~~~
\\
D_{\rm R}^{(-4)} &:=&
\frac{1}{48}D^{(-\bad) ij}D^{(-\bad)}_{ij}~,~~~
D^{(-\bad)}_{ij}:=D^{(-\bau)\g}_{i}D_{\g j}^{(-\bau)}
~,~~~
D_\a^{(-\bau)i}:=\frac{u_\bai}{(v_\rR,u_\rR)}D_\a^{i\bai}~.~~~~~~
\eea
\esubeq
The functionals  (\ref{Action-left-flat}) and (\ref{Action-right-flat}) 
are the 3D versions \cite{KPT-MvU}
of the 4D projective-superspace action \cite{KLR}.
The flat-superspace limit of the hybrid action (\ref{hybrid-Ac1}) is
\bea
S_{\rm hybrid} (L^{(0,{0}) }) &=& \frac{1}{(2\p)^2}
\oint {(v_\rL, \rd v_\rL)} 
\oint  {(v_\rR, \rd v_\rR)}
\int \rd^3 x \,  D_{\rm H}^{(-2,{-2})} L^{(0,{0})}\Big|_{\q =0}
~.~~~~~~
\label{hybrid-Ac1-flat}
\eea
where
\bea
D_{\rm H}^{(-2,{-2})}&:=&
-\frac{\ri}{64}(D^{\a(-1,-{1})}D_\a^{(-1,-{1})})
(D^{\b(1,-{1})}D_\b^{(1,-{1})})
(D^{\g(-1,{1})}D_\g^{(-1,{1})})
~,
\non \\
D_\a^{(-1, -{1})}&:=&\frac{u_iu_\bai}{(v_\rL,u_\rL)(v_\rR,u_\rR)}D_\a^{i\bai}
~,~ \\
D_\a^{(1, -{1})}&:=&\frac{v_iu_\bai}{(v_\rR,u_\rR)}D_\a^{i\bai}
~,\qquad
D_\a^{(-1, {1})}:=\frac{u_iv_\bai}{(v_\rL,u_\rL)}D_\a^{i\bai}
~.~~~~~~~~~
\non
\eea
The hybrid action (\ref{hybrid-Ac1-flat}) proves to be invariant 
under two types of  projective transformations, left and right ones. 
The left transformations have the form:
\be
(u_\rL \,,\,v_\rL)~\to~(u_\rL\,,\, v_\rL )\,F~, \qquad F=
\left(\begin{array}{cc}a~&0\\ b~&c~\end{array}\right)\,\in\,{\rm GL(2,\mathbb{C})}~.
\label{projectiveGaugeVar}
\ee
The right projective transformations are defined similarly.
Since $\{ D^{i \bar i}_\a , D^{ j \bar j}_\b \} \propto \pa_{\a \b}$, 
the left/right actions  (\ref{Action-left-flat}) and (\ref{Action-right-flat})
generate two derivatives at the component level, while the hybrid action (\ref{hybrid-Ac1-flat}) 
gives rise to three derivatives.
To the best of our knowledge, 
the hybrid projective action has been presented here for the first time.

The flat-superspace hybrid action (\ref{hybrid-Ac1-flat}) can also be rewritten 
in the following forms:
\begin{subequations}
\bea
S_{\rm hybrid} (L^{(0,{0}) }) &=& S_{\rm left} ({\frak L}_{\rm L}^{(2)}  )
~, \quad 
{\frak L}_\rL^{(2)} := \frac{\ri}{8\p} \oint  {(v_\rR, \rd v_\rR)} \, D^{\a(1,-{1})}D_\a^{(1,-{1})}
L^{(0,{0}) }~; \\
S_{\rm hybrid}(L^{(0,{0}) }) &=& S_{\rm right} ({\frak L}_{\rm R}^{(2)}  )
~, \quad 
{\frak L}_\rR^{(2)} := \frac{\ri}{8\p} \oint  {(v_\rL, \rd v_\rL)} \,D^{\a(-1,{1})}D_\a^{(-1,{1})}
L^{(0,{0}) }~.~~~~~~~
\eea
\end{subequations}

\subsection{Vector multiplet prepotentials}

The field strengths of two inequivalent vector multiplets are described 
by  left and right linear multiplets, $W^{(2)}_\rL$ and $W^{(2)}_\rR$  
subject to the constraints (\ref{6.24a}) and (\ref{6.24b}).
These constraints can be solved in terms of covariant weight-zero tropical multiplets. 
It suffices to restrict our analysis to the right linear multiplets 
$W^{(2)}_\rR =W^{\bar i \bar j} v_{\bar i} v_{\bar j}$.

A general solution to the constraint (\ref{6.24b}) is 
\bea
W_\rR^{(2)}(v_\rR)&=&
\frac{\ri}{4} \Big(\cD^{(\bad) i  j}
-4\ri\cS^{(\bad)i j}\Big)
\oint \frac{(v_\rL, \rd v_\rL)}{2\pi}
\frac{u_i u_j}{(v_\rL ,u_\rL)^2}
V_\rL(v_\rL) ~,
\label{6.79}
\eea
where $V_\rL(v_\rL) $ is a left tropical multiplet of weight zero.
The right-hand side of (\ref{6.79}) involves a constant isospinor $u_\rL = u^i$ constrained 
by the only condition 
$(v_\rL,  u_\rL) \neq 0$. It can be shown that (\ref{6.79}) is invariant under an arbitrary infinitesimal 
variation of $u_\rL$, that is  $\d u_\rL = \a u_\rL + \b v_\rL $, with $\a, \b \in {\mathbb C}$.
Thus $W_\rR^{(2)}(v_\rR)$ is independent of $u_\rL$.

The relation (\ref{6.79}) demonstrates a remarkable interplay between left and right 
projective multiplets.
The left-hand side of (\ref{6.79}) is the right $\cO(2)$ multiplet $W_\rR^{(2)}$, 
while the right-hand side 
is given in term of the left tropical prepotential $V_\rL$.
The above result can be represented in a slightly different form: 
\bea
W^{\bar i \bar j}&=&\frac{\ri}{4 }\oint \frac{(v_\rL, \rd v_\rL)}{2\p}
\Big(\cD^{(-2)\bar i \bar j}
-4\ri\cS^{(-2)\bar i \bar j}
\Big)V_\rL(v_\rL) ~.
\label{6.80}
\eea
This representation can be used to show that $W^{\bar i \bar j}$ is invariant under 
gauge transformations 
\bea
\d  V_\rL = \l_\rL + \breve{\l}_\rL~,
\label{left-gauge}
\eea
where the gauge parameter $\l_\rL$ is an arbitrary left arctic multiplet of weight zero.

Let us represent $V_\rL$ in terms of an unconstrained left isotwistor superfield 
$T^{(-4)}_\rL (v_\rL)$,
\bea 
V_\rL(v_\rL) = \D^{(4)}_\rL T^{(-4)}_\rL (v_\rL)~,
\label{6.81}
\eea
with the super-Weyl transformation law
\bea
\d_\s T^{(-4)}_\rL =-2 \s T^{(-4)}_\rL~.
\eea

As remarked at the end of subsection \ref{CovProjOp}, a left isotwistor superfield
$T^{(-4)}_\rL(v_\rL)$ can be represented in terms of a weight-$(-4,-2)$ isotwistor superfield 
$T^{(-4,-2)}_\rL(v_\rL,{v}_\rR)$
through the integral equation (\ref{hybrid-integral}).
It then appears  that (\ref{6.79}) is equivalent to 
\bea
W_\rR^{(2)}({v}_\rR)\equiv
\D^{(4)}_\rR \oint \frac{(v_\rL, \rd v_\rL)}{ 2\pi} 
\oint \frac{(\hat{v}_\rR, \rd \hat{v}_\rR)}{2\pi(v_\rR,\hat{v}_\rR)^2} 
\D^{(2,\hat{2})}T^{(-4,-2)}_\rL(v_\rL,\hat{v}_\rR) 
~,
\label{6.84}
\eea
with the operator $\D^{(2,\hat{2})}$ given by
\bea
\D^{(2,\hat{2})}:=
\frac{\ri}{4}
v_iv_j\hat{v}_\bai\hat{v}_\baj
\Big(\cD^{\a i\bai}\cD_\a^{j\baj}
-4\ri\cS^{ij\bai\baj}
\Big)
~.
\eea
In the form of equation (\ref{6.84}) it becomes manifest that $W_\rR^{(2)}({v}_\rR)$, 
originally defined by
(\ref{6.79}), satisfies the right analyticity constraint. 

The proof of (\ref{6.84}) is achieved in few steps.
By using (\ref{6.81}), (\ref{hybrid-integral}) and (\ref{proj2L}), the relation (\ref{6.79}) 
can be equivalently written as
\bea
W_\rR^{(2)}(v_\rR)&=&
-\oint \frac{(v_\rL, \rd v_\rL)}{ 2\pi} 
\oint \frac{(\hat{v}_\rR, \rd \hat{v}_\rR)}{2\pi} \,
\D^{(-2,{2})}\D^{(2,-\hat{2})}\D^{(2,\hat{2})}\,
T^{(-4,-2)}_\rL(v_\rL,\hat{v}_\rR) 
~.
\label{proof-1}
\eea
Here we have introduced the operators
\bea
\D^{(-2,2)}:=\frac{\ri}{4}\frac{u_iu_j}{(v_\rL,u_\rL)^2}
\Big(\cD^{(\bad) ij}
-4\ri\cS^{(\bad)ij}
\Big)
~,~~~
\D^{(2,-\hat{2})}:=\frac{\ri}{4}\frac{\hat{u}_\bai\hat{u}_\baj}{(\hat{v}_\rR,\hat{u}_\rR)^2}
\Big(\cD^{(2)\bai\baj}
-4\ri\cS^{(2)\bai\baj}
\Big)~.~~~~~~
\eea
In (\ref{proof-1}) we have the freedom 
to choose\footnote{We assume that the contour integral 
in the isotwistor variable $\hat{v}_\rR$ is such that 
$(\hat{v}_\rR,v_\rR)\ne0$.} $\hat{u}_\rR={v}_\rR$
and obtain
\bea
W_\rR^{(2)}(v_\rR)&=&
-\oint \frac{(v_\rL, \rd v_\rL)}{ 2\pi} \,
\D^{(-2,{2})}\D^{(2,{2})}\oint \frac{(\hat{v}_\rR, \rd \hat{v}_\rR)}{2\pi(\hat{v}_\rR,v_\rR)} \,
\D^{(2,\hat{2})}\,
T^{(-4,-2)}_\rL(v_\rL,\hat{v}_\rR) 
~.
\label{proof-2}
\eea
Now,  making use of eq. (\ref{proj2R}), we readily arrive at (\ref{6.84}).

\subsection{Poincar\'e supergravity}

To describe $\cN=4$ Poincar\'e, we need two compensators coupled to conformal supergravity.
In the role of compensators we can choose a left linear multiplet $W^{ij}$ and a right linear 
multiplet $W^{\bar i \bar j}$ such that 
\bea
W_\rL:=\sqrt{W^{ij}W_{ij}}\neq 0 ~, \qquad W_\rR:=\sqrt{W^{\bai\baj}W_{\bai\baj}} \neq 0~.
\eea
These scalar superfields are characterized by  the super-Weyl transformation laws
\bea
\d_\s W_\rL=\s W_\rL~,  \qquad \d_\s W_\rR=\s W_\rR~.
\eea
These scalars turn out to have interesting properties.
The superfield $W_\rL$ satisfies the equation
\bea
\Big(\cD^{k(\bai}_{\a}\cD^{\baj)}_{\b k}-4\ri C_{\a\b}^{\bai\baj}\Big)(W_\rL)^{-1}=0
~,
\eea
which can be derived 
using the identity
\bea
\cD^{(\bak}_{\a k}\cD_{\b l}^{\bal)}W_{ij}&=&
-\frac{1}{6}\ve_{\a\b}\ve_{k(i}\ve_{j)l}\cD^{\g(\bak}_p\cD_{\g q}^{\bal)}W^{pq}
-{4\ri}C_{\a\b}^{\bak\bal}W_{k(i}\ve_{j)l}
+{2\ri}\ve_{\a\b}S_{kl}{}^{\bak\bal}W_{ij}
\non\\
&&
-{2\ri}\ve_{\a\b}S_{ij}{}^{\bak\bal}W_{kl}
-{2\ri}\ve_{\a\b}\ve_{kl}S_{(i}{}^{p}{}^{\bak\bal}W_{j)p}
~.
\eea
Similarly one can derive the equation
\bea
\Big(\cD^{(i \bak}_{\a}\cD^{j)}_{\b \bak}-4\ri B_{\a\b}^{ij}\Big)(W_\rR)^{-1}=0
~.
\eea

Poincar\'e supergravity is described by two Lagrangians, left and right ones, 
which can be chosen as 
\begin{subequations}
\bea
\cL^{(2)}_{\rm SUGRA, left } &=& 
\frac{1}{\k^2} \,W^{(2)}_\rL \ln \frac{W^{(2)}_\rL }{\ri \Upsilon^{(1)}_\rL \breve\Upsilon^{(1)}_\rL } 
+\frac{\x_\rL }{\k^2}\, V_\rL W^{(2)}_\rL~, \\
\cL^{(2)}_{\rm SUGRA, right } &=& 
\frac{1}{\k^2} \,W^{(2)}_\rR \ln \frac{W^{(2)}_\rR }{\ri \Upsilon^{(1)}_\rR \breve\Upsilon^{(1)}_\rR } 
+\frac{\x_\rR}{\k^2}\, V_\rR W^{(2)}_\rR~.
\eea
\end{subequations}
Here $V_\rL$ is the tropical prepotential for $W^{(2)}_\rR$, see equation (\ref{6.79}), while 
$V_\rR$ is the tropical prepotential for $W^{(2)}_\rL$, in particular
\bea
W_\rL^{(2)}(v_\rL)&=&
\frac{\ri}{4} \Big(\cD^{(2) \bai  \baj}
-4\ri\cS^{(2)\bai \baj}\Big)
\oint \frac{(v_\rR, \rd v_\rR)}{2\pi}
\frac{u_\bai u_\baj}{(v_\rR ,u_\rR)^2}
V_\rR(v_\rR) ~.
\eea 
The action is invariant under left and right gauge transformations, 
the left one being given by eq. (\ref{left-gauge}).
The cosmological term
is described by two BF-couplings. 
Using the representation (\ref{6.84}), it can be shown that the action does not change if
the BF coupling constants are modified as 
\bea 
\x_\rL \to \x_\rL + a~, \qquad \x_\rR \to \x_\rR - a~,
\label{6.97}
\eea
for any real constant $a$.
Moreover,  using eq. (\ref{6.84}), integration by parts and the relations (\ref{proj2L})--(\ref{proj2R}), 
the reader can prove the following important results:
\bea
S_{\rm left}(V_\rL W_\rL^{(2)} )
=S_{\rm right}(V_\rR W_\rR^{(2)} )
=- S_{\rm hybrid}( V_\rL V_\rR)~.
\eea
Note that the freedom (\ref{6.97}) 
is absent if the theory is required to be mirror invariant, 
for then $\x_\rL = \x_\rR \equiv \x/2$.

There exists a dual off-shell formulation for Poincar\'e supergravity with two 
compensators, a vector multiplet and a hypermultiplet.
Let us use the freedom (\ref{6.97}) to set $\x_\rR=0$, 
\begin{subequations}
\bea
\cL^{(2)}_{\rm SUGRA, left } &=& 
\frac{1}{\k^2} \,W^{(2)}_\rL \ln \frac{W^{(2)}_\rL }{\ri \Upsilon^{(1)}_\rL \breve\Upsilon^{(1)}_\rL } 
+\frac{\x }{\k^2}\, V_\rL W^{(2)}_\rL
= \frac{1}{\k^2} \,W^{(2)}_\rL \ln \frac{W^{(2)}_\rL }{\ri \Upsilon^{(1)}_\rL {\rm e}^{-\x V_\rL}
\breve\Upsilon^{(1)}_\rL } 
~, ~~~~~~~~~~~ 
\label{6.98a} \\
\cL^{(2)}_{\rm SUGRA, right } &=& 
\frac{1}{\k^2} \,W^{(2)}_\rR \ln \frac{W^{(2)}_\rR }{\ri \Upsilon^{(1)}_\rR \breve\Upsilon^{(1)}_\rR } 
~.
\eea
\end{subequations}
The left model (\ref{6.98a}) can now be dualized in the same fashion as it was done in subsection 
\ref{subsection5.6}. As a result, we arrive at the following formulation 
\begin{subequations}
\bea
\cL^{(2)}_{\rm SUGRA, left } &=& 
- \frac{\ri}{\k^2} 
\breve\Upsilon^{(1)}_\rL {\rm e}^{-\x V_\rL} \Upsilon^{(1)}_\rL~, 
\label{6.99a}  \\
\cL^{(2)}_{\rm SUGRA, right } &=& 
\frac{1}{\k^2} \,W^{(2)}_\rR \ln \frac{W^{(2)}_\rR }{\ri \Upsilon^{(1)}_\rR \breve\Upsilon^{(1)}_\rR } 
~.
\eea
\end{subequations}
The theory  is invariant under 
 the gauge transformations (\ref{left-gauge}) provided the hypermultiplet transforms as 
 \be
 \d \U^{(1)}_\rL = \x \l \U^{(1)}_\rL~.
 \ee
In the case of supergravity without cosmological term, $\x =0$, we can also dualize $W^{(2)}_\rR$
into a right weight-one arctic multiplet $ \U^{(1)}_\rR$ and its conjugate $\breve{ \U}^{(1)}_\rR$.

It is instructive to see explicitly how the compensators can be used to obtain 
Poincar\'e supergravity from the conformal one by a process 
known as ``de-gauging'' \cite{Howe} (or, equivalently, fixing 
the conformal gauge).  We will use the formulation with two vector multiplets, 
left and right ones, as the compensators. 
First of all, we note that the super-Weyl freedom 
can be completely fixed by choosing the gauge condition
\bea
W_\rL=1~.
\label{6.101}
\eea
Let $w_{ij}$ denote the field strength $W_{ij}$ in this gauge.
An important observation is that, because the superfield $w^{ij}$ is analytic 
$\cD_\a^{(i\bai}w^{kl)}=0$, from $\cD_\a^{i\bai}(w^{kl}w_{kl})=0$ one can obtain that $w_{ij}$
is annihilated by the spinor covariant derivatives,
\bea
\cD_\a^{i\bai}w^{kl}=0
~.
\eea
This condition implies nontrivial constraints on the geometry.
In particular, 
the consistency condition
\bea
0=\{\cD_\a^{i\bai},\cD_\b^{j\baj }\}w^{kl}&=&
2\ri\ve^{ij}\ve^{\bai \baj }(\g^c)_{\a\b}\cD_cw^{kl}
+{2\ri}\ve_{\a\b}\ve^{\bai \baj }(2\cS+X)(\ve^{k(i}w^{j)l}+\ve^{l(i}w^{j)k})
\non\\
&&
-4\ri\ve_{\a\b}\ve^{ij}\cS^{(k}{}_{p}{}^{\bai\baj}w^{l)p}
+4\ri C_{\a\b}{}^{\bai\baj}(\ve^{k(i}w^{j)l}+\ve^{l(i}w^{j)k})
\eea
is equivalent to
\bea
\cD_aw^{kl}=0
~,~~~
X&=&-2\cS~,~~~
\cS^{ij\bai\baj}=w^{ij}\cS^{\bai\baj}
~,~~~
C_{\a\b}^{\bai\baj}=0~,
\label{6.105}
\eea
for some right $\cO(2)$ multiplet $\cS^{\bar i \bar j}$, 
\bea
\cD^{i (\bar i}_\a \cS^{\bar j \bar k)} =0~.
\label{6.106}
\eea
As a result, $w_{ij}$ is covariantly constant in the super-Weyl gauge (\ref{6.101}).
All the relations in (\ref{6.105}) and (\ref{6.106}) are artifacts  of the same super-Weyl gauge fixing.
The algebra of covariant derivatives  reduces to
\bea
\{\cD_\a^{i\bai},\cD_\b^{j\baj }\}&=&
2\ri\ve^{ij}\ve^{\bai \baj }\cD_{\a\b}
-2\ri\ve_{\a\b}\ve^{ij}\cS^{\bai \baj}w^{kl}\bL_{kl}
+8\ri\ve_{\a\b}\ve^{ij}\cS\bR^{\bai \baj }
-2\ri\ve_{\a\b}\ve^{\bai \baj }w^{ij}\cS^{\bak\bal}\bR_{\bak\bal}
\non\\
&&
+4\ri B_{\a\b}{}^{ij}\bR^{\bai \baj}
+2\ri\ve_{\a\b}\ve^{\bai \baj }B^{\g\d}{}^{ij}\cM_{\g\d}
-4\ri w^{ij}\cS^{\bai \baj }\cM_{\a\b}
-4\ri\ve^{ij}\ve^{\bai \baj }\cS\cM_{\a\b}
~.~~~~~~
\eea
It is clear that the super-Weyl  gauge condition has broken the mirror symmetry.
The structure group is still ${\rm SL}(2,\mathbb{R}) \times {\rm SU}(2)_\rL
\times {\rm SU}(2)_\rR$. However, 
the SU(2)$_\rL$ curvature can be seen to take its values in 
a one-dimensional subalgebra of ${\frak su}(2)$ 
generated by 
$w^{kl}\bL_{kl}$. 
Therefore, the SU(2)$_\rL$ gauge freedom can be partially fixed by choosing 
 the SU(2)$_\rL$ connection as 
\bea
(\Phi_\rL)_A=\Phi_A \bL~,\qquad
\bL:=w^{kl}\bL_{kl}
~.
\eea
As a result, in the left sector  we stay with a residual 
gauge group ${\rm U}(1)_\rL \subset {\rm SU}(2)_\rL$  generated by $\bL$. 
 The condition of covariant constancy, $\cD_A w^{ij}=0$, 
now means 
that $w^{ij}$ is constant,
$\pa_Mw^{ij}=0$.

Using the second compensator, $W^{\bar i \bar j}$, allows us to partially fix the gauge group 
$ {\rm SU}(2)_\rR$ by imposing a condition $W^{\bar i \bar j} \propto \d^{\bar i \bar j}$, in complete 
analogy with 4D $\cN=2$ supergravity \cite{deWPV}. In this gauge,  we stay with a residual 
local group ${\rm U}(1)_\rR \subset {\rm SU}(2)_\rR$.

\subsection{Dynamical systems}
All the $\cN=3$ locally supersymmetric sigma-models considered in subsection 
\ref{subsection5.7} can be readily generalized to off-shell $\cN=4$ theories 
described by either left projective multiplets or right ones.
The nontrivial new feature of $\cN=4$ supersymmetry is that it allows
off-shell couplings that mix left and right projective multiplets.
To illustrate this idea, it suffices to consider dynamical systems 
involving  left and right vector multiplets.

Consider several vector multiplets described by right tropical prepotentials $V^{( \bar 0)}_{ I}$
and left tropical prepotential $V^{( 0)}_{\bar I}$, and let $W^{(2)}_I $ and $W^{(\bar 2)}_{\bar I} $
be the corresponding left and right $\cO(2)$  field strengths.
A gauge-invariant 
action functional is generated by three Lagrangians (left, right and hybrid)  of the form:
\begin{subequations}
\bea
\cL^{(2)}_\rL &=& 
\cF_\rL (W^{(2)}_I ) + m^{I \bar J } W^{(2)}_I V^{( 0)}_{\bar J}~,
\qquad m^{I \bar J} = ( m^{I \bar J} )^* = {\rm const}
\label{6.109a}  \\
 \cL^{(2)}_\rR &=& 
\cF_\rR (W^{(\bar 2)}_{\bar I} ) + m^{{\bar I} J } W^{(\bar 2)}_{\bar I} V^{( \bar 0)}_{ J}~,
\qquad m^{ \bar I J} = ( m^{\bar I J} )^* = {\rm const}
\label{6.109b} \\
\cL^{(0,0)} &=& 
\cH  ( W^{2)}_I , W^{(\bar 2)}_{\bar J} ) + \m^{I \bar J } V^{(\bar 0)}_I V^{(0)}_{\bar J}~, 
\qquad \m^{I \bar J} = ( \m^{I \bar J} )^* = {\rm const}~.~~~
\label{6.109c}
\eea
\end{subequations}
The kinetic terms should obey the following homogeneity conditions:
\begin{subequations}
\bea
W^{(2)}_I \frac{\pa }{\pa W^{(2)}_I} \cF_\rL &=&\cF_\rL~,\\
W^{(\bar 2)}_{\bar I} \frac{\pa }{\pa W^{(\bar 2)}_{\bar I}} \cF_\rR &=&\cF_\rR~,\\
\Big( W^{(2)}_I \frac{\pa }{\pa W^{(2)}_I}  
+ W^{(\bar 2)}_{\bar J} \frac{\pa }{\pa W^{(\bar 2)}_{\bar J}}  \Big)
\cH &=&0~.
\eea
\end{subequations}
The $m$- and $\m$-terms in (\ref{6.109a})--(\ref{6.109c}) are three different forms 
of the BF couplings.

\section{Conclusion}
\setcounter{equation}{0}

As is well known,  off-shell supergravity-matter couplings in diverse dimensions
may be  conveniently derived starting from a superconformal perspective. 
In this paper we have developed the superspace geometry of $\cN$-extended conformal 
supergravity in three space-time dimensions. Using this geometric setup, we 
have constructed general off-shell supergravity-matter couplings for $\cN \leq 4$.
In the most interesting and previously  unexplored  cases $\cN=3$ and $\cN=4$, 
we have proposed new off-shell supermultiplets coupled to conformal supergravity, 
in terms of which both the supergravity and matter actions are given.

It should be emphasized that the conventional constraints 
on $\cN$-extended superspace geometry,  
eqs. (\ref{constr-0-4-N})--(\ref{constr-1-4-N}),
were introduced fifteen  years ago  in \cite{HIPT}. However, the corresponding Bianchi identities
were not been solved by Howe {\it et a}l. Moreover, the issue of constructing 
supergravity-matter couplings 
or even a superfield supergravity action in the case $\cN=3, \,4$ was not  addressed in \cite{HIPT}.

Our approach to the three-dimensional $\cN=3\,,4$ supergravity theories 
is a natural extension of  the projective-superspace formulations 
for general 5D $\cN=1$  and 4D $\cN=2$ supergravity-matter theories which were 
developed in \cite{KT-M5D-1,KT-M5D-2,KLRT-M1,K-08,KLRT-M2}.
More specifically, this is true for $\cN=3$ supergravity. In the $\cN=4$ case, however, 
we have discovered a new theoretical phenomenon as compared with the situation 
in higher dimensions. 
It is the existence of three types of covariant off-shell projective supermultiplets 
(left, right and hybrid ones) in terms of which the general matter couplings are constructed. 

In this paper, the supergravity-matter couplings are formulated  using
superspace and superfields. Of course, many applications require a reformulation 
in terms of component
fields. In four dimensions, techniques have been developed \cite{KT-M2,BK,GKT-M}
to reduce   the 4D $\cN=2$ supergravity-matter actions of \cite{KLRT-M1,K-08,KLRT-M2}
to components.  Similar techniques can be developed in three dimensions 
for the theories constructed above. This  issue of component reduction will be addressed 
in a separate publication.
\\

\noindent
{\bf Acknowledgements:}\\
SMK is grateful to the Department of Physics and Astronomy at
Uppsala University for hospitality at the initial stage of this project.
GT-M acknowledges the hospitality and support of the School of Physics at the University 
of Western Australia during the final stage of this project. 
The work  of SMK and UL is supported in part by the Australian Research Council.
The work of UL is supported by VR-grant 621-2009-4066.
The work of GT-M is supported by the European 
Commission, Marie Curie Intra-European Fellowships under contract No.
PIEF-GA-2009-236454.  

~\\
\noindent
{\bf Note added in proof:} After this paper had been accepted for publication,Ê
one of us (UL) was informed of a paper by Howe and Sezgin \cite{Howe:2004ib}Ê
in which some aspects of 3D $\cN=8$ superconformal geometry were elaboratedÊ
following \cite{HIPT}. Even in this special case, our results obtained in section 2Ê
are more complete.

\appendix

\section{Notation and conventions}
\label{3Dconventions}
\setcounter{equation}{0}

Our conventions for spinors
in three space-time dimensions (3D) are compatible with
the 4D two-component spinor
formalism used in \cite{WB,Ideas}.
More specifically, the starting point for setting up  our 3D spinor formalism  is
the 4D sigma-matrices
\bea
(\s_{\mun } )_{\a \dt \b}:= ({\mathbbm 1}, \vec{\s} ) ~, \qquad
(\tilde{\s}_{\mun } )^{{\dt \a}  \b}:= ({\mathbbm 1}, - \vec{\s} ) ~, \qquad {\mun }=0,1,2,3
~,
\label{4DsigmaM}
\eea
where $\vec{\s}=(\s_1,\s_2,\s_3)$ are the Pauli matrices.
By deleting the matrices with space index $\mun =2$ we obtain the 3D gamma-matrices
\begin{subequations}
\bea
(\s_{\mun } )_{\a \dt \b}\quad & \longrightarrow & \quad (\g_m )_{\a  \b} = (\g_m)_{\b\a} ~
=({\mathbbm 1}, \s_1, \s_3) ~,\\
(\tilde{\s}_{\mun } )^{\dt \a  \b}\quad & \longrightarrow & \quad (\g_m )^{\a  \b} = (\g_m)^{\b\a}
=\ve^{\a \g} \ve^{\b \d} (\g_m)_{\g \d} ~,
\eea
\end{subequations}
where the spinor indices are  raised and lowered using
the SL(2,${\mathbb R}$) invariant tensors
\bea
\ve_{\a\b}=\left(\begin{array}{cc}0~&-1\\1~&0\end{array}\right)~,\qquad
\ve^{\a\b}=\left(\begin{array}{cc}0~&1\\-1~&0\end{array}\right)~,\qquad
\ve^{\a\g}\ve_{\g\b}=\d^\a_\b
\eea
as follows:
\bea
\psi^{\a}=\ve^{\a\b}\psi_\b~, \qquad \psi_{\a}=\ve_{\a\b}\psi^\b~.
\eea
By construction, the matrices $ (\g_m )_{\a  \b} $ and $ (\g_m )^{\a  \b} $ are {\it real} 
and symmetric.
Using the properties of the 4D sigma-matrices, we can immediately read off the properties
of the  3D gamma-matrices. In particular, for the
matrices
\be
\g_m:=(\g_m)_\a{}^{\b}=\ve^{\b\g}(\g_m)_{\a\g}
\ee
we readily  obtain the relations
\bsubeq
\bea
&\{\g_m,\g_n\}=2\eta_{mn}{\mathbbm 1}~,
\\
&\g_m\g_n=\eta_{mn}{\mathbbm 1}+\ve_{mnp}\g^p~,
\eea
\esubeq
where the 3D Minkowski metric is $\eta_{mn}=\eta^{mn}={\rm diag}(-1,1,1)$, 
and the Levi-Civita tensor is normalized as $\ve_{012}=-\ve^{012}=-1$.
As usual, the 3D vector indices are labeled by values $m=0,\,1,\,2$.
Some useful relations involving $\g$-matrices are 
\bsubeq
\bea
(\g^a)_{\a \b} (\g_a)_{\g \d} &=& 2\ve_{\a(\g}  \ve_{\d)\b} ~,
\\
\ve_{abc}(\g^b)_{\a\b}(\g^c)_{\g\d}&=&
\ve_{\g(\a}(\g_a)_{\b)\d}
+\ve_{\d(\a}(\g_a)_{\b)\g}
~,
\\
\tr[\g_a\g_b\g_{c}\g_d]&=&
2\eta_{ab}\eta_{cd}
-2\eta_{ac}\eta_{db}
+2\eta_{ad}\eta_{bc}
~.
\eea
\esubeq

Given a three-vector $V_m$, it can equivalently be realized as a symmetric spinor 
$V_{\a\b} =V_{\b \a}$.
The relationship between $V_m$ and $V_{\a \b}$ is as follows:
\bea
V_{\a\b}:=(\g^a)_{\a\b}V_a=V_{\b\a}~,\qquad
V_a=-\hf(\g_a)^{\a\b}V_{\a\b}~.
\label{vector-rule}
\eea
In three-dimensions an
antisymmetric tensor $F_{ab}=-F_{ba}$ is Hodge-dual to a three-vector $F_a$, 
specifically
\bea
F_a=\hf\ve_{abc}F^{bc}~,\qquad
F_{ab}=-\ve_{abc}F^c~.
\label{hodge-1}
\eea
Then, the symmetric spinor $F_{\a\b} =F_{\b\a}$, which is associated with $F_a$, can 
equivalently be defined in terms of  $F_{ab}$: 
\bea
F_{\a\b}:=(\g^a)_{\a\b}F_a=\hf(\g^a)_{\a\b}\ve_{abc}F^{bc}
~.
\label{hodge-2}
\eea
These three algebraic objects, $F_a$, $F_{ab}$ and $F_{\a \b}$, 
are in one-to-one correspondence to each other, 
$F_a \leftrightarrow F_{ab} \leftrightarrow F_{\a\b}$.
The corresponding inner products are related to each other as follows:
\bea
-F^aG_a=
\hf F^{ab}G_{ab}=\hf F^{\a\b}G_{\a\b}
~.
\eea

Let $\cM_{ab}=-\cM_{ba}$  be the Lorentz generators. 
They act on  a vector $V_a$ as 
\bea
\cM_{ab}V_c=2\eta_{c[a}V_{b]}~,
\eea
and on a spinor $\j_\a$ as  
\bea
\cM_{ab}\psi_\a
=\hf\ve_{abc}(\g^c)_\a{}^\b\psi_\b
~.
\eea
In accordance with
(\ref{vector-rule})--(\ref{hodge-2}), the Lorentz generators can also be realized as the vector 
$\cM_a$ or the symmetric spinor $\cM_{\a\b}$ such that 
\bea
\cM_{a}\psi_\a
=- \hf (\g_a)_\a{}^\b\psi_\b~, \qquad 
\cM_{\a\b}\psi_\g
=\ve_{\g(\a}\psi_{\b)}
~.
\eea
As is clear from the explicit form of the $\g$-matrices, 
we are using a Majorana representation in which all the $\g$-matrices are real,
and any Majorana spinor $\j^\a$ is real,  
\bea
(\psi^{\a})^* = \psi^{\a} ~,\qquad (\psi_{\a})^*= \psi_{\a}  ~.
\eea

In this paper we often make use of the group isomorphisms 
${\rm SO}(3) \cong {\rm SU}(2)/{\mathbb Z}_2$
and ${\rm SO}(4) \cong  \big( {\rm SU}(2)_{\rL}\times {\rm SU}(2)_{\rR} \big)/{\mathbb Z}_2$
in order to convert 
each SO(3) and SO(4) vector index into a pair of SU(2) ones. 
In the case of SO(4), the $R$-symmetry group of $\cN=4$ supersymmetry,  
we first  introduce the following $\S$-matrices
\bea
(\S_I)_{i\bar{i}}=({\mathbbm 1},\ri\s_1,\ri\s_2,\ri\s_3)
~,~~~~~~
I={\bf 1},\cdots,{\bf 4}~,~~~
i=1,2~,~~
\bar{i}=\bar{1},\bar{2}
\label{A.16}
\eea
which 
can be compared 
to the 4D Minkowski-space $\s$-matrices (\ref{4DsigmaM}).
The index $I$ is an SO(4) vector one, while the indices $i$ and  $\bar{i}$  are, respectively,
SU(2)$_{\rL}$ and SU(2)$_{\rR}$ spinor indices.
Under complex conjugation the $\S$-matrices satisfy the reality property
\bea
\big((\S_I)_{i\bar{i}}\big)^*=(\S_I)^{i\bar{i}} =\ve^{ij} \ve^{\bar i \bar j}(\S_I)_{j\bar{j}}~.
\eea
Given SU(2)$_{\rL}$ and SU(2)$_{\rR}$ spinors $\psi_i$ and $\chi_{\bar{i}}$, respectively,
we raise and lower their indices by using the antisymmetric  
tensors $\ve^{ij},\ve_{ij}$ and $\ve^{{\bar i}{\bar j}},\ve_{{\bar i}{\bar j}}$
($\ve^{12}=\ve_{21}=\ve^{{\bar 1}{\bar 2}}=\ve_{{\bar 2}{\bar 1}}=1$) according to the rules:
\bea
\psi^{i}=\ve^{ij}\psi_j~,~~~
\psi_{i}=\ve_{ij}\psi^j
~,~~~~~~
\chi^{{\bar i}}=\ve^{\bai\baj}\chi_\baj~,~~~
\chi_{\bai}=\ve_{\bai\baj}\chi^\baj
~.
\eea
For practical calculations, it is useful to  introduce the $\t$-matrices
\bea
(\t_I)_{i\bar{i}}:=\frac{1}{\sqrt{2}}(\S_{I})_{i\bar{i}}
\eea
which have the following properties:
\bsubeq
\bea
(\t_{(I})_{i\bar{j}}(\t_{J)})^{j\bar{j}}=\hf\d_{IJ}\d_i^j
~&,&~~~
(\t_{(I})_{j\bar{i}}(\t_{J)})^{j\bar{j}}=\hf\d_{IJ}\d_{\bar{i}}^{\bar{j}}
~,
\\
(\t_I)_{i\bar{i}}(\t^I)_{j\bar{j}}=\ve_{ij}\ve_{\bar{i}\bar{j}}~&,&~~~
(\t_I)_{i\bar{i}}(\t_J)^{i\bar{i}}=\d_{IJ}
~.
\eea
\esubeq
It is the $\t$-matrices which are used in the paper to convert 
each SO(4) vector index  to a pair of isospinor ones,  $I\,\to\,i\bar{i}$.
Associated with  an SO(4) vector $A_{I}$ 
is the second-rank isospinor $A_{i\bar{i}}$ 
defined by
\bea
A_{i\bar{i}}:=(\t_I)_{i\bar{i}}A^{I} \quad \longleftrightarrow \quad
A_{I}=(\t_I)^{i\bar{i}}A_{i\bar{i}}
~.
\eea
With the normalization chosen for the $\t$-matrices, it holds that 
\bea
\d_I^J\,\to\,\d_i^j\d_{\bar{i}}^{\bar{j}}
~,~~~~~~
A_{I}B^{I}= A_{i\bar{i}}B^{i\bar{i}}
~.
\eea

In the case of $\cN=3$ supersymmetry,  the 
$R$-symmetry group is ${\rm SO}(3) \cong {\rm SU}(2)/{\mathbb Z}_2$.
The corresponding $\S$-matrices are 
\bea
(\S_I)_{i j}=({\mathbbm 1},\ri\s_1, \ri\s_3) =(\S_I)_{ji}
~,~~~~~~
I={\bf 1}, {\bf 2},{\bf 3}~,~~~
i,j=1,2~.
\eea 
They are obtained from the SO(4) 
$\S$-matrices, eq.  (\ref{A.16}),
by removing $(\S_{\bf 3})_{i\bar{i}}$.
The remaining matrices $(\S_I)_{ij}$
are symmetric in $i,j$.
The symmetric $\t$-matrices are defined as $(\t_I)_{ij }:=\frac{1}{\sqrt{2}}(\S_{I})_{ij} =(\t_I)_{ji}$.
 Their properties are
\bea
(\t_I)_{ij}(\t^I)_{kl}=-\ve_{i(k}\ve_{l)j}~&,&~~~
(\t_I)_{ij}(\t_J)^{ij}=\d_{IJ}~.
\eea

More relations involving $\cN=3,4$ isospinors are described in the main body of the paper.


\section{Derivation of the left projection operator}
\label{AppB}
\setcounter{equation}{0}

In this Appendix we derive the  left covariant projection operator $\D^{(4)}_\rL$ used in subsection 
\ref{CovProjOp}.
The expression for the right covariant projection operator $\D^{(4)}_\rR$ 
follows using  the mirror map.

The covariant projector operator is a fourth order differential operator 
$\D_\rL^{(4)}$ such that given any weight-$(n-4)$ left isotwistor superfield $U_\rL^{(n-4)}$,
\bea
Q_\rL^{(n)}=\D_\rL^{(4)}U_\rL^{(n-4)}
\eea
is a weight-$n$ projective superfield.
The projector will be of the form
\bea
\D_\rL^{(4)}=\frac{1}{48}\Big(\cD^{(2)\bai\baj}\cD^{(2)}_{\bai\baj}+\cdots\Big)
~,~~~~~~
\cD^{(2)\bai\baj}:=\cD^{(1)\g (\bai}\cD_\g^{(1)\baj)}
~,
\eea
with the first term being the  flat superspace limit and  the dots 
denoting curvature dependent terms. 
A systematic, albeit  time consuming, way 
to construct the full projector is to act with $\cD_\a^{(1)\bai}$ on
$\cD^{(2)\baj\bak}\cD^{(2)}_{\baj\bak}$; 
the result is a function of  curvature terms and covariant derivatives which vanish in the flat limit.
One then iteratively adds curvature dependent 
terms to complete $\cD^{(2)\bai\baj}\cD^{(2)}_{\bai\baj}$ to the full $\D_\rL^{(4)}$.
Instead, we  use a short cut
and derive the  projector  using known results together with some simple observations. 
Let us  list the steps:

(i)
 A crucial observation is that, when acting on weight-$n$ left isotwistor superfields,
eventually carrying also Lorentz and SU(2)$_\rR$ indices, 
the algebra 
of $\cD_\a^{(1)\bai}$ derivatives (\ref{N=4alg}) becomes as follows
\bea
\{\cD_\a^{(1)\bai},\cD_\b^{(1)\baj}\}U^{(n)}_{\g_1\cdots\g_p}{}{}_{\bak_1\cdots \bak_q}
&=&
\Big(-2\ri\ve_{\a\b}\ve^{\bai\baj}\cS^{(2)}{}^{\bak\bal}\bR_{\bak\bal}
+4\ri B^{(2)}_{\a\b}\bR^{\bai\baj}
-4\ri \cS^{(2)}{}^{\bai\baj}\cM_{\a\b}
\non\\
&&~~~
+2\ri\ve_{\a\b}\ve^{\bai\baj}B^{(2)\g\d}\cM_{\g\d}
\Big)U^{(n)}_{\g_1\cdots\g_p}{}{}_{\bak_1\cdots \bak_q}
~.
\label{alg+3D}
\eea

(ii)
We next note an analogue to the superspace geometry of 4D, $\cN=2$ conformal supergravity
as formulated in \cite{Howe}, where the structure group is SL(2,$\mathbb{C})\times$U(2).
When acting on a superfield $U_{\a_1\cdots\a_p}{}_{i_1\cdots i_q}$ 
with $p$ undotted spinor indices and $q$ SU(2) indices,
the undotted 
spinor covariant derivatives algebra 
reduces to\footnote{We use the 4D notations and the algebra of \cite{KLRT-M2}.}\bea
\{\cD_\a^i,\cD_\b^j\}U_{\a_1\cdots\a_p}{}_{i_1\cdots i_q}&=&
\Big(\,2 \ve_{\a\b}\ve^{ij}S^{kl}J_{kl}
+4 Y_{\a\b}J^{ij}
+4S^{ij}M_{\a\b}
\non\\
&&~~~
+2\ve_{\a\b}\ve^{ij}Y^{\g\d}M_{\g\d}\Big)U_{\a_1\cdots\a_p}{}_{i_1\cdots i_q}
~.~~~~~~~~
\label{alg-4D}
\eea
Here, $M_{\a\b},\bar{M}_{\ad\bd}$ are the 4D Lorentz generators in spinor notations and
$J^{ij}$ is the SU(2) generator.\footnote{Note that for this sector of the geometry
the ${\rm U}(1)$ generator never appears and we can forget about it in our considerations.}
Some relevant dimension-3/2 Bianchi identities are
\bea
\cD_{\a}^{(i}S^{jk)}=\cD_{(\a}^{i}Y_{\b\g)}=0~,~~~~~~
\cD_{\a}^{i}S_{ij}+\cD^{\b}_{j}Y_{\b\a}=0~.
\label{3/2-4D}
\eea

(iii)
Since the action on spinor and isospinor indices
of the 3D generators $\cM_{\a\b}$ and $\bR_{\bai\baj}$, see eqs. (\ref{acM}) and (\ref{acL-R}), 
are formally equivalent to the ones of 
$M_{\a\b}$ and $J_{ij}$, the 3D 
$\{\cD_\a^{(1)\bai},\cD_\b^{(1)\baj}\}$ algebra in (\ref{alg+3D}) becomes equivalent to
the $\{\cD_\a^{i},\cD_\b^{j}\}$  algebra in (\ref{alg-4D}) if we identify
\bea
\cD_\a^i\leftrightarrow \cD_\a^{(1)\bai}~,~~~~~~
S_{ij}\leftrightarrow -\ri \cS^{(2)}_{\bai\baj}~,~~~~~~
Y_{\a\b}\leftrightarrow \ri B^{(2)}_{\a\b}~.
\label{trick1}
\eea
This correspondence holds also at higher mass-dimensions due 
 to the 3D  dimension-3/2 Bianchi identities
\bea
\cD^{(1)}_{\a(\bai }\cS^{(2)}_{\baj\bak)}=\cD_{(\a}^{(1)\bai}B^{(2)}_{\b\g)}=0~,~~~~~~
-\cD^{(1)\baj}_{\a}\cS^{(2)}_{\bai\baj}+\cD^{(1)\b}_\bai B^{(2)}_{\a\b}=0
~.
\eea

(iv)
The antichiral projector  in 4D $\cN=2$ supergravity
\cite{Muller,KT-M2} is
\bsubeq
\bea
\D^4&=&\frac{1}{96}\Big(
(\cD^{ij}+16S^{ij})\cD_{ij}
-(\cD^{\a\b}-16Y^{\a\b})\cD_{\a\b}
\Big)
~,
\label{chiralP1}
\\
&=&\frac{1}{96}\Big(
\cD^{ij}(\cD_{ij}+16S_{ij})
-\cD^{\a\b}(\cD_{\a\b}-16Y_{\a\b})
\Big)
~,
\label{chiralP2}
\eea
\esubeq
with
\bea
\cD_{ij}:=\cD^\g_{(i}\cD_{\g j)}~,~~~
\cD_{\a\b}:=\cD^k_{(\a}\cD_{\b) k}
~.
\eea
${}$From the previous discussion it  follows that we may now find the left projection operator
(\ref{6.54a}), (\ref{6.54b})  using the identifications (\ref{trick1}) in (\ref{chiralP1}), 
(\ref{chiralP2}).

Another important property of the left projection operator is (\ref{sWU-L})--(\ref{sWQ-L-2}).
To show those equations
we use the super-Weyl transformation rules of the $\cD_\a^{(1)\bai}$
derivatives (\ref{sWD1-L}) along with those of the dimension-1 superfields $\cS^{(2)\bai\baj}$ and
$B_{\a\b}^{(2)}$, as well as the transformation rules  of the following 
dimension-3/2 superfield $\cT_\a^{(3)\bai}$
\bsubeq
\bea
\cT_\a^{(3)\bai}&:=&v_iv_jv_k\cT_\a^{ijk\bai}=
-\frac{1}{3}\cD^{(1)}_{\a \baj} \cS^{(2)\bai\baj}
=\frac{1}{3}\cD^{(1)\b\bai}B_{\a\b}^{(2)}
~,
\\
\cD_\a^{(1)\bai} \cS^{(2)\baj\bak}&=&
2\cT_\a^{(3)(\baj}\ve^{\bak)\bai}
~,~~~
\cD_{\a}^{(1)\bai}B_{\b\g}^{(2)}
=
2\ve_{\a(\b}\cT_{\g)}{}^{(3)\bai}
~.
\eea
\esubeq
We have
\bea
\d_\s \cS^{(2)}_{\bai\baj}&=&
\s\cS^{(2)}_{\bai\baj}
-{\ri\over 4}(\cD^{(2)}_{\bai\baj}\s)
~,~~~~
\d_\s B_{\a\b}^{(2)}=
\s B_{\a\b}^{(2)}
+{\ri\over 4}(\cD_{\a\b}^{(2)}\s)
~,~~~
\eea
and
\bsubeq
\bea
\d_\s\cT_\a^{(3)\bai}&=&
\frac{3}{2}\s\cT_\a^{(3)\bai}
+\frac{\ri}{12}(\cD_{\a \baj}^{(1)} \cD^{(2)\bai\baj}\s)
-\frac{2}{3}\cS^{(2)\bai\baj}(\cD_{\a \baj}^{(1)} \s)
~,
\\
&=&
\frac{3}{2}\s\cT_\a^{(3)\bai}
+\frac{\ri}{12}(\cD^{(1) \b \bai}\cD^{(2)}_{\a\b}\s)
-\frac{2}{3}B^{(2)}_{\a\b}(\cD^{(1)\b \bai}\s)
~.
\eea
\esubeq
To check eq. (\ref{sWQ-L-2}), 
the reader may also use the following equation
\bea
\cD_\a^{(1)\baj}\cD^{(2)}_{\bai\baj}U_\rL^{(n-4)}&=&
\Big(-\cD^{(1)\b}_{\bai}\cD^{(2)}_{\a\b}
-8\ri \cS^{(2)}_{\bai\baj}\cD_{\a}^{(1) \baj}
-8\ri B^{(2)}_{\a\b}\cD^{(1)\b}_{\bak}\Big)U_\rL^{(n-4)}
~,
\eea
together with (\ref{four-four-L}).

We conclude this appendix by proving that the right-hand side of (\ref{proj2L}) is:
(i)  independent of the isospinors $u_\rR = u^\bai$;
and (ii) obeys the left analyticity constraint (\ref{ana-L}).
The derivation is completely analogous to the 4D $\cN=2$ analysis given in appendix C of
\cite{KT-M2}. It is instructive, however, to repeat the computation in the 3D $\cN=4$ case.

To prove the independence of 
(\ref{proj2L}) from $u_\bai$
it is sufficient to prove its invariance under infinitesimal projective transformations
of the form 
\be
u_\bai~\to~ u_\bai +\d u_\bai~, \qquad
\d u_\bai\,=\,\a(t)\,u_\bai+\b(t)\,v_\bai(t)~.
\label{delta-u-}
\ee
Here the time $t$ is the integration variable of the contour integral.
Since both $u_\bai$ and $\d {u}_\bai$ are required to  be time-independent, 
the transformation parameters  should obey the equations:
\bea
\dt{\a}=\b\,{(\dt{v}_\rR,u_\rR)\over (v_\rR,v_\rR)}~, \qquad 
\dt{\b}=-\b\,{(\dt{v}_\rR,u_\rR)\over (v_\rR,v_\rR)}~.
\label{ode}
\eea
Equation (\ref{proj2L})
is manifestly invariant 
under the $\a$-transformations. 
It remains to check invariance under $\b$-transformations 
(\ref{delta-u-}).
Applying the $\b$-transformation gives
\bea
&&\d 
\Big( \cD^{(2,-2)} -4\ri\cS^{(2,-2)}\Big) 
\D^{(2,2)} U^{(n,-2)}
=-
\frac{16\b}{(v_\rR,u_\rR)}{\bm\pa}_\rR^{(-2)}
\cS^{(2,2)}
\D^{(2,2)} U^{(n,-2)}
~.~~~~~~~~~
\label{D1}
\eea
${}$From \cite{KT-M2}
\bea
\b\frac{(\dt{v}_\rR, v_\rR)}{(v_\rR,u_\rR)}{\bm\pa}_\rR^{(-2)}V^{(n+4,2)}&=&
-{\rd\over\rd t}\Big({b\over (v_\rR,u_\rR)}
V^{(n+4,2)}\Big)
~,
\eea
 for any isotwistor superfield $V^{(n+4,2)}$ of weight $(n+4,2)$,  
such as
$(\cS^{(2,2)}\D^{(2,2)} U^{(n,-2)})$
appearing in (\ref{D1}).
Using this we find that  the right hand side of (\ref{proj2L}) is independent of $u_\rR$.

Now let us prove that the right hand side of (\ref{proj2L}) obeys the left analyticity 
constraint (\ref{ana-L}).
{}First of all, consider a
weight-$(n+2,0)$  hybrid superfield $P^{(n+2,0)} (z,v_\rL,v_\rR)$,
as for example the superfield 
$\D^{(2,2)} U^{(n,-2)}$.
Using the identities
\bsubeq
\bea
\cD_\a^{(1,-1)}\Big(\cD^{(2,-2)}-4\ri \cS^{(2,-{2}) }\Big)P^{(n+2,0)}&=&
4\ri(v_\rR,u_\rR)B_{\a\b}^{ ({2}) } \cD^{\b(1,-1)} {\bm\pa}^{-2}_\rR P^{(n+2,0)}
\non\\
&&
+2\ri(v_\rR,u_\rR)(\cD^{(1,1)}\cS^{ ({2},-2) }) {\bm\pa}^{-2}_\rR P^{(n+2,0)}
~,~~~~~~
\eea\bea
{[}\cD_\a^{(1,1)},\cD^{(-2,2)}{]}P^{({n+2},0)}&=&
\Big(
-4\ri B_{\a\b}^{(2)}\cD^{\b(1,-1)}
-4\ri (\cD_\a^{(1,-1)}\cS^{(2,0)})
\non\\
&&
+{\bm\pa}_\rR^{(-2)}\Big(4\ri\cS^{(2,2)}\cD_\a^{(1,-1)}
+2\ri(\cD_\a^{(1,-1)}\cS^{(2,2)})\Big)
\Big)P^{({m+2},0)}
~, ~~~~~~~~~
\eea
\esubeq
one can show that 
\bea
\cD_{\a \bak}^{(1)}\Big( \cD^{(2,-2)} -4\ri\cS^{(2,-2)}\Big) 
P^{({n+2},0)}&=&
-{u_\bak\over(v_\rR,u_\rR)}\cD_\a^{(1,1)}\Big( \cD^{(2,-2)} -4\ri\cS^{(2,-2)}\Big) P^{({n+2},0)}
\non\\
&&
+v_\bak\cD_\a^{(1,-1)}\Big( \cD^{(2,-2)} -4\ri\cS^{(2,-2)}\Big) P^{({n+2},0)}
\non\\
\cD_{\a \bak}^{(1)}\Big( \cD^{(2,-2)} -4\ri\cS^{(2,-2)}\Big) 
P^{({n+2},0)}&=&
2\ri{\bm\pa}_\rR^{(-2)}
\Big{[}
\frac{u_\bak}{(v_\rR,u_\rR)}\Big(-2\cS^{(2,2)}\cD_\a^{(1,-1)}
-(\cD_\a^{(1,-1)}\cS^{(2,2)})\Big)
\non\\
&&
+ v_k\Big(
2 B_{\a\b}^{ ({2}) } \cD^{\b(1,-1)}
+(\cD_{\a}^{(1,1)}\cS^{(2,-2)}) 
 \Big) 
\Big{]}
P^{(n+2,0)}
~.~~~~~~~~~
\label{D.8}
\eea
It is shown in  \cite{KT-M2}  that the right-hand side of the last equation is zero when integrated
over a closed contour.
Thus we have shown that 
\bea
&&
\cD_{\a \bak}^{(1)}\,\frac{1}{16}  \oint (v_\rR, \rd v_\rR)
\Big( \cD^{(2,-2)} -4\ri\cS^{(2,-2)}\Big) 
\D^{(2,2)}
U^{({m},0)}=0
~.
\eea
As a result,  the right hand side of (\ref{proj2L}) indeed obeys the left analyticity 
constraint (\ref{ana-L}).


\begin{footnotesize}

\end{footnotesize}

\end{document}